\documentclass[twocolumn]{aastex631}
\usepackage{amsmath}
\usepackage{amssymb}
\usepackage{threeparttable}
\usepackage{CJK}

\def\rs{r_{\rm S}}
\def\rsi{R_{\rm SI}}
\def\rp{r_{\rm P}}

\def\kappas{\kappa_{s}}
\def\kappar{\kappa_{R}}
\def\kappap{\kappa_{P}}
\def\medd{\dot{M}_{\rm Edd}}
\def\mfb{\dot{M}_{\rm fb}}
\def\macc{\dot{M}_{\rm acc}}
\def\mout{\dot{M}_{\rm outflow}}
\def\alpharey{\alpha_{\rm Re}}

\begin{document}
\begin{CJK*}{UTF8}{gbsn}
\title{X-ray Variability and Photosphere Evolution during Accretion Disk Formation in Tidal Disruption Events}
\author[0000-0003-2868-489X]{Xiaoshan Huang (黄小珊)}
\affiliation{California Institute of Technology, TAPIR, Pasadena, CA 91125, USA}
\author[0009-0002-8615-6403]{Maria Renee Meza}
\affiliation{Department of Astronomy, University of Virginia, 530 McCormick Road, Charlottesville, VA 22904, USA}
\author[0009-0000-4440-155X]{Sol Bin Yun}
\affiliation{Department of Astronomy, California Institute of Technology,  Pasadena, CA 91125, USA}
\author[0000-0001-6350-8168]{Brenna Mockler}
\affiliation{The Observatories of the Carnegie Institution for Science, Pasadena, CA 91101, USA}
\affiliation{Department of Physics and Astronomy, University of California, Davis, CA 95616, USA}
\author[0000-0001-7488-4468]{Shane W. Davis}
\affiliation{Department of Astronomy, University of Virginia, 530 McCormick Road, Charlottesville, VA 22904, USA}
\affiliation{Virginia Institute for Theoretical Astronomy, University of Virginia, Charlottesville, VA 22904, USA}
\author[0000-0002-2624-3399]{Yan-fei Jiang (姜燕飞)}
\affiliation{Center for Computational Astrophysics, Flatiron Institute, 162 Fifth Avenue, New York, NY 10010, USA}

\begin{abstract}
The early time emission in tidal disruption events (TDEs) originates from both accretion and shocks, producing photons that eventually emerge from an inhomogeneous photosphere. We model disk formation following debris stream self-intersection in a TDE using three-dimensional, frequency-integrated and multi-group radiation hydrodynamic simulations. We find a more circularized disk forms about 24 days following the stream-stream collision, once the mass fallback rate passes its peak and the debris stream density decreases. Despite the absence of a circularized disk at early times, various shocks and the asymmetric photosphere are sufficient to drive a wide range of x-ray-to-optical ratios and soft-X-ray variability. We find that with strong apsidal precession, the first light is from the stream-stream collision. It launches an optically-thick outflow, but only produces modest prompt emission. The subsequent optical and ultraviolet (UV) light curve rise is mainly powered by shocks in the turbulent accretion flow close to the black hole. The optical-UV luminosity peaks roughly when the disk forms and shock-driven outflows subside. Radiation pressure clears the polar region and creates optically-thin channels. We obtain the broadband spectral energy distribution (SED) by post-processing multi-group simulations with 16-20 frequency groups. The SED has a black body component that peaks in the extreme UV. The soft X-ray component either resembles a thermal tail, or can be described by a power law associated with bulk Compton scattering. The blackbody parameters are broadly consistent with observed optical TDEs and vary weakly with viewing angle, but the soft X-ray emission is highly angle-dependent.
\end{abstract}

\section{Introduction}\label{sec:introduction}

Stars are occasionally scattered to the galactic center that hosts a supermassive black hole, where the black hole's strong tidal force can fully or partially disrupt the star. Accretion occurs when the stellar debris returns to the black hole from the orbital apocenter, producing transient flares that peak on the timescales of weeks to months and last for several years \citep{rees1988tidal,phinney1989manifestations}. Tidal disruption events (TDEs) are primarily discovered in optical and X-ray surveys. Compared to the classic theory that predicts soft X-ray emission from a compact disk assembled by low angular momentum stellar debris, the optical TDEs show surprisingly large photosphere sizes $R_{\rm BB}\sim10^{14-15}$ cm and low photosphere temperatures $T_{\rm BB}\sim10^{4}$K \citep{gezari2021tidal}. This dependency has sometimes been described as X-ray and optical emission in TDEs are rarely simultaneous.  

Recent observations reveal that TDEs are multi-wavelength emitters. More TDEs have been observed in optical-ultraviolet(UV) \citep{van2011optical,holoien2019ps18kh, nicholl2020outflow,hinkle2021discovery, van2021seventeen,hammerstein2022final,yao2023tidal} and X-rays \citep{saxton2017xmmsl1, saxton2020x, yao2022tidal, guolo2023systematic, ho2025luminous, hajela2025eight, wu2025early, lin2025delayed}. Recent works also find $\sim40\%$ of optically selected candidates show delayed radio emission \citep{alexander2020radio, goodwin2023radio, somalwar2023vlass,cendes2024ubiquitous, goodwin2025systematic}. Some TDEs that are partially or largely obscured by dust can lead to infrared emission \citep{jiang2021mid, dodd2023mid, masterson2024new, lin2025insights}. Notably, recent works suggest that a few TDEs have an ``optical precursor''. Their optical light curves show a ``bump" before the main optical peak. The separation of such precursors and main optical peak in AT2019azh \citep{liu2022uv,faris2024light}, AT2023lli \citep{huang20242023lli} and AT2024kmq \citep{ho2025luminous} are all on the order of a few weeks. These optical bumps are discussed as a signal from shocks prior to disk formation, hinting at an emission source other than direct accretion.

Several theoretical pictures are proposed to explain the multi-band emissions. For example, TDEs can be modeled as a central emitter that is covered by an optically thick reprocessing layer \citep{strubbe2009optical,metzger2016bright,roth2016x}. Variations in the emitting source and the photosphere can introduce a wide range of X-ray-to-optical ratios. A natural central emitter is the accretion disk formed from the stellar debris, and various processes are discussed to generate the optically thick layer. For example, the TDE disk accreting at a super-Eddington rate itself is optically and geometrically thick, often including optically thin funnel regions \citep{roth2016x, dai2018unified}. Radiation pressure also drives disk winds to further extend the photosphere. High energy photons are produced in the hot inner disk from accretion, those that diffuse through the wind result in optical-UV emission, and those that propagate through the funnel give rise to X-ray emission, leading to angular-dependent spectral energy distribution (SED). Recent radiation transfer simulations support such viewing-angle models and can reproduce optical bright or X-ray bright TDEs by changing lines of sight \citep{thomsen2022dynamical,parkinson2025multidimensional,qiao2025early,mockler2026light}. 

Other optical-UV emission mechanisms do not require immediate formation of a compact accretion disk, arising instead from how the stellar debris feeds the black hole. For example, when the debris stream passes the pericenter, apsidal precession can lead to stream self-intersection \citep[e.g.][]{dai2015soft, piran2015disk, lu2020self}. When the apsidal precession is weak, the stream self-intersection occurs near the apocenter $\rsi\sim10^{2-3}\rs$ and is sometimes referred to as ``apocenter shock'' \citep[e.g.]{guillochon2014ps1, piran2015disk, shiokawa2015general, bonnerot2016disc, ryu2023shocks}. The shock continuously powers optical-UV emission, dissipating energy and leading to debris circularization. In contrast, when the apsidal precession is strong, the stream self-intersection is instead close to the black hole near $\rsi\lesssim100\rs$, leading to strong ``stream-stream collisions'' that drive significant outflows \citep{dai2015soft, jiang2016prompt, lu2020self, piro2020wind, bonnerot2021first}. The outflows are optically thick and can reprocess hot post-shock photons generated near the black hole to the optical-UV bands. 

Alternatively, regardless of the initial dissipation processes such as stream self-intersection, with nearly zero orbital energy, the stellar debris may undergo a phase where they form a quasi-spherical structure around the black hole when cooling is inefficient. The debris can settle into a quasi-isotropic envelope \citep{metzger2022cooling,price2024eddington} or quasi-spherical accretion flow \citep[][“Zero-Bernoulli accretion flows”]{coughlin2014hyperaccretion, wu2018super, eyles2022simulated} instead of directly arriving at the black hole. The nozzle shock is another frequently-discussed but yet debated emission source. It originates from the vertical compression of the debris stream at the pericenter, which can dissipate energy and substantially expand the stream \citep{rosswog2009tidal,shiokawa2015general,sadowski2016magnetohydrodynamical,liptai2019disc,curd2021global,andalman2022tidal}. However, recent theoretical works suggest that the compression is nearly reversible, so the dissipation is minimal \citep{bonnerot2022nozzle}. Simulation works found numerical dissipation can be predominant when the debris stream is under-resolved \citep[e.g. see resolution study][]{steinberg2022origins,huang2024pre,hu2025converged,kubli2025tidal}. Hydrogen recombination are also discussed to expand the stream \citep{kochanek1994aftermath, kasen2010optical,steinberg2022origins,coughlin2023dynamics, andalman2025resolving}, adopting a more realistic equation of state is important to model the debris stream evolution.

Despite the complex early dynamics, modeling of X-ray and UV observations indicates that an accretion disk nonetheless forms in many TDEs \citep[e.g.][]{auchettl2017new,jonker2020implications,yao2024subrelativistic,mummery2025optical}. Canonically, the X-ray emissions originate from the inner disk, therefore, their detections are often thought to mark the formation of an accretion disk \citep[e.g.][]{lodato2011multiband, dai2018unified}. Indeed, many TDEs show X-ray emission after the optical peak, and these delayed late-time X-ray detections reveal accretion disk evolution \citep{shen2014evolution,miller2015disk,piro2025late,alush2025late}. However, for a subset of optically discovered TDEs, the X-ray emission is detected near the optical peak \citep[e.g.][]{auchettl2017new, guolo2023systematic, malyali2023transient, cao2024tidal}. The first X-ray detection can be closer to the optical peak than the viscous time (with an assumed typical disk $\alpha\sim0.1$ at circularization radius). Such near-peak X-ray emission raises the question of how rapidly the debris can arrive at the black hole and form a circularized disk, which is still a debated topic \citep[e.g.][]{steinberg2022origins,ryu2023shocks}. In parallel, recent works propose that shocks, instead of accretion, play a central role in producing early optical-UV light curves and shaping the dynamics \citep{steinberg2022origins,ryu2023shocks,curd2025jet,meza2025radiation}. These previous works and discussions motivate us to study multi-band emission and its dependence on viewing angle during TDE disk formation. Here we further show that if accretion is the secondary process for the early optical-UV emission, it can be subdominant relative to various shocks in near-peak X-ray emission too. 

In this work, we focus on the disk formation process following the debris stream self-intersection around a spin-less $M_{\rm BH}=3\times10^{6}M_{\odot}$ with an impact parameter $\beta=1.73$. The debris stream self-intersection occurs at $\rsi=76.2\rs$, which is significantly smaller than the apocenter radius of the assumed stellar debris orbit $r_{\rm apo}\sim10^{3}\rs$. This differs from an ``apocenter shock'' scenario for smaller black hole $M_{\rm BH}\sim10^{4-6}M_{\odot}$ or shallower events $\beta\sim1$, where the $\rsi\sim r_{\rm apo}\sim10^{2-3}\rs$ \citep[e.g.][]{piran2015disk,steinberg2022origins,ryu2023shocks,martire2025wind}. Both scenarios originate from the apsidal precession, with important differences in the collision dynamics. For example,  when $\rsi\ll r_{\rm apo}$ like in this work, the collision velocity is on the order of $v\sim0.1c$. The returning stream also expands less and maintains a high density when colliding with fallback stream. These lead to more violent collisions \citep[e.g.][]{lu2020self,bonnerot2021first,huang2023bright}. The post-shock gas temperature reaches $T\sim10^{5}$K, which is hotter than the photosphere temperatures of optical TDEs due to the larger kinetic energy reservoir. Many previous works suggest that the collision also drives a strong outflow or wind that can reprocess the hotter downstream photons to lower energy, yielding photosphere temperatures consistent with observations \citep[e.g.][]{lu2020self,piro2020wind}. 

However, for the ``apocenter shock'' scenario where $\rsi\sim r_{\rm apo}$, the collision velocity is on the order of $v\sim0.01c$. The post-collision gas and radiation temperature $T\sim10^{4}$K matches photosphere temperatures of optical TDEs. The large $\rsi$ also naturally yield a photosphere size comparable to optical TDEs \citep{piran2015disk}, so these apocenter shocks can potentially dominate the optical emission. Meanwhile, if the post-collision gas could arrive at black hole on the free fall timescale, delaying the disk formation \citep{dai2015soft, wong2022revisiting, guo2025reverberation}. Numerical simulation results support this picture and find that the stream self-intersection shock and subsequent stream-disk shock create optical-UV emission region between the circularization radius and the apocenter \citep{steinberg2022origins,ryu2023shocks}

In the rest of the paper, we introduce the numerical setup in Section~\ref{sec:method_set-up} and Section~\ref{subsec:mg_setup}, where we discuss the prescribed debris stream fallback rate (as well as limitations of our idealized initialization). Section~\ref{sec:result} describes the overall dynamics from the frequency-integrated radiation hydrodynamic (RHD) simulation. We also show results from a series of short post-processing multi-group RHD simulations that derived from frequency-integrated simulation in Section~\ref{sec:mg}, which focus on band-dependent light curves, broad band SEDs and the viewing angle dependency. We discuss the role of shocks, the photosphere properties, the soft X-ray origin and variability in Section~\ref{sec:discussion}. We also compare our simulations with previous works and discuss caveats in Section~\ref{sec:discussion}. We highlight several comparisons between light curves derived from simulations and from observation in Appendix~\ref{appendix:observation_comparison}. We discuss cases where simulation can roughly reproduce the optical-UV color evolution and X-ray light curve (AT2018hyz, AT2019azh, AT2020upj) and cases where simulation cannot explain the early multi-band light curve (AT2019dsg, AT2019qiz).

\section{Simulation Set-up}\label{sec:method_set-up}

This work includes a long-term, frequency-integrated radiation hydrodynamics (RHD) simulation and a series of short-term multi-group radiation hydrodynamics simulations. Hereafter, we refer to the long-term simulation as the ``gray'' simulation, and the short-term multi-group RHD simulations as ``multi-group'' simulations. 

We assume that a solar type star ($M_{*}=M_{\odot},~R_{*}=R_{\odot}$) is disrupted by a black hole with mass $M_{\rm BH}=3\times10^{6}M_{\odot}$. So that the tidal radius $r_{\rm T}=R_{*}(M_{\rm BH}/M_{*})^{1/3}\approx11.3\rs$. We adopt impact parameter $\beta=r_{\rm T}/r_{\rm P}=1.73$. We do not model the disruption of the star, but instead inject a thin, cold stream with fixed angular momentum from the simulation domain boundary to approximate the fallback debris stream. We obtain the ballistic trajectories of the debris using the general Newtonian potential \citep{tejeda2013accurate}, which sets the angular momentum of injected stream. This injection method is generally similar to our previous method in \citet{huang2023bright}, however, we allow the mass fallback rate to vary over time in this work. The mass fallback rate is informed by STARS \citep{law2020stellar}, which is used to set the gas density of the injected stream. The numerical method is elaborated in the following sections.

\subsection{Radiation Hydrodynamics, Opacity, and Scaling}\label{subsec:method_rhd}
We solve the radiation hydrodynamic equations in Athena++ with the explicit, multi-group radiation transfer module. The treatment of angular discretization, frequency-dependent radiation transfer, and gas-radiation coupling are introduced in \citet{jiang2021implicit,jiang2022multigroup}. In the code, we solve the unit-less equations with the scaling of density unit $\rho_{0}=10^{-10}\rm g~cm^{-3}$, velocity unit $v_{0}=0.0005c$ and length unit $l_{0}=r_{\rm s}=2GM_{\rm BH}/c^2$. In the rest of the paper, we report quantities in c.g.s units unless otherwise explicitly specified. The angular resolution for the radiation field is set to 80 angles for the total $4\pi$ solid angles. 

In the gray simulation, we adopt the frequency-integrated opacity from \citet{zhu2021global}, which extends the OPAL Opacity Table \citep{iglesias1996updated} to temperature $\sim10^{3}$K by including dust opacity. For multi-group simulations, we adopt multi-group opacity from TOPS Opacity database \citep{colgan2016new}. The Rosseland mean $\kappar$ and Planck mean opacity $\kappap$ from TOPS are averaged over frequency within each photon frequency group. The returned Rosseland mean opacity is the sum of Rosseland mean opacity and electron scattering opacity $\kappar+\kappa_{\rm es}$. We approximate electron scattering opacity by Thompson opacity $\kappa_{\rm es}=\kappa_{\rm Th}$, and subtract it from the total returned Rosseland mean opacity to calculate $\kappar$. 

In the simulations, the opacity of each cell is based on the OPAL or TOPS opacity data and linearly interpolated in the temperature and density grid. We tested extended or restricted density grids $10^{-17\pm2}-10^{-6\pm2}\,\rm g~cm^{-3}$ and did not find a significant impact. As will be elaborated in Section~\ref{sec:mg}, a subset of multi-group simulations adopts a larger photon frequency grid to include higher energy photons from the inner accretion disk after the disk formation.

In the low density region, such as gas with density $\rho\lesssim10^{-14}\rm g~cm^{-3}$, the local thermal equilibrium (LTE) assumption may not be valid.  The tabulated opacity lacks data in both the high temperature end $T\gtrsim10^{5}$K and low temperature end $T\lesssim10^{4}$K. For the high temperature, we replace the absorption opacity by free-free opacity. For the low temperature and low density ambient gas, we simply use the smaller value between the free-free opacity and the first available temperature point for the density. Our approach may overestimate the absorption opacity, thus risking artificially coupling of the ambient gas and radiation. Adopting more realistic non-LTE opacity will be explored in future works \citep[e.g.][]{mockler2026light}.

\subsection{Initialize Stream, Resolution, and Boundary Conditions}\label{subsec:method_init}

The simulations are performed in spherical polar coordinates. The calculation domain is $(2.7\rs,~400.0\rs)\times(0,\pi)\times(0,2\pi)$ in R-, $\theta$- $\phi$-direction, the R-direction grid is logarithmically spaced. The lowest resolution in each direction is $64\times32\times64$. We add six levels of adaptive mesh refinement (AMR) to follow the high gas density regions and the regions near the black hole. The refinement scheme yields $\Delta R\sim R\Delta\theta\approx0.01\rs$ when the stream first passes pericenter, equivalent to $\Delta R\approx0.1R_{\odot}$. 

We test maximum refinement level ranges from four to six and find a similar dependency as in our previous works \citep{huang2024pre, meza2025radiation}. Here we reiterate the following relevant findings: first, the stream width is still resolution limited in the vertical ($\theta$-) direction and transverse direction, considering the stream can be significantly compressed near the pericenter \citep{bonnerot2022pericenter, bonnerot2022nozzle, coughlin2023dynamics}. Second, resolving the vertical optical depth of the stream is essential to converge the pericenter stream thermal structure during the stream's first pericenter passage. In this work, we allow the fallback rate to vary over time, which introduces a range of gas density at the pericenter. The optical depth depends on stream density, therefore, we further increase the maximum resolution compare to \citet{huang2024pre} to ensure the convergence of the radiation and thermal structure. We also refer the readers to recent works \citet{hu2025converged, kubli2025tidal} on spatial resolution's impact on pericenter stream dissipation, where smooth particle codes found $10^{10-12}$ particles are required to reach convergence in stream width. \citet{nixon2026origin} showed that the numerical dissipation is associated with the velocity field distribution near pericenter, and is sensitive to numerical treatment of viscosity. Exploring the convergence with a Eulerian code like Athena++ will be a goal of future work. 

We approximate the stellar debris stream as a thin, cold stream injected from the simulation domain boundary with constant angular momentum. In each simulation, the stream is injected at $r=r_{\rm inj}=400\,\rs$, $\phi=\phi_{\rm inj}$ and $\theta_{\rm inj}=\pi/2$ at the boundary. The stream temperature is assumed to be $T_{\rm inj}=10^{4}$ K. The energy density of the stream is dominated by kinetic energy, internal energy and radiation energy are orders of magnitude smaller when injected. We set the density injection rate $\rho_{\rm inj}$ and velocity $\textbf{v}_{\rm inj}=(v_{\rm r,inj}, ~0, ~v_{\rm \phi, inj})$ of the ghost cells in the injection region, which is defined as the neighboring four cells of $\phi_{\rm inj}$ and $\theta=\pi/2$ in $\phi-$ and $\theta-$ directions. We set the injection region cells to be identical in density $\rho_{\rm inj}$ and velocity $\textbf{v}_{\rm inj}$. After entering the simulation domain, the uniformly injected stream will be compressed by the tidal force of the black hole and develop a non-uniform density and velocity structure. 

With such implementation, we do not capture the angular momentum distribution and vertical velocity field of a realistic debris stream, which affects the nozzle shock and gas distribution soon after the first pericenter passage \citep[e.g.][]{kochanek1994aftermath, coughlin2016structure, bonnerot2022pericenter, steinberg2022origins, ryu2023shocks, martire2025wind}. 

There are two main effects that introduce deviations from injecting a stream with fixed location and angular momentum. First, when the star is disrupted near its pericenter orbit, its physical size leads to a spread in orbital energy, but our chosen injecting stream only represents the mass center of the star. To assess the impact of this assumption we ballistically evolved a set of bound trajectories. These orbits share the same eccentricity but have an orbital energy spread corresponding to the size of $R_{\odot}$ at pericenter. The spread of location at the injection point is about $\phi_{\rm inj,spread}\approx\pi/2048$, which is smaller than grid resolution in $\Delta\phi=\pi/1024$. But near the self-intersection radius, the spread in initial orbital energy introduces a range of $\Delta R_{\rm SI, spread}\approx7.6\rs$. Second, the bound debris gains a spread of angular momentum during disruption, and they will arrive at the injection radius at slightly different $\phi$ and time, which is also neglected with the simplified injection method in this work. By performing an additional stellar disruption simulation that is similar to \citet{guillochon2014ps1} with parameters matching this work, we estimate the shift of injection radius at is about $R_{\rm inj,spread}\approx2.9\rs$, corresponding to an angular size that would be resolved by only two injection cells. Obtaining more self-consistent initial debris stream properties that are informed by stellar disruption and debris fallback process will be important in future work.

We set the injection velocity $(v_{\rm r,inj}, ~0, ~v_{\rm \phi, inj})$ and injection cells location $(r_{\rm inj}, \pi/2.0, \phi_{\rm inj})$ based on ballistic trajectories from integrating the equations of motion using the general Newtonian potential in \citet{tejeda2013accurate}. The approximated potential is specialized in capturing the apsidal precession for a spin-less black hole, and there is zero vertical offset of returning and fallback stream. We assume the stream represents the most bound material, and estimate the orbit eccentricity $e_{\rm orb}$ as in \citet{dai2015soft}. The orbit integration initial velocity $(0.0, 0.0, v_{\phi, 0})$ and location $(r_{0}, \pi/2.0, 0.0)$ are approximated by a Keplerian orbit with $r_{0}=\rp(1+e_{\rm orb})/(1-e_{\rm orb})$ and $v_{\phi,0}=\sqrt{GM_{\rm BH}/r_{0}}$ at apocenter. This yields the injection location and velocity $r_{\rm inj}=400.0\rs,~\phi_{\rm inj}=0.22$, $\textbf{v}_{\rm inj}=(-0.043c,~0,~0.0064c)$. With the prescribed stream orbit and accounting for apsidal precession, the pericenter radius is $r_{\rm p}\approx5.07\rs$ and circularization radius is $r_{\rm circ}=(1+e_{\rm orb})r_{\rm p}\approx2r_{\rm p}=10.1\rs$.

The low density background gas in the simulation domain is initially set to $\rho_{\rm init}=2\times10^{-9}$ and $P_{\rm init}'=5\times10^{-12}$ as unit-less code variables. We set the density and pressure floor for the hydrodynamic Riemann solver to be $\rho_{\rm floor}'=2\times10^{-9}$ and $P_{\rm floor}'=2\times10^{-12}$. The temperature floor for the radiation transfer module is set to be $T_{\rm floor}'=2\times10^{-3}$.

The fallback rate $\dot{M}_{\rm fb}$ sets the total mass flux that is injected into the simulation domain, so that $\rho_{\rm inj}=\dot{M}_{\rm fb}/(\mathbf{v}_{\rm inj}\cdot\mathbf{A}_{\rm inj})$, where $\textbf{A}_{\rm inj}$ is the total injection area. We retrieve the mass fallback rate from stellar disruption simulations in STARs \citet{law2020stellar} with corresponding stellar mass, radius, and impact parameter. However, we do not interpolate in black hole mass. STARS assumes a $10^{6}M_{\odot}$ black hole, which is slightly lower than what we assumed $M_{\rm BH}=3\times10^{6}M_{\odot}$. The fallback rate from disruption by a smaller mass black hole is motivated by both faster $\dot{M}_{\rm fb}$ evolution to save computation cost, and avoiding black hole mass interpolation. For further demonstration, we compare mass fallback rates from STARS for different black hole masses, and also compare them with mass fallback rate from the MOSFiT backend \citep{mockler2019weighing} in Appendix~\ref{appendix:mdotfallback}. STARS assumes stellar profiles informed by MESA, while MOSFiT assumes polytropic stellar structures. In Figure~\ref{fig:mdotfb_STARS_mosfit} Appendix\ref{appendix:mdotfallback}, they show similar mass fallback rates. We estimate that the main effect is an artificial speed-up of the fallback rate evolution, but the effect may be moderate. The best-fit Gaussian for the mass fallback rate rise is 
$\dot{M}_{\rm fb}\approx\dot{M}_{\rm fb, peak}\exp(-(t-t_{\rm peak})^{2}/2t_{\rm rise}^{2})$, with $\dot{M}_{\rm fb, peak}=2.15M_{\odot}~\rm yr^{-1}$ and $t_{\rm rise}=10.4$ days.

We set the $\theta$ direction boundary to be outflow, the $\phi$ direction boundary condition is periodic. The $r$ direction boundaries are set to be single-direction outflow for hydrodynamical variables, which copies all the values from the first active cells but sets any velocity that enters the calculation domain to zero. The radiation boundaries in the $r$ direction is ``vacuum'' radiation boundaries, which copy all the intensities with $\textbf{n}$ pointing outward, but set all intensities with $\textbf{n}$ pointing inward to zero.

\section{Overall Dynamics}\label{sec:result}

\begin{figure*}
    \centering
    \includegraphics[width=0.9\linewidth]{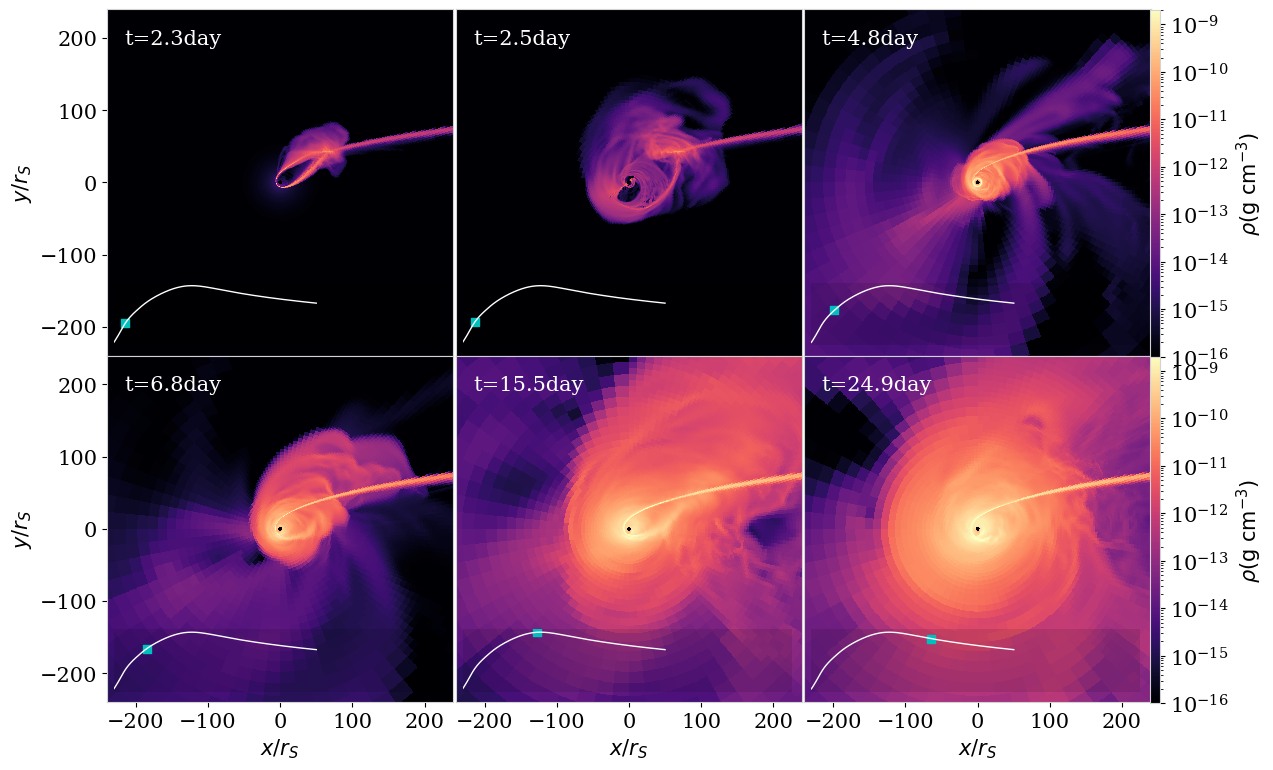}
    \caption{Snapshots at the orbital-plane, which show gas density averaged for  $\theta=90^{\circ}\pm10^{\circ}$. From left to right, and top to bottom, the times are t=2.3, 2.5, 4.8, 6.8, 15.5 and 24.9 days from the beginning of simulation, indicated by the white text in each panel. The stream-stream collision happens at around t=2.2 days. The orbital period at the injected stream pericenter radius is  $P_{\rm peri}\approx0.88$ hours. In each panel, the white curve in the inset plot shows the shape of the fallback rate, and the blue marker labels the current time.}
    \label{fig:timeseries_dens_horizontal}
\end{figure*}

\begin{figure}
    \centering
    \includegraphics[width=\linewidth]{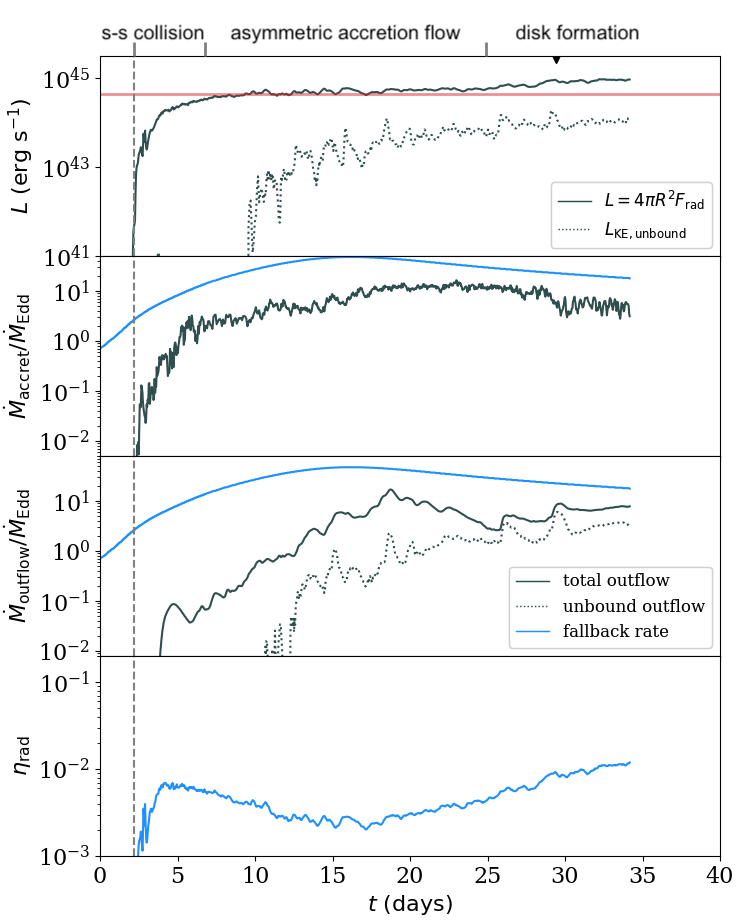}
    \caption{First row: frequency-integrated luminosity from gray simulation. The red horizontal line labels the Eddington luminosity for $M_{\rm bh}=3\times10^{6}M_{\odot}$ assuming $10\%$ efficiency. The dotted line shows the kinetic energy flux carried by unbound gas. In all panels, the vertical dashed line marks the first stream-stream collision time. On the top, we label the three dynamical stages discussed in Section~\ref{sec:result} : stream-stream collision, asymmetric accretion flow, disk formation . The small black triangle roughly labels the time when the polar region is cleared by radiation pressure after disk formation. Second row: the black solid line is the accretion rate estimated by the mass flux measured at ISCO. The blue solid line shows the mass fallback rate of the injected stream. In the third row, the solid black line is the total mass flux measured at the outer boundary radius, the dotted black line is the unbound mass flux. The blue solid line shows the mass fallback rate of the injected stream. The mass fluxes are normalized to the Eddington accretion rate. The fourth row shows radiation efficiency (Equation~\ref{eq:efficiency}). In the second and third row, we assume a radiation efficiency of 0.1 to calculate Eddington accretion rate, note that it is not equivalent to the observed radiation efficiency in the fourth row.}
    \label{fig:gray_lum_mdot}
\end{figure}

\begin{figure*}
    \centering
    \includegraphics[width=0.9\linewidth]{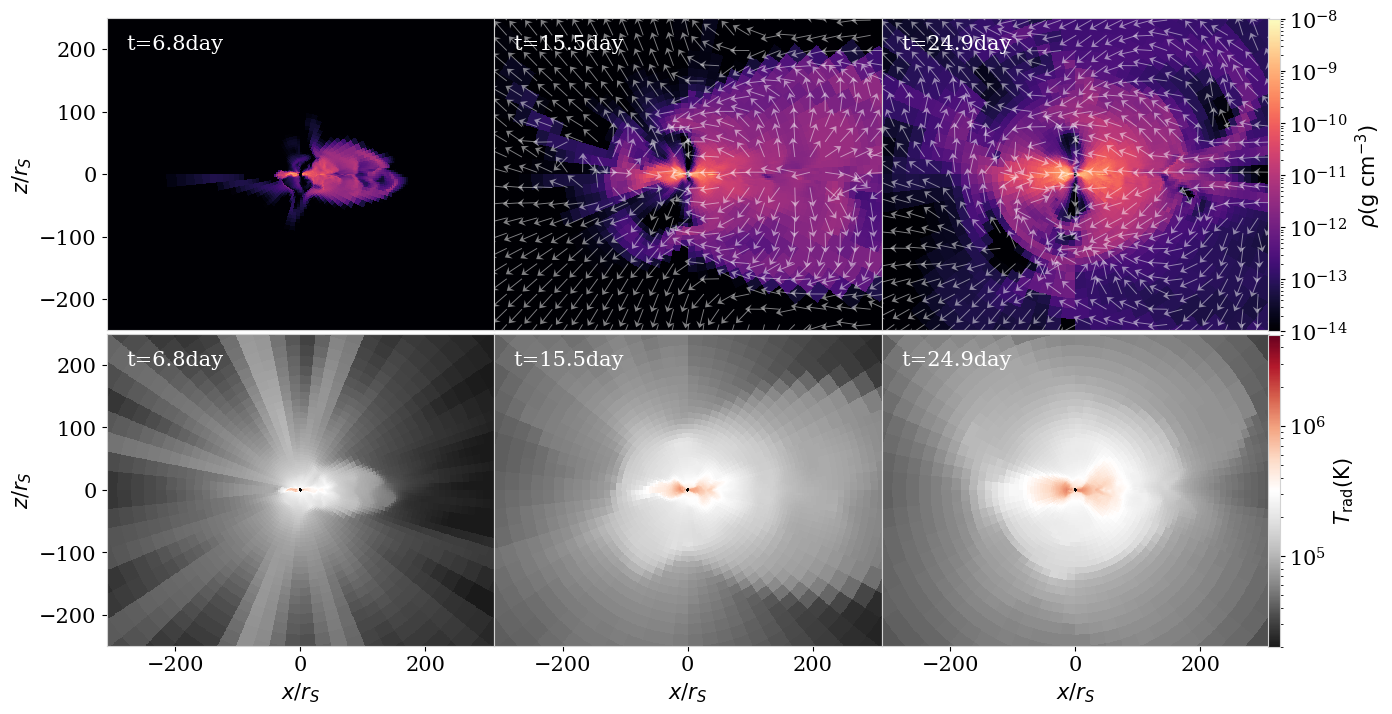}
    \caption{Gas density (upper panels) and radiation temperature (lower panels) snapshots. The plotted variables are averaged for $\phi=180^{\circ}\pm10^{\circ}$. The first, second, third columns are t=6.8, 15.5 and 24.9 days. The stream-stream collision happens at around t=2.2 days, the orbital period at the injected stream pericenter radius is $P_{\rm peri}\approx0.88$ hours. In the upper panel plots, the white arrows show the direction of velocity field, the velocity arrows are uniform and not scaled to their magnitude. In the lower panel, the beam pattern in t=6.8 day radiation temperature is due to the ray effect in the low optical depth region, which is associated with angular discretization of the radiation field. }
    \label{fig:timeseries_denser_vertical}
\end{figure*}

\subsection{Stream-stream collision}\label{subsec:dynamic_sscollision}
Around $t=2.2$ days and soon after the first pericenter passage, the injected debris stream intersects itself due to apsidal precession, which is commonly referred to as the ``stream-stream collision''. The first panel of Figure~\ref{fig:timeseries_dens_horizontal} shows gas density soon after the encounter at $t=2.3$ days. With the assumed stream orbit of $\beta=1.73$ and black hole mass $M_{\rm BH}=3\times10^{6}M_{\odot}$, the collision occurs at the radius $\rsi=76.2\rs$, which is beyond the circularization radius $r_{\rm circ}\approx2r_{\rm p}=10.1\rs$. At $t=2.5$ days. The returning stream briefly disrupts the fallback stream, a fraction of the post-collision gas is dispersed beyond $\rsi$, others accumulate near the black hole. Afterward, the returning stream expands and keep impacting the thin fallback stream. 

During the collision, a fraction of stream kinetic energy is converted to thermal energy, increasing local gas temperature. The collision region is optically-thick, and the thermal equilibrium timescale can be estimated as $t_{\rm eq}\approx (\rho k_{\rm B}T/\mu m_{\rm p})/(\kappap\rho caT^{4})\approx10^{-3}s(T/10^{5}\rm K)^{-4}(\kappap/0.32\rm cm^{2}~g^{-1})^{-1}$, which is short relative to the dynamical timescale.  The gas temperature and radiation temperature are promptly in equilibrium, resulting in radiation pressure dominated post-shock gas, with temperature $T_{\rm gas}\gtrsim5\times10^{5}$K. The hot photons are trapped in the post-shock flow, until they are able to diffuse out of the photosphere. 

In Figure~\ref{fig:gray_lum_mdot}, we show an estimated bolometric luminosity in the first row, which is obtained by integrating the radial direction radiation flux over a sphere of radius $R=400\rs$ at the outer simulation boundary. The accretion rate (second row) $\dot{M}_{\rm acc}$ is the mass flux through the innermost stable circular orbit (ISCO) for the spin-less black hole at $R=3\rs$. The outflow rate (third row) $\dot{M}_{\rm outflow}$ is the mass flux through the outer boundary. In the last row, we define an observed radiation efficiencies as the ratio between the emitting luminosity from photosphere and mass fallback rate:
\begin{equation}\label{eq:efficiency}
    \eta_{\rm rad,fb}=L/\dot{M}_{\rm fb}c^{2}
\end{equation}

The stream-stream collision produces a prompt luminosity rise.  In less than one day, the luminosity rises to $L\approx10^{43}\rm erg~s^{-1}$.  When the collision happens, the mass fallback rate is $\mfb\approx2.5\medd$, giving $\eta_{\rm rad,fb}\approx10^{-3}$, where $\medd=40\pi GM_{\rm BH}/\kappa c$ assuming $10\%$ efficiency. The immediate accretion rate is low $\macc\approx0.1\medd\approx0.04\mfb$, suggesting that the stream-stream collision itself does not lead to immediate super-Eddington accretion when the fallback rate is only at the order of the Eddington accretion rate. 

The collision launches an initial outflow. The outflow rate rises to $\mout\approx0.1\medd$ about two days after the collision, when it reaches $R=400\rs$. This suggests the average collision induced outflow velocity is on the order $v_{\rm outflow}\approx0.05c$. The stream-stream collision velocity is roughly $v_{\rm SI}=0.1c$, the fraction of fallback stream kinetic energy that is converted to outflow kinetic energy is only on the order of $\eta_{KE}=\mout v_{\rm outflow}^{2}/\mfb v_{\rm SI}^{2}\approx10^{-2}$. The majority of mass and kinetic energy carried by the fallback stream is dispersed to the region near the black hole and remains within the simulation domain.

In this simulation, the stream-stream collision does not recur. In our previous works, we found that such a one-time collision scenario happens when the $\rsi\gtrsim20\rs$ and $\dot{M}_{\rm fb}\lesssim\dot{M}_{\rm Edd}$, which is common for most spin-less $M_{\rm BH}\sim10^{6}M_{\odot}$ black holes. For a larger or spinning  black hole with deeper impact (larger $\beta$), $\rsi/\rs$ is small,  and the collision can repeat a few times \citep[e.g.][]{andalman2022tidal,huang2024pre}. In these regimes, the TDE debris stream has a small pericenter radius and low orbital angular momentum, the GR precession (both apsidal and Lense-Thirring) strongly shapes early dissipation, the resulting emission depends sensitively on black hole mass and spin \citep[e.g. and more][]{guillochon2015dark,dai2015soft,jiang2016prompt,lu2020self,jankovic2023spininduced}.

In the simulation, a one-time stream-stream collision does not power the entire light curve rise. The prompt initial rise to $L\sim10^{43}\rm erg~s^{-1}$ is due to the short diffusion time in fresh post-shock gas. At $t=2.5$day, we estimate the average post-shock gas density is $\bar{\rho}\approx5\times10^{-13}\rm g~cm^{-3}$, extending about $50\rs$ around the collision site. This gives a diffusion time at the order of 0.1 day. The luminosity from expanding post-shock gas lasts until $t=4.8$ days. Afterward, the gas distribution transits to the next stage of an asymmetric, eccentric accretion flow. As we will discuss in the following section, the majority of the optical light curve rise is instead powered by photons produced by shocks near the black hole and reprocessed by optically-thick layer extending beyond $r_{\rm circ}$.

\subsection{Asymmetric Accretion Flow before Circularization}\label{subsec:dynamic_asymmetric}

Around $t=6.8$ days (Figure~\ref{fig:timeseries_dens_horizontal}), the post stream-stream collision gas accumulates near the black hole $r\lesssim\rsi$, increasing the average density. In Figure~\ref{fig:timeseries_denser_vertical} before t=24.9 days, the gas distribution is more extended on the stream-stream collision side, where the thin fallback stream continuously interacts with the accretion flow, forming an azimuthal- and vertically-asymmetric reprocessing layer. 

The accretion rate through the ISCO reaches Eddington when the fallback rate is $\dot{M}_{\rm fb}=11.6\dot{M}_{\rm Edd}$ at $t=6.2$ days. At this time, the mass flux through the simulation outer boundary is only $\dot{M}_{\rm outflow}= 0.05M_{\rm Edd}$ with a negligible fraction of unbound gas, implying that the majority of mass being carried by the fallback stream remains within the simulation domain. Here we use the subscript of ``outflow'', but it is the measurement of total mass flux including both bound and unbound fractions.

\begin{figure}
    \centering
    \includegraphics[width=\linewidth]{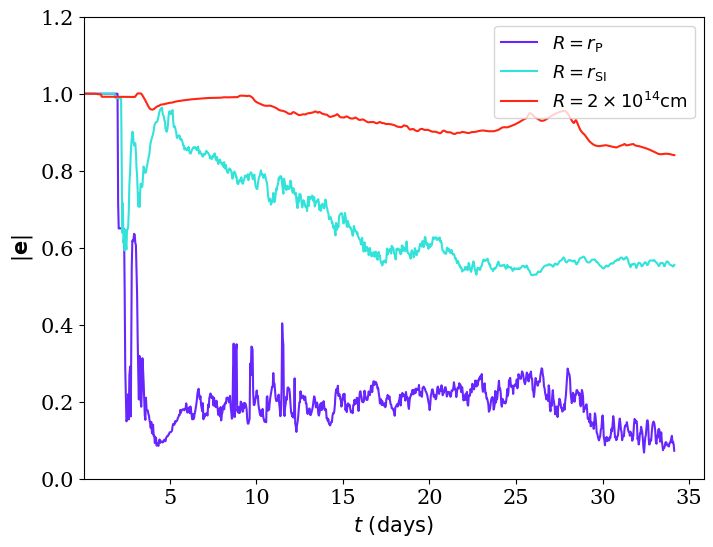}
    \caption{The eccentricity (Equation~\ref{eq:ecc}) calculated at the pericenter radius $R=r_{\rm P}=5.07\rs$ (the blue solid line), the stream-stream collision radius $R=r_{\rm SI}=76.2\rs$ (the cyan solid line) and an outer radius of $R=250\rs=2\times10^{14}$cm (the red solid line). With the defined $|\mathbf{e}|$, the non-negligible $\theta$ and $\phi$ velocity components can lead to eccentricity greater than unity. The stream-stream collision and post-collision happen about $t=2.2-6.8$ day, the asymmetric accretion flow stage lasts $t=6.8-24.9$ days, and after which the flow evolves into the circularization accreting stage. }
    \label{fig:ecc_time}
\end{figure}

We define the eccentricity of flow 
\begin{equation}\label{eq:ecc}
    \mathbf{e}=\frac{1}{GM_{\rm bh}}\mathbf{v}\times(\mathbf{r}\times\mathbf{v})-\mathbf{\hat{r}},
\end{equation}
so that the magnitude of eccentricity is $|\mathbf{e}|=0$ for a circular Keplerian flow. However, with the adopted gravity potential, $|\mathbf{e}|$ is non-zero for circular orbits at small radii, such as $r<10\rs$. The injected stream eccentricity is $|\mathbf{e}|\approx1.0$ at the injection radius and decreases to its minimum $|\mathbf{e}|\approx0.6$ at $\rp=5.3\rs$. 

In Figure~\ref{fig:ecc_time}, we calculate mass weighted average eccentricity $|\mathbf{e}|$ for gas in the $|\theta-90^{\circ}|\leq 30^{\circ}$ disk region at three radii: the pericenter radius $R=r_{\rm P}$, the stream-stream collision radius $R=r_{\rm SI}$ and an outer radius that is similar to typical optical TDE photosphere size: $R=2\times10^{14}$cm . The eccentricity at $r_{\rm P}$ drop significantly soon after the stream-stream collision $t=2.2$ days. The eccentricity at $r_{\rm SI}$ is dominated by the dense stream, it drops promptly as the fallback stream is disrupted during the self intersection but increases again after it is resupplied. Around $t=6.8$ days, eccentricity at $r_{\rm P}$ (blue line) becomes relatively stable $|\mathbf{e}|\approx0.2$, while the eccentricity at $r_{\rm SI}$ starts to continuously decrease as accretion flow density increases and becomes less dominated by the stream. The fallback rate $\dot{M}_{\rm fb}$ peaks near $t=15.5$ days, after which the eccentricity at $r_{\rm SI}$ stabilizes near $|\mathbf{e}|\approx0.6$. The accretion flow is still largely asymmetric despite the super-Eddington accretion rate (Figure~\ref{fig:timeseries_dens_horizontal}). Such a geometry dominates the early TDE dynamics, from the rise to about a week after the peak fallback rate ($t=6.8-24.9$ days). 

\begin{figure}
    \centering
    \includegraphics[width=\linewidth]{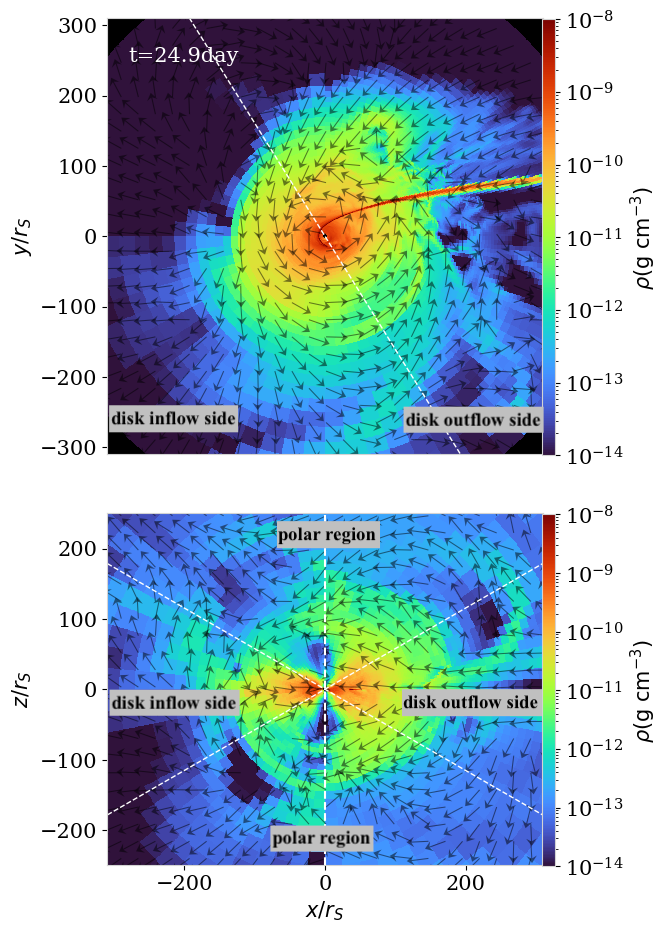}
    \caption{Schematic annotation of the azimuthal- and vertical- asymmetric accretion flow structure. The snapshots are gas density, the arrows show the projected direction of flow velocity. The upper panel is the ``face-on'' view averaged over $\theta=90^{\circ}\pm10^{\circ}$, lower panel is an ``edge-on'' view averaged over $\phi=\phi_{\rm SI}\pm10^{\circ}$, where $\phi_{\rm SI}$ is the initial stream self-intersection angle. In the upper panel, the dashed line shows $\phi_{\rm SI}+90^\circ$. We defined the region of $\phi_{\rm SI}\pm90^\circ$ as the ``outflow region'', where the interaction of incoming fallback stream and accretion flow drives outflow. Opposite to the outflow region is the inflow region, where the outflow converges to the orbital plane. In the lower panel, the tilted dashed lines correspond to $\theta=90\pm30^{\circ}$, which roughly separates the disk and polar region. The $\theta$ and $\phi$ range for each angular section is summarized in Table~\ref{tab:angular_sec} }
    \label{fig:schematic_winddisk}
\end{figure}

\begin{figure}
    \centering
    \includegraphics[width=\linewidth]{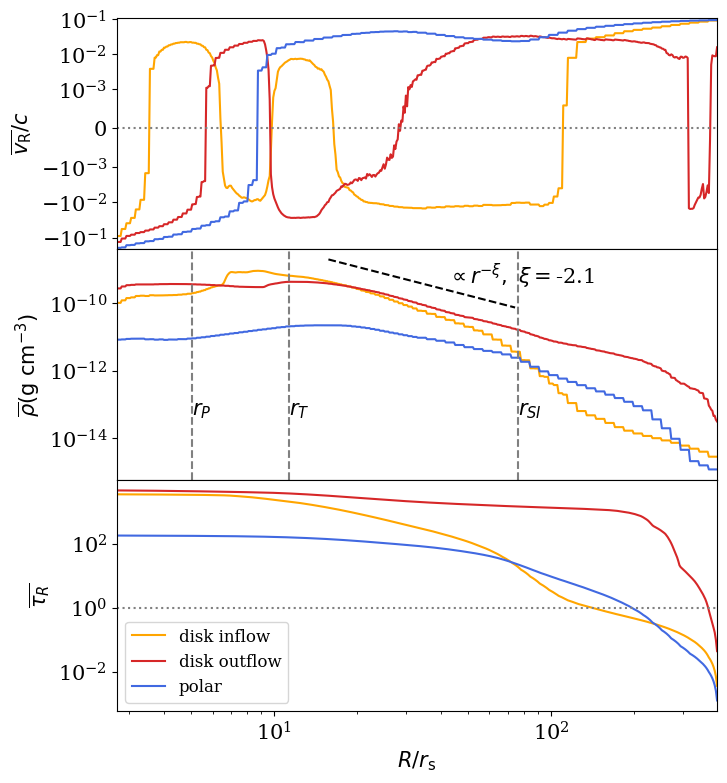}
    \caption{Azimuthal- and vertical- averaged one-dimensional radial profiles at t=15.5 days during the asymmetric accretion flow stage. The x-axis is distance to the black hole. From the first to the third row: mass-weighted average radial velocity, gas density, and optical depth (Eq~\ref{eq:tau_R}). In each panel, the red, blue and yellow lines correspond to disk outflow , disk inflow and polar region including both outflow- and inflow-side (Table~\ref{tab:angular_sec}). In the first row, positive (negative) velocity is pointing away from (towards) black hole, representing outflow (inflow) velocity. In the second row, the vertical dashed line labels the pericenter radius of the injected debris stream $r_{\rm P}$ and the tidal radius $r_{\rm T}$. The dashed black line shows a $\rho\propto r^{-2.1}$ for reference.}
    \label{fig:radial_profile_wind}
\end{figure}

Figure~\ref{fig:schematic_winddisk} shows density snapshots of the typical flow structure with schematic annotations. We divide the flow into four angular sections: in the vertical ($\theta$ direction), there is an eccentric, geometrically and optically thick accretion flow in the region with $\theta\lesssim 90^{\circ}\pm30^{\circ}$, which we refer to as ``disk'' and label by the white dashed lines. The regions $|\theta-90^{\circ}|\gtrsim 30^{\circ}$ show lower density and vary over time, which we refer to as the ``polar region". They are summarized in Table~\ref{tab:angular_sec} and consistently used in the rest of this work.

The flow velocity arrows in Figure~\ref{fig:schematic_winddisk} reveal that such geometry is driven by dynamics: the lower density eccentric flow keeps be impacted by the higher density fallback stream, diverting the gas away from the midplane and forming a fountain of bound vertical outflow. These gas gains off-midplane velocity and flows across the polar region, and further colliding with each other, forming secondary shocks near the midplane on the opposite side of $\phi_{\rm SI}$. 

In the $\phi$ direction, we refer to the ``outflow side'' as $|\phi-\phi_{\rm SI}|\lesssim 90^{\circ}$, which is near the interaction region of accretion flow and the fallback stream. The rest of $|\phi-\phi_{\rm SI}|\gtrsim90^{\circ}$ is the ``inflow side'', where the bound outflow colliding with each other near the midplane, channeling mass flux radially towards the black hole. Therefore, in Figure~\ref{fig:schematic_winddisk}, we have four angular sections: polar region and disk region near the inflow and outflow side accordingly (Table~\ref{tab:angular_sec}). At this stage, the polar regions near the inflow and outflow sides are not significantly different. In the remainder of this section, we do not distinguish them unless explicitly specified. 

\begin{table}
\centering
\caption{Angular Section Definition}
\label{tab:angular_sec}
\begin{threeparttable}
\begin{tabular}{lcc}
\hline
Name & $\phi$ range & $\theta$ range \\ 
\hline
disk outflow & $|\theta-90^{\circ}|<30^{\circ}$ & $|\phi-\phi_{\rm SI}|<90^\circ$\\
\hline
disk inflow & $|\theta-90^{\circ}|<30^{\circ}$ & $|\phi-\phi_{\rm SI}|\geq90^\circ$ \\
\hline
polar outflow & $|\theta-90^{\circ}|\geq30^{\circ}$ & $|\phi-\phi_{\rm SI}|<90^\circ$\\
\hline
polar inflow & $|\theta-90^{\circ}|\geq30^{\circ}$ & $|\phi-\phi_{\rm SI}|\geq 90^\circ$ \\
\hline
\end{tabular}
\end{threeparttable}
\end{table}

The polar region, disk inflow, and disk outflow region show distinct average density and velocity. In Figure~\ref{fig:radial_profile_wind}, the average radial velocity is the highest in the polar region $v_{\rm R}\approx0.05c$ for $R\gtrsim r_{\rm T}$. The disk inflow side has an average inflow velocity of $v_{\rm R}\approx-0.01c$ for $r_{\rm T}\lesssim R\lesssim r_{\rm SI}$. Outside $r_{\rm SI}$, the disk inflow shows positive radial velocity due to low density gas that is moving outward (also seen in the velocity field Figure~\ref{fig:schematic_winddisk}). The disk outflow side is dominated by the stream-accretion flow interaction, yielding outflow velocities $v_{\rm R}\approx0.01c-0.1c$ for gas in $R\gtrsim r_{\rm SI}$. Because of the mass weighting, the region $R\lesssim r_{\rm T}$ is more affected by the dense stream and can be noisy. Similarly, the negative $v_{\rm R}$ at $R\gtrsim200\rs$ is also dominated by the injected stream.

The second row of Figure~\ref{fig:radial_profile_wind} are the average density of the polar, disk inflow and outflow regions. The majority of mass is distributed between $r_{\rm P}\lesssim R\lesssim r_{\rm SI}$.  A simple estimation of the average flow density can be found by assuming spherical distribution:
\begin{equation}\label{eq:rho_wind_avg}
\begin{split}
    \rho_{\rm flow}&\approx\frac{\dot{M}_{\rm fb}-\dot{M}_{\rm acc}}{4\pi R_{\rm flow}^{2}v_{\rm flow}}\\
    &=1.3\times10^{-11}\rm g~cm^{-3}\times\\
    &\left(\frac{\dot{M}_{\rm fb}-\dot{M}_{\rm acc}}{50\medd-5\medd}\right)\left(\frac{R_{\rm flow}}{r_{\rm SI}}\right)^{-2}\left(\frac{v_{\rm flow}}{0.01c}\right),
\end{split}
\end{equation}
where we adopt the fallback rate $\dot{M}_{\rm fb}\approx50\medd$, accretion rate $\dot{M}_{\rm acc}\approx5\medd$ from simulation (Figure~\ref{fig:gray_lum_mdot}) at $t=15.5$ days. We assume the gas is spherically distributed up to $R=r_{\rm SI}$, with average flow velocity $v\sim0.01c$.

The average density of the disk outflow and inflow sides ranges from $\rho
\sim10^{-10}-10^{-12}\rm g~cm^{-3}$ for $R\lesssim R_{\rm SI}$, roughly comparable to the above estimation. The best-fit power law shows an index of $\rho\propto r^{-2.1}$. The disk outflow side is more extended than the disk inflow side. The polar region follows a similar power law, but the average density is about one to two orders of magnitude lower than the disk regions.

The averagely lower density in the polar region suggests that it is more optically-thin. The third row shows the estimated optical depth:
\begin{equation}\label{eq:tau_R}
    \tau_{\rm R}(R)=\int_{R}^{R_{\rm out}}(\kappas+\kappar)\rho dR
\end{equation}
The optical depth in the polar region is significantly lower than in the disk region, but remains larger than unity until $R\sim200\rs\approx1.7\times10^{14}$cm, suggesting less reprocessing and potential soft X-ray emission. However, the polar region is often obscured by the optically-thick gas dispersed by shocks in the asymmetric accretion flow. Therefore, the X-ray flux emerging from the polar region can show potential variability due to the dynamical obscuration (Section~\ref{subsec:mg_xray}). As will be discussed later, when a more circularized disk forms, the outflow subsides, the polar region becomes free from obscuration and the X-ray flux may be less variable.

With the highest average density, the disk outflow region is the main reprocessing layer. Its average optical depth is larger than unity until $R\sim300\rs\approx2.7\times10^{14}$cm. The optical depth of the disk inflow region drops below unity in $R\sim100\rs\approx8.9\times10^{13}$cm. Nevertheless, such geometry of a more optically-thin polar region and denser disk region is broadly consistent with the picture of viewing angle dependent emission \citep{roth2016x,dai2018unified}.  We note that in Figure~\ref{fig:radial_profile_wind}, we average the polar and disk region properties over a relatively large solid angle range. The optical depth should not be directly interpreted as the photosphere size. More discussions of photosphere size are in Section~\ref{sec:mg} with multi-group simulations.

\begin{figure}
    \centering
    \includegraphics[width=\linewidth]{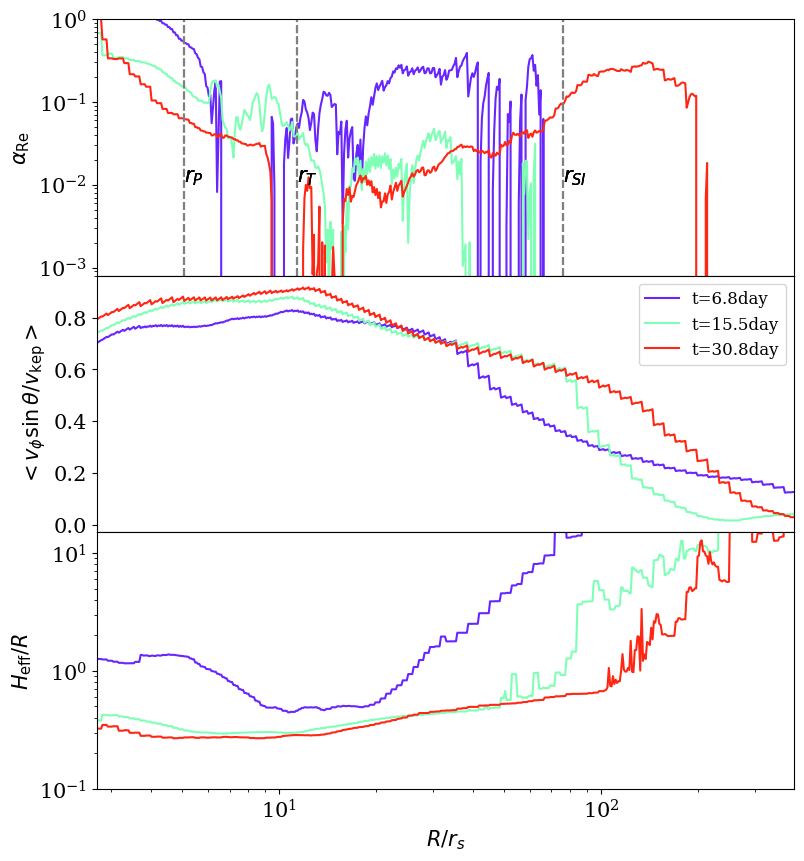}
    \caption{Normalized Reynold stress (first row), the ratio between $v_{\phi}$ and Keplerian velocity (second row) and the estimated scale height (third row) for $t=6.8,~15.5,~30.8$ days (blue, green, red lines). In the first row, the three vertical dashed lines mark the pericenter radius $r_{\rm P}$, the tidal radius $r_{\rm T}$ and the stream-stream collision radius $r_{\rm SI}$. }
    \label{fig:radial_profile_reynoldstress}
\end{figure}

We next consider the angular momentum transport in asymmetric accretion flow. Despite lacking a well-circularized disk, we follow canonical definitions for an axisymmetric accretion disk to get an estimate of the stress. We define the hydrodynamic Reynold stress as:
\begin{equation}\label{eq:alpha_reynold}
    \alpharey \equiv \frac{\left<\rho(v_{\rm R}-<v_{\rm R}>)(v_{\phi}\sin\theta-<v_{\phi}\sin\theta>)\right>}{\left<P_{\rm gas}+P_{\rm rad}\right>},
\end{equation}
where the  $<X>$ represents averaging in azimuthal and vertical directions $<X>=\int_{0}^{2\pi}\int_{\theta-}^{\theta+}Xd\theta d\phi$. $P_{\rm gas}$ and $P_{\rm rad}$ are gas and radiation pressure. Note that the Reynold stress defined here differs from the purely turbulent Reynold stress in a canonical accretion disk. It is instead the total fluctuating hydrodynamic stress in the angular momentum equation \citep{meza2025radiation}. The scale height is estimated as:
\begin{equation}\label{eq:h_eff}
    H_{\rm eff}\equiv\frac{c_{\rm s,eff}}{v_{\phi}\sin\theta/R}, \quad c_{\rm s,eff}\equiv\sqrt{\frac{5P_{\rm gas}+4P_{\rm rad}}{3\rho}}
\end{equation}
Figure~\ref{fig:radial_profile_reynoldstress} shows $\alpharey$ and scale height $H_{\rm eff}$ of the asymmetric accretion flow, with the vertical range $\theta^\pm =90^{\circ}\pm 30^{\circ}$ to be consistent with the defined disk regions. 

The disk mass is mainly distributed between $r_{T}$ and $r_{\rm SI}$ during asymmetric accretion flow stage. At $t=6.8$ days, following the stream-stream collision, the shock and off-orbital plane outflow drive the high hydrodynamic Reynold stress $\alpharey\sim10^{-1}$. When the fallback rate peaks $t=15.5$ days, the Reynold stress decreases to $\alpharey\sim10^{-2}$ as the flow eccentricity drops. The $v_{\phi}\sin\theta$ component is slightly sub-Keplerian at all times, consistent with the non-zero eccentricity. Beyond $R\gtrsim R_{\rm SI}$, the flow velocity is primarily radially outward, resulting in significantly smaller $v_{\phi}\sin\theta/v_{\rm Kep}$. The $\alpharey$ discontinuities near $r_{\rm P}\lesssim R\lesssim r_{\rm T}$ are related to the dense stream and a pair of local acoustic spiral shock near the black hole, which we discuss in Section~\ref{subsec:discussion_shock} . The effective scale height decreases to $H_{\rm eff}/R\sim0.3$ since $t=6.8$ days. We also tested averaging over a smaller vertical range $\theta^\pm=90 \pm 10^{\circ}$ and found qualitatively similar results.

\begin{figure}
    \centering
    \includegraphics[width=\linewidth]{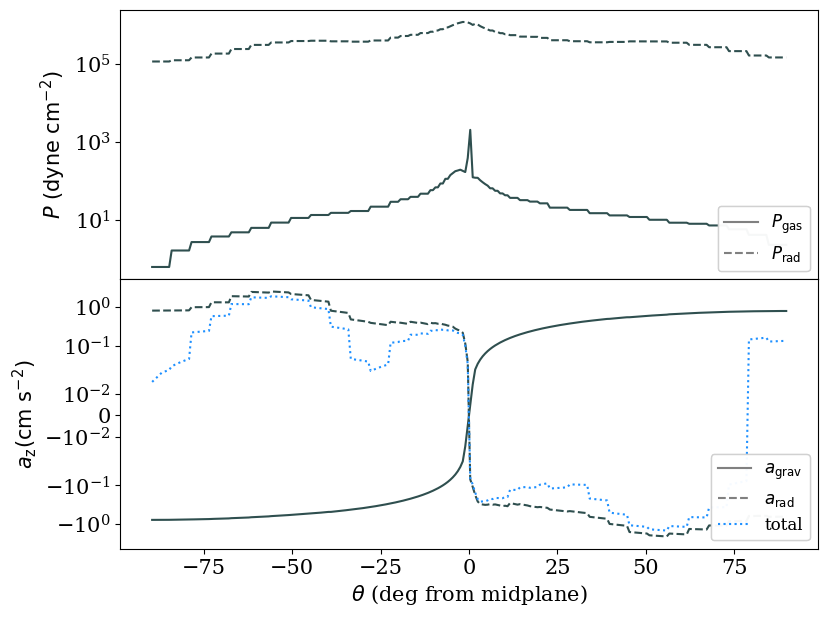}
    \caption{The first row is the vertical pressure component at $t=15.5$ days. The variables are averaged in $\phi$ direction at $R=100\rs$. The solid line is gas pressure, the dashed line is radiation pressure. The spike of gas pressure in the orbital plane is due to the dense fallback stream. The second row shows the vertical radiation force (dashed line) and gravitational force (solid line) at the same time. The blue dotted line is the sum of radiation and gravity force. The gas pressure gradient is significantly smaller than radiation and gravitational force, thus not shown in the plot.}
    \label{fig:acc_vertical_t15.5}
\end{figure}

Vertically, the asymmetric accretion flow is supported by the radiation pressure. We show in Figure~\ref{fig:acc_vertical_t15.5} that the vertical radiation pressure is a few orders of magnitude larger than the gas pressure. The pressure gradient, which measures the vertical force, is also dominated by radiation. In the lower panel, we show the gravitational acceleration and the radiation acceleration projected in the $z$ direction, estimated by $a_{\rm rad}=(\kappas+\kappar)\rho F_{\rm rad, z}/c$, where $F_{\rm rad, z}$ is the $z$ direction radiation flux. For negative (positive) $\theta$ in Figure~\ref{fig:acc_vertical_t15.5}, positive (negative) $a_{\rm z}$ is pointing away from the mid-plane, so that gravitational force is towards the mid-plane.  At this radius ($R=100\rs$), the radiation force exceeds gravity and drives gas away from midplane. At larger radius (e.g. $R=250,~350\,\rs$) we find that the gravitational force is larger than the radiation force, so the net acceleration is pointing towards the midplane. Such dominance of the radiation pressure is also found in our companion work with magnetic fields in pre-peak time \citep{meza2025radiation}. 

\subsection{Formation of Accretion Disk}\label{subsec:dynamic_diskformation}

\begin{figure*}
    \centering
    \includegraphics[width=\textwidth]{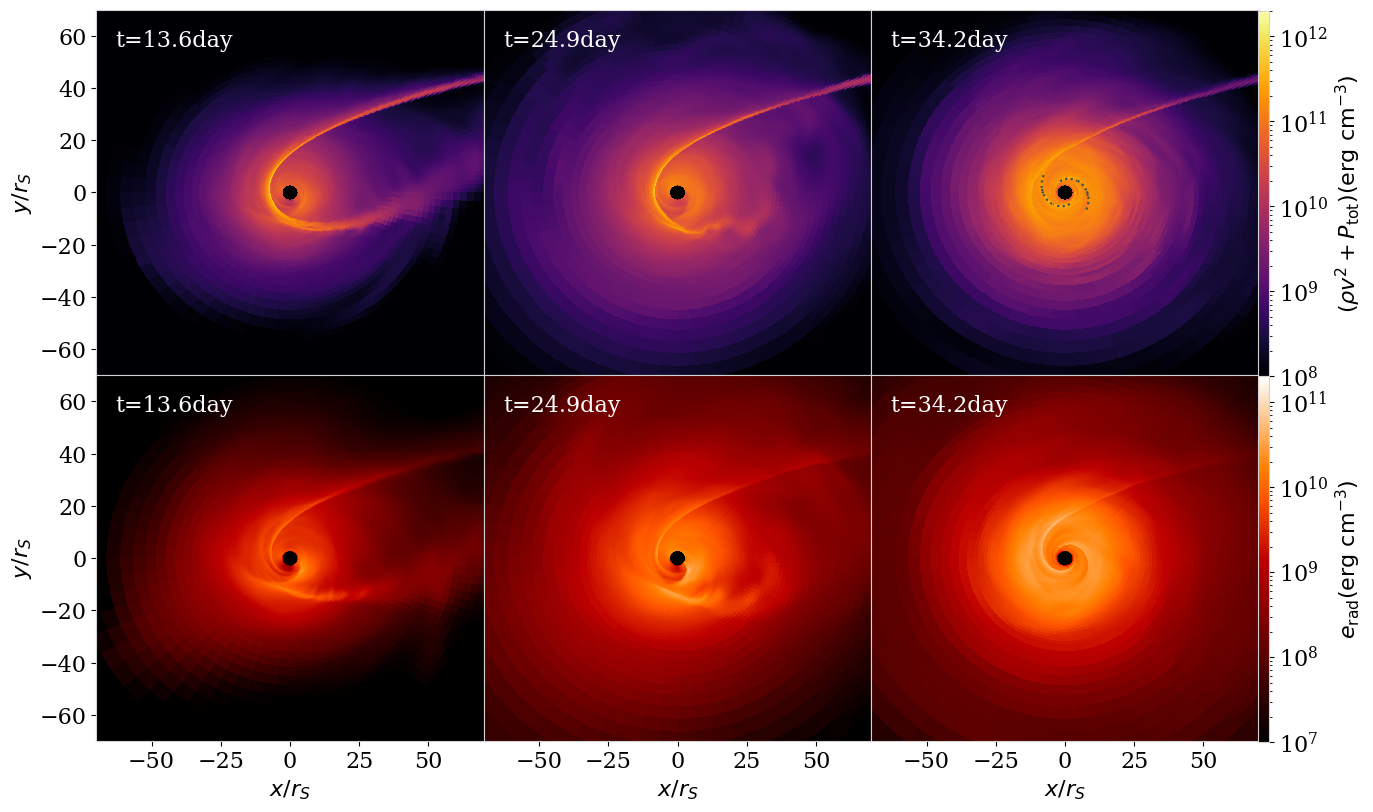}
    \caption{The sum of ram pressure and total pressure $\rho v^{2}+P_{\rm tot}$ (upper panel) and radiation density (lower panel) snapshots, the variables are averaged within $\theta=90^{\circ}\pm10^{\circ}$. The total pressure includes gas internal pressure and radiation pressure, both are subdominant relative to the ram pressure. In the lower right panel, the dotted black lines show the pitch angle of the spiral shock estimated from the average Mach number near the black hole (Section~\ref{subsec:discussion_shock}). The top and lower panels also compare the specific kinetic energy content  and radiation energy content. }
    \label{fig:errhovv_streamdisk}
\end{figure*}

After $t\approx24.9$ days, the fallback rate drops below $\dot{M}_{\rm fb}\lesssim10\dot{M}_{\rm Edd}$. The stream density decreases, the orbital plane ram pressure $\rho v^{2}$ becomes comparable in the accretion flow and in the stream. The stream dissolves in the accretion flow instead of transporting high eccentricity gas that impacts the accretion flow. We show the sum of ram pressure and total pressure at $t=13.2,~24.9,~34.2$ days in Figure~\ref{fig:errhovv_streamdisk}, corresponding to the asymmetric accretion flow stage, the beginning of accretion disk formation, and the end of the simulation. 

Both the disk and the stream are kinetic energy dominated, equivalently, the ram pressure (or momentum flux) is larger than the radiation pressure and internal pressure (thermal pressure): $\rho v^{2}> P_{\rm rad}\gg P_{\rm gas}$. This is also consistent with the radiation energy density $e_{\rm rad}$ snapshots in the bottom row, which is roughly related to radiation pressure by $P_{\rm rad}\sim e_{\rm rad}/3$ in optically thick region. In the bottom row, $e_{\rm rad}$ is orders of magnitude smaller than $\rho v^{2}+P_{\rm rad}+P_{\rm gas}$, so the flow is kinetic energy dominated. At $t=34.2$ days, the stream ram pressure is similar to the disk ram pressure, in strong contrast with the two earlier snapshots when the high eccentricity stream pierced through the accretion flow. The decreasing fallback rate and increasing accretion disk density are essential to the formation of such a more circularized disk.

The pericenter eccentricity further decreases to $|\mathbf{e}|\approx0.1$ near $t=30$ days (Figure~\ref{fig:ecc_time}), but the eccentricity at $R_{\rm SI}$ remains $|\mathbf{e}|\approx0.5$. The disk is geometrically thick, with scale height $H_{\rm eff}/R\sim0.3$ (Figure~\ref{fig:radial_profile_reynoldstress}) for $R\lesssim100\rs$. As the flow becomes more circularized and velocity dispersion decreases, the hydrodynamic Reynold stress $\alpharey$ decreases. For gas between $r_{\rm T}\lesssim R\lesssim \rsi$, the stress $\alpha_{\rm Re}\sim10^{-2}$ at disk formation stage is about one order of magnitude lower compared to $t=6.8$ days. 

\begin{figure}
    \centering
    \includegraphics[width=\linewidth]{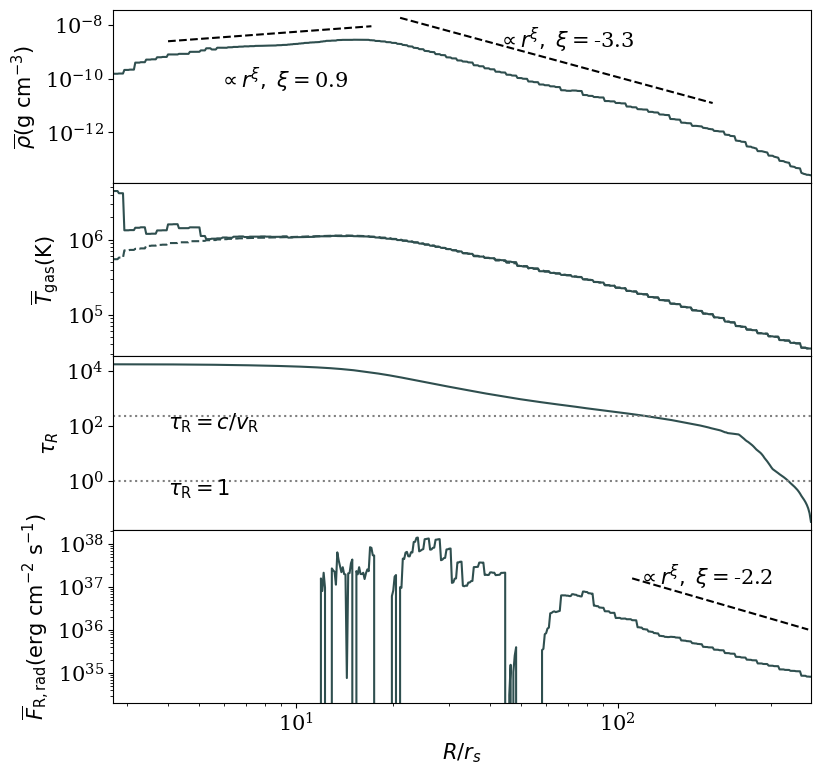}
    \caption{Azimuthal- and vertical- averaged one-dimensional disk radial profiles att=34.2 days, when the accretion disk forms. The x-axis is the distance to the black hole. The vertical range spans $|\theta-90^{\circ}|<30^{\circ}$. In the first row, the average density profile can be approximated by two power laws (the dashed lines), with best-fit index $\rho\propto r^{0.9}$ for the inner compact disk $R_{\rm ISCO}\lesssim R\lesssim r_{\rm T}$, and $\rho\propto r^{-3.3}$ for the outer disk that extends to $R\approx200\rs$. The second row is average gas temperature (the solid line) and radiation temperature (the dashed line). In the third row, the solid line is the average optical depth (Equation~\ref{eq:tau_R}), the two horizontal dotted lines are where the average $\tau_{\rm R}$ equals the average $c/v_{\rm R}$ and equals  unity. The last row is the average lab-frame radiation flux in the radial direction, which roughly declines as $F_{\rm rad}\propto R^{-2}$ when $\tau_{\rm R}\lesssim c/v_{\rm R}$ (the dashed line). }
    \label{fig:radial_disk_profile}
\end{figure}

In Figure~\ref{fig:radial_disk_profile}, we show the average disk density profile at $t=34.2$ days. The density can be described by a broken power law, with an extended outer region of $r_{\rm T}\lesssim R\lesssim 200\rs$ where density decreases as $\rho\propto R^{-3.3}$, and a compact inner region of $R\lesssim r_{\rm T}$ with density profile $\rho\propto R^{0.9}$. The disk size reaches about $200\rs$, the Reynold stress increases to $\alpharey\sim10^{-1}$ in the outermost disk. Compared to $t=15.5$ days, when the flow is more eccentric (Figure~\ref{fig:radial_profile_wind}), the outer disk density profiles become steeper. The inner disk shows a density profile that increases with radius, which can be related to the efficient accretion driven by the acoustic spiral wave. We discuss this further in Section~\ref{subsec:discussion_shock}.

The average disk temperature is relatively constant in the inner compact disk $\overline{T}_{\rm gas}\approx10^{6}$K except for the region within the ISCO. We do not accurately model the structure in the proximity of the ISCO by ignoring GR effect. The deviation of gas temperature from radiation temperature suggests they decouple locally. In the outer disk, the temperature decreases to $\overline{T}_{\rm gas}\approx10^{5}$K. For $R\gtrsim 200\rs$, the gas temperature is roughly consistent with the observed photosphere temperature of optical TDEs $\overline{T}_{\rm gas}\approx10^{4}$K. 

The temperature and density range suggest that the bound-bound and bound-free opacity can be important in the disk when assuming solar abundance and not fully ionized gas. As a result, the disk is optically thick in both radial and vertical directions. In the third row of Figure~\ref{fig:radial_disk_profile}, we find the average optical depth $\tau_{\rm R}>1$ (Equation~\ref{eq:tau_R}) until $R\approx300\rs$. The location of the ``photon trapping'' radius \citep{strubbe2009optical,piro2020wind}, defined here by $\tau_{\rm R}=c/v$ is interior to the $\tau_{\rm R}=1$. Consistently, the bolometric radiation flux in the radial direction decays approximately $F_{R,\rm rad}\propto R^{-2}$ outside the photon trapping radius, so that the bolometric luminosity roughly stays constant. If instead using geometric mean opacity in the optical depth integration $\sqrt{(\kappa_{\rm s}+\kappa_{R})\kappa_{R}}$, we find lower effective optical depth in the inner disk due to $\kappa_{R}<\kappa_{\rm s}$ up to $R\simeq100\rs$, where $T_{\rm gas}\gtrsim2\times10^{5}$K. However, the location of $\tau_{\rm R}=c/v$ and $\tau_{\rm R}=1$ remains similar in the outer disk, where the temperature is lower. In the outer disk, $\kappa_{R}>\kappa_{\rm s}$, the two estimations of ``average'' opacity are thus comparable: $\sqrt{(\kappa_{\rm s}+\kappa_{R})\kappa_{R}}\approx\kappa_{R}$ and $(\kappa_{\rm s}+\kappa_{R})\approx\kappa_{R}$. In the Appendix~\ref{appendix:bbfitting}, we show that outside of ``trapping radius'', the shapes of SEDs from multi-group simulations can still vary, but the total luminosity is roughly unchanged.

\begin{figure}
    \centering
    \includegraphics[width=\linewidth]{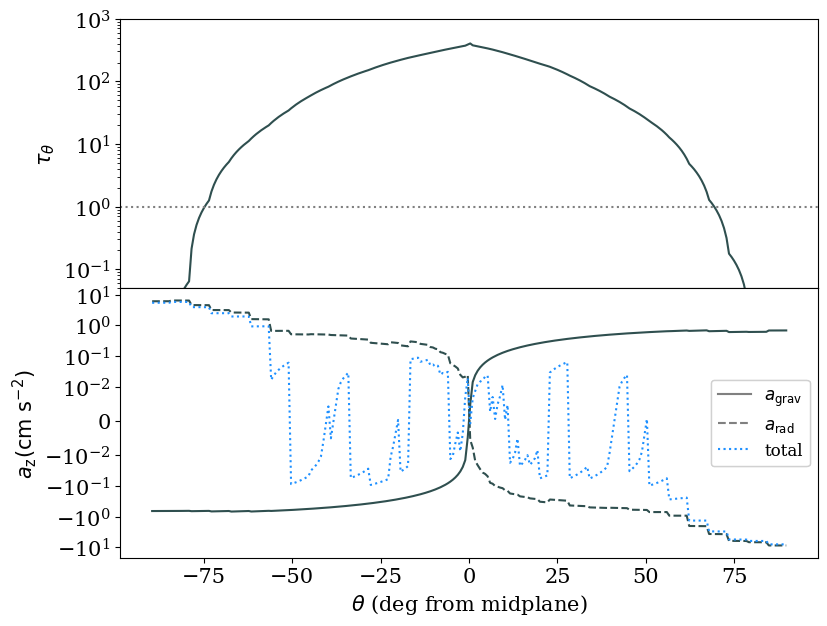}
    \caption{Disk vertical profiles that are similar to Figure~\ref{fig:acc_vertical_t15.5} lower panel but at t=30.2 days, averaging  when the accretion disk forms. The first row is disk vertical ($\theta$ direction) optical depth, similar to the definition in Equation~\ref{eq:tau_R} but integrated in the vertical direction. The second row shows the vertical radiation force (dashed line) and gravity force (solid line), the blue dotted line is the sum of radiation and gravitational acceleration.}
    \label{fig:vertical_disk_acc}
\end{figure}

Vertically, the disk is also optically thick. We estimate the optical depth in the $\theta$ direction similar to in Equation~\ref{eq:tau_R}, but integrating from the disk midplane to a $\Delta\theta$ range $\tau_{\theta}=\int_{90^{\circ}-\Delta\theta}^{90^{\circ}+\Delta\theta}(\kappa_{\rm s}+\kappa_{R})\rho Rd\theta$ for a fixed radius. Figure~\ref{fig:vertical_disk_acc}'s first row shows $\tau_{\theta}$ at $R=100\rs$, the optical depth drops below unity for $\Delta\theta\gtrsim70^{\circ}$. The $\theta$ direction average velocity is less uniform. 

The vertical pressure is dominated by radiation pressure, similar to earlier times. The force balance between radiation force and gravity in $z$ direction is shown in the lower panel of Figure~\ref{fig:vertical_disk_acc}. The estimated average radiation force points away from gravity. However, the total force fluctuates within $\Delta\theta\lesssim50^{\circ}$, where $\tau_{\theta}\gg1$ and noticeable radiation flux is advected with gas motion. This leads to the variations in radiation acceleration where the gas motion is turbulent in the $\theta$ or $z$ direction. 

As the interaction of the fallback stream and accretion flow weakens, the amount of gas flowing across the pole significantly decreases. As a result, the average density and optical depth of the polar region become lower. When the high energy photons propagate through the optically thin polar region, the radiation pressure further drives the remaining low density gas away. Around $t=27$ days, radiation pressure clears the polar region that can sometimes be obscured previously. The polar region density approaches the numerical floor value, forming an optically thin ``channel''. However, this channel does not precisely align with the z-directions ($\theta=0^{\circ}$ or $\theta=180^{\circ}$).

We summarize the mass budget as follows: at the beginning of the disk formation stage at $t=24.9$ days, the total fallback mass is $M_{\rm fb}=1.48\times10^{-1}M_{\odot}$, the accreted amount is $M_{\rm acc}=3.08\times10^{-2}M_{\odot}$ and the outflow through the simulation boundary is $M_{\rm out}=1.79\times10^{-2}M_{\odot}$. Closer to the end of the simulation, when the disk forms, the total fallback mass is $M_{\rm fb}=1.80\times10^{-1}M_{\odot}$, the accreted amount is $M_{\rm acc}=4.20\times10^{-2}M_{\odot}$ and the outflow through the simulation boundary is $M_{\rm out}=2.42\times10^{-2}M_{\odot}$.  The disk mass, if roughly estimated by total gas mass within $R=200\rs$ and $|\theta-90^\circ|<30^{\circ}$ is $M_{\rm disk}=6.20\times10^{-2}M_{\odot}$ at $t=24.9$ days and $M_{\rm disk}=8.04\times10^{-2}M_{\odot}$ at $t=30.8$ days.  

\section{Emission Property and SED evolution}\label{sec:mg}

\subsection{Multi-group Simulations}\label{subsec:mg_setup}
To obtain the SED and broad-band light curve, we perform 11 multi-group simulations derived from the gray RHD simulation at $t=2.4,~2.5,~3.8,~6.5,~13.4,~17.2,~19.2,~23.6,~28.4,~30.8,~33.9$ days. Due to high computational cost, each multi-group run lasts only for $10-30$ minutes in physical time, which is significantly shorter than the typical photon diffusion timescale in the optically thick disk (e.g. Figure~\ref{fig:radial_disk_profile}), but still longer than the light travel time from the typical photosphere radius to the outer simulation boundary. Equivalently, the emerging radiation flux represent the photosphere properties. Therefore, they are considered to be post-processing simulations and do not fully capture the thermal evolution. They are, however, different from typical post-processing approaches that fix the hydrodynamic and thermodynamic variables, since we still evolve the dynamical coupling between radiation and gas, but are limited in runtime by the computational cost.

We adopt multi-group opacity from the TOPS Opacity database \citep{colgan2016new}.  We retrieve multi-group opacities for 16 photon frequency groups ranging from $h\nu=10^{-3}$ keV to $h\nu=5.6$ keV. This corresponds to webpage entries of 17 photon frequencies with lower and upper bound of $10^{-3}$ keV and $10$ keV. The first group represents $0$ keV to $10^{-3}$ keV; the last group represents $5.6$ keV to infinity. We request 100 density points from $10^{-17}-10^{-6}\,\rm g~cm^{-3}$ , 69 default non-uniform temperature points from $5\times10^{-4}$ keV to $10$ keV. For the last four multi-group runs, when the accretion disk forms, we adopt a larger photon energy grid for the opacity including 20 photon frequency groups ranging from $h\nu=10^{-3}$ keV to $h\nu=56.2$ keV, with the same density and temperature grid as described above.

We assume standard solar abundance X=0.735, Y=0.248, Z=0.017 \citep{grevesse1998standard} for the mixture. In the simulations, the opacity of each cell is linearly interpolated in temperature and density grid of TOPs data. We replace the absorption opacity by free-free opacity in low density and high-temperature regions where TOPS does not provide multi-group opacity data, usually because assuming local thermal equilibrium (LTE) is not typically well justified. Incorporating multi-group non-LTE calculations will be important for future work to model the radiation transfer in these low density regions.

For each multi-group run, we map the gray RHD simulation at a certain time as the initial condition. The states of density, velocity, gas and radiation temperatures are used to initialize the multi-group simulations. Then, the comoving frame intensities correspond to a blackbody spectrum at the radiation temperature $T_{\rm rad}$, set by the gray RHD simulation. The intensities inherit the anisotropy from the gray simulation in the comoving frame \citep{jiang2022multigroup}. We tested initializing the intensities to be isotropic in lab frame and did not find significant impact to the emission properties at the photosphere. These multi-group simulations then evolve the hydrodynamics and the coupled radiation transfer. 

We adopt outflow boundary conditions for the hydrodynamic variables in all multi-group runs. The boundary conditions for radiation variables are single-direction outflow, which copies the outward intensities from the last active zone and sets inward intensities to zero. 

\subsection{Optical to UV emission}\label{subsec:mg_opticaluv}
To calculate the SED from the simulations, we collect the radiation flux in the radial direction for each photon energy group $\nu_{i}$ that is integrated over a spherical surface at radius $R$:
\begin{equation}
\begin{split}
    \nu L_{\nu}&=\int_{\nu,i}^{\nu, i+1}L_{\nu}d\nu\\
    &=\int_{\nu,i}^{\nu, i+1}\int_{\phi-}^{\phi+}\int_{\theta-}^{\theta+}F_{\nu, R}R^{2}\sin\theta d\theta d\phi d\nu
\end{split}
\end{equation}
We use the same definition of angular sections as previous sections, and list the $\theta$ and $\phi$ range in Table~\ref{tab:angular_sec}. 

\begin{figure*}
    \centering
    \includegraphics[width=\linewidth]{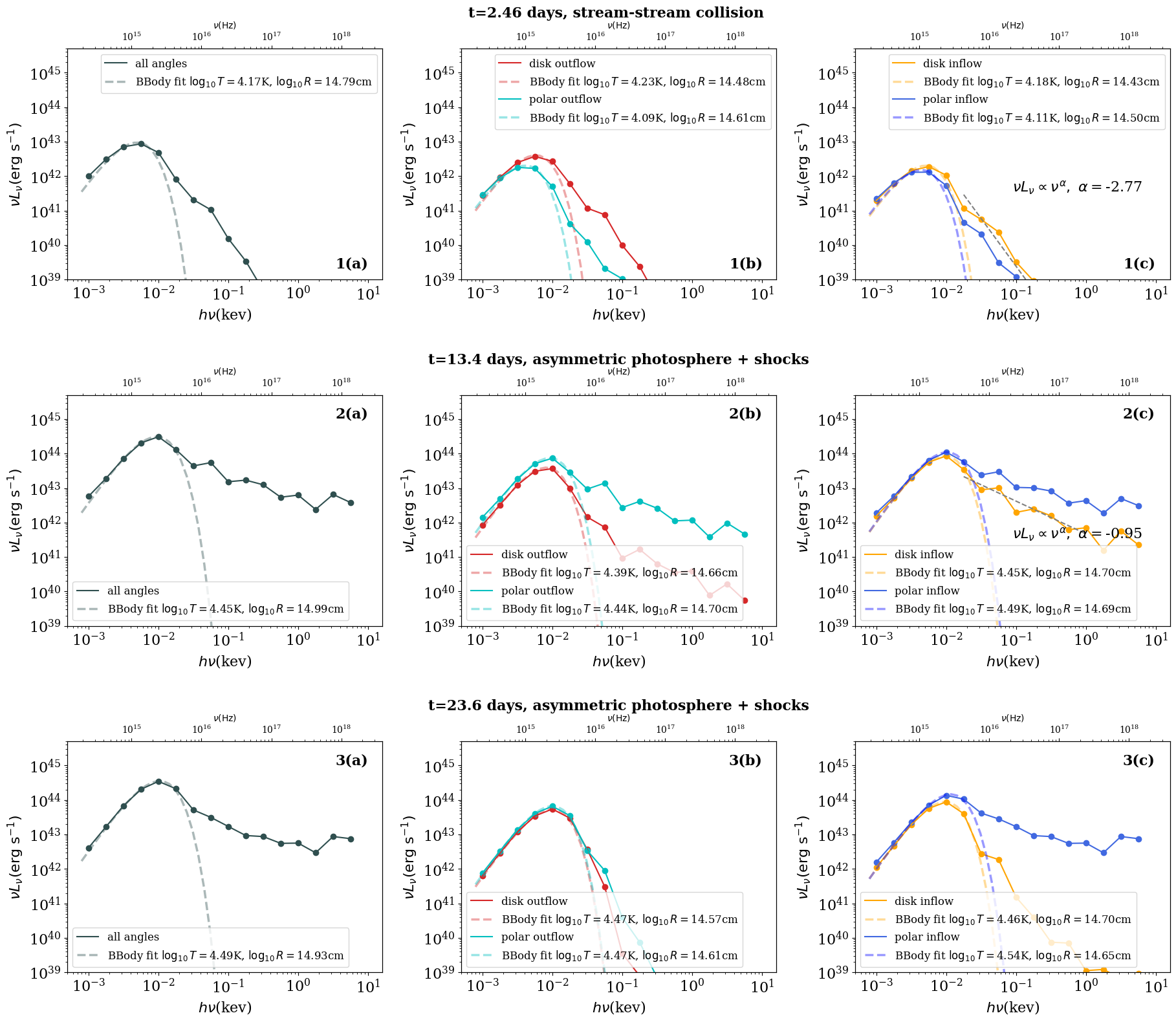}
    \caption{Early time SEDs before the disk formation, mainly driven by shocks and the changing photosphere. From the first row to the third row: at t=2.5, 13.4, 23.6 days. In each row, the left, middle, and right plot are total SED from all angles, the two angular sections near the outflow side, and the two angular sections near the inflow side (defined in Table~\ref{tab:angular_sec} and Figure~\ref{fig:schematic_winddisk}). In each plot, the dashed line with corresponding color is the best-fit black body spectrum using frequency groups from $1.0$eV-$31.6$eV ($2.42\times10^{14}$Hz-$7.64\times10^{15}$Hz), the fitted temperature and size are indicated in the legend. In the first two rows of t=2.5 and t=13.4 days, we also fit an additional power law with $\nu L_{\nu}\propto\nu^{-2.8}, ~\nu^{-0.95}$ for UV-soft X-ray part of SED respectively.}
    \label{fig:multi_spec_time}
\end{figure*}

We show example early SEDs calculated at $R=350\rs=3.1\times10^{14}\rm cm$ in Figure~\ref{fig:multi_spec_time}. The first to third rows correspond to three different times: the stream-stream collision ($t=2.5$ days, first row) and the asymmetric accretion outflow stage ($t=13.4,~23.6$ days, second and third rows) . The first column is the SED calculated with all angles $\theta_{\pm}=0^{\circ}-180^{\circ},~\phi_{\pm}=0^{\circ}-360^{\circ}$, the second and third columns are the two inflow regions and the two outflow regions defined in Table~\ref{tab:angular_sec}. We find that the SED peaks at extreme UV. The SED between $h\nu=1-17.7$eV can be well-fit by a black body component, the soft X-ray tail can extend up to $h\nu=5.6$keV in some angular sections. 

In the rest of this section, we focus on the properties and evolution of the black body component that covers the optical to UV emission. At $t=2.5$ days, soon after the stream-stream collision (Figure~\ref{fig:timeseries_dens_horizontal}), the fallback stream is disrupted, and the post-shock gas is redistributed around $R=\rsi$. In the simulation, the collision velocity is $v_{\rm coll}\approx0.1c$, the stream density is $\rho_{\rm coll}\approx10^{-10}\rm g~cm^{3}$ and the collision angle is $\theta_{\rm coll}\approx145^{\circ}$. A rough estimation of the post-shock temperature is $T_{\rm coll}\approx[\rho_{\rm coll}(v_{\rm coll}\cos(\theta_{\rm coll}/2))^{2}/a_{\rm R}]^{1/4}\approx4.2\times10^{5}$K, assuming all the kinetic energy of the head-on direction velocity is dissipated into radiation energy. We find the post-shock radiation and gas temperature near the collision site is consistent with this estimation within a factor of unity, suggesting the collision site is optically thick.

However, the SED we obtain (the first row of Figure~\ref{fig:multi_spec_time}) suggests a black body temperature $T_{\rm BB}=10^{4.1-4.2}$K, which is lower than the collision site radiation temperature $T_{\rm coll}\sim10^{5}$K. The black body size is on the order of the collision radius $R_{\rm BB}=10^{14.4-14.5}\rm cm\approx3.7\rsi$, suggesting that the radiation emerging from the photosphere is already reprocessed. In our simulation, the strong apsidal precession and near-Eddington mass fallback rate drive the strong stream-stream collision. The reprocessing layer formation is rapid. Thus we do not capture any pre-peak cooling. 

If the reprocessing is slow, for example, when the mass fallback rate is sub-Eddington and thus the stream-stream collision is weaker, the first photosphere temperature may be higher. Compared to the gray simulation, this photosphere size is closer to the frequency-integrated Planck mean photosphere, the frequency-integrated Rosseland mean photosphere is closer to $\rsi$. As discussed in Section~\ref{subsec:mg_setup}, the free-free and Thompson opacity assumption in the low density ambient gas can potentially overestimate the optical depth. The  effects of non-LTE radiation transfer in low density gas will be explored in future work.
\begin{figure}
    \centering
    \includegraphics[width=\linewidth]{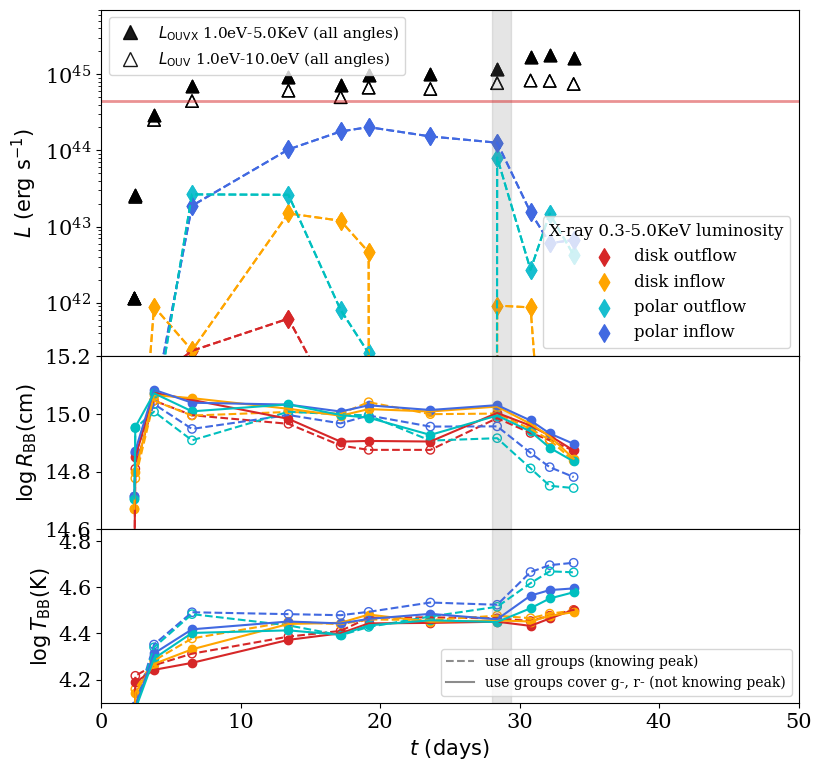}
    \caption{Band-dependent luminosity from multigroup simulations. The first row: the colored diamond data points show 0.3-5.0keV X-ray luminosity in four viewing angle sections defined in Table~\ref{tab:angular_sec}. The filled triangle data points are ``bolometric'' 1.0eV-5keV luminosity, unfilled triangle data points are 1.0eV-10.0eV optical luminosity including all angles. The red horizontal line shows the Eddington luminosity for the black hole. The second and third row: the filled data points and solid line show the best-fit black body size (second row) and temperature (third row) using photon frequency groups of $h\nu=$1.78eV-5.62eV ($4.31\times10^{14}$Hz-$1.36\times10^{15}$Hz, or $\lambda=696.40$nm$-220.57$nm), which do not include the thermal peak. The unfilled circle data points and dashed line show the best-fit black body parameters using all photon frequency groups that cover the thermal peak.}
    \label{fig:multi_lum_bbfit}
\end{figure}

At $t=13.4$ days  (the second row of Figure~\ref{fig:multi_spec_time}), the mass fallback rate approaches the peak, the interaction of the fallback stream and the accretion flow continuously disperses gas. The outflow rate measured at the simulation boundary becomes super-Eddington. The outflow extends the photosphere to $R_{\rm BB}=10^{14.6-14.7}\rm cm$, the black body temperature increases to $T_{\rm BB}=10^{4.3-4.5}$K in all angular sections. The increment of $T_{\rm BB}$ is associated with the increasing average gas and radiation temperature in the eccentric accretion flow. The optical-UV SED of the four angular sections are similar, suggesting that despite the asymmetric geometry, the optical photosphere is relatively independent of viewing angle. The black body component fits the SED up to $h\nu\approx20$eV, above which the SED deviates from the Wien tail. In contrast to the roughly isotropic optical-UV SED, the soft X-ray emission strongly depends on viewing angle and can be described as a power law $ L_{\nu}/\nu\propto\nu^{-\Gamma_{X}}$. We discuss the origin and evolution of this non-blackbody soft X-ray emission in the next section. 

\begin{figure*}
    \centering
    \includegraphics[width=0.8\linewidth]{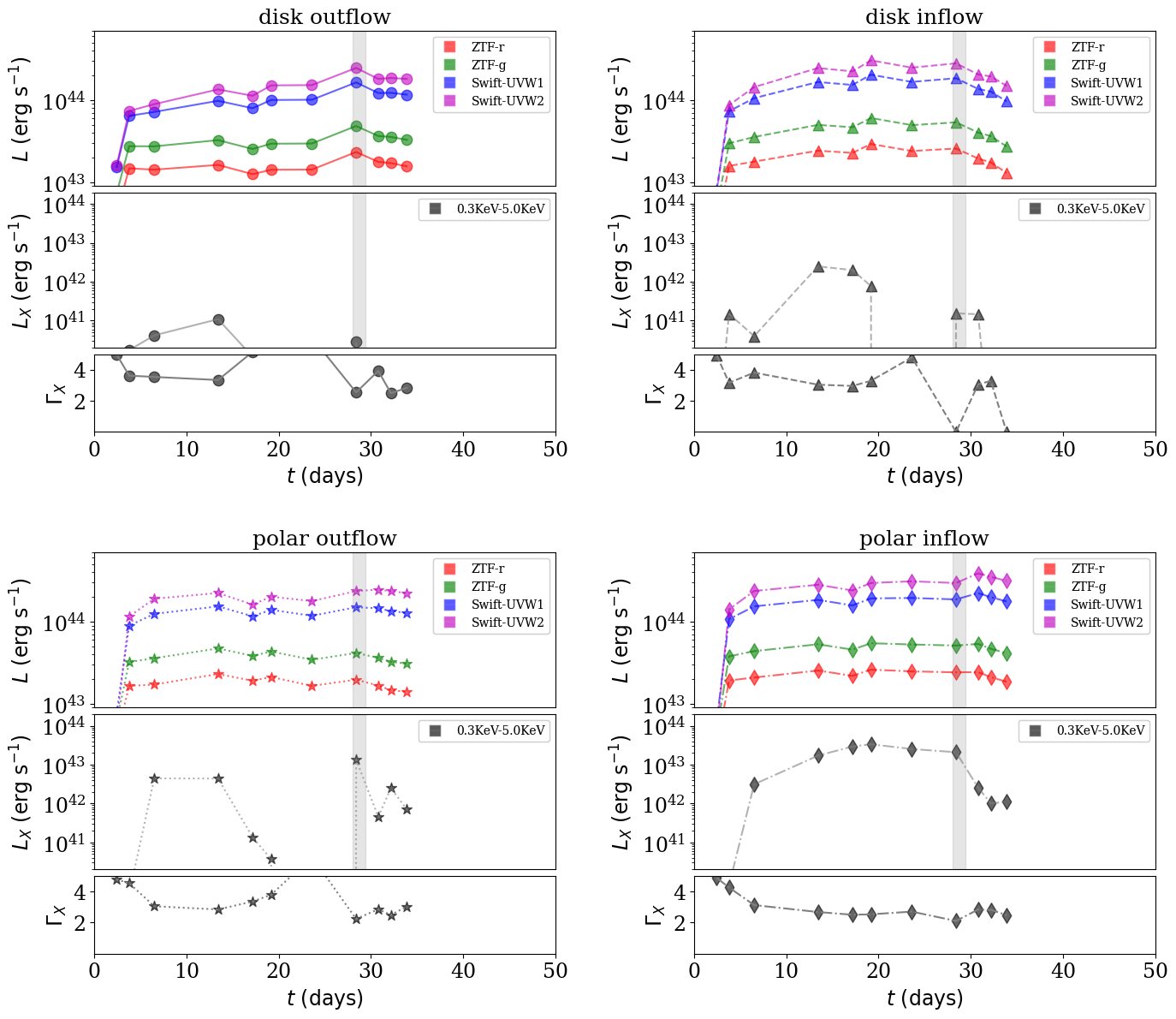}
    \caption{Estimated band-dependent luminosity from the best-fit black body parameters using multi-group simulation SED. In each panel, the first row is UV and optical luminosity: ZTF-r, ZTF-g, Swift UVOT UVW1, UVOT UVW2, calculated from multi-group simulation SEDs with assumed bands information in Table~\ref{tab:band_params}. The second row is 0.3keV-5.6keV soft X-ray luminosity, and the third row is the best-fit photon index $\Gamma_{X}$ of the 0.3keV-5.6keV SED, assuming a power law $F_{\nu}/\nu\propto \nu^{-\Gamma_{X}}$. The four panels correspond to four viewing angle sections, indicated by the title. In each panel, the gray shaded region in the top panel corresponds to the time $t=28.0-29.4$ days, when the polar region is cleared by radiation pressure.  }
    \label{fig:multi_lum_angle}
\end{figure*}

At $t=23.6$ days, the mass fallback rate declines after the peak,  the stream-accretion flow interaction weakens. As a result, the outflow rate decreases, a fraction of bound gas that was previously dispersed to $\gtrsim200\rs$ falls back to the black hole. The photosphere $R_{\rm BB}$ recedes slightly compared to $t=13.4$ days. In the two outflow angular sections (third row of Figure~\ref{fig:multi_spec_time}), the SED can be well fit by blackbody components with similar $T_{\rm BB}=10^{4.5}$K and $R_{\rm BB}=10^{14.5}$cm, the soft X-ray emission decreases below a detectable luminosity.

The disk inflow region shows similar black body temperature and photosphere size as the disk outflow, and the soft X-ray emission also disappears. The polar inflow region shows slightly higher $T_{\rm BB}$, the SED is relatively unchanged compared to $t=13.4$ days.  This is consistent with the geometry that the hot inner flow with temperature $T_{\rm gas}\approx T_{\rm rad}\approx10^{5-6}$K at $R\lesssim\rsi$ is largely obscured by the optically thick disk and outflow, the photons are reprocessed to lower energy at most angles, leaving a small amount of soft X-ray flux leakage from some optically-thin channels. Due to the large range $\Delta\theta$ we chose for angular sections, the polar sections ($|\theta-90^{\circ}|\geq 30^{\circ}$) also include optically thick outflow, yielding the black body component in the SED.

Before an accretion disk forms, all angular sections show optical-UV emission with $T_{\rm BB}\sim10^{4-4.5}$K and $R_{\rm BB}\sim10^{14-14.5}$cm that lack significant temperature variation, which is consistent with rapid photon reprocessing, and the variations within photosphere are smoothed by the long photon diffusion time. By dispersing gas beyond the circularization radius, the stream-stream collision and stream-accretion flow interaction are essential to such rapid reprocessing layer formation. 

The optical-UV part of the SED after a $t=29.4-33.9$ day remains consistent with a black body spectrum. In Figure~\ref{fig:multi_lum_bbfit}, the gray shaded region marks the expulsion of polar region gas and transition to a more circularized disk around $t=29.4$ day. After which, $T_{\rm BB}$ increases and $R_{\rm BB}$ decreases  in all angular sections, suggesting the photosphere is receding on a timescale of $\sim4.5$ days in the simulation. As eccentricity decreases and the polar regions are cleared by radiation pressure, the  temperature in the polar region increases more significantly than in the disk region. The SED peak shifts to the frequency group centered on $h\nu=17.8$eV from its previous peak around the $h\nu=10$eV group. 

The photosphere recession that is captured in the multi-group simulations is also consistent with the gray simulation. It is driven by a combination of subsiding outflows and increasing disk temperature. In the gray simulation, as the outflow weakens and gas falls back to the accretion disk, the disk density profile steepens. When the fallback rate is near the peak and the outflows are strong, the average outflow density profile in Figure~\ref{fig:radial_profile_wind} is roughly $\rho\propto R^{-2.1}$. However, when outflows weaken and a more circularized disk forms, the disk density profile changes to $\rho\propto R^{-3.3}$ in Figure~\ref{fig:radial_disk_profile}. 

We find that during its formation stage, the disk's average density and temperature increase. Therefore, the absorption opacity decreases due to less contribution from photon-ionization opacity in higher temperature. The average photon diffusion time estimated with $t_{\rm diff}\approx \tau_{R}R/c$ decreases a few times in between $100\rs\lesssim R\lesssim 300\rs$. For example, at $R=200\rs$, which is in between the radius of $\tau_{\rm R}=c/v_{R}$ and the radius of $\tau_{\rm R}=1$ (Figure~\ref{fig:radial_disk_profile}), the diffusion time drops from $t_{\rm diff}\approx15.8$ days to $t_{\rm diff}\approx5.3$ days between $t=24.9-34.2$ days during the disk circularization. 

\begin{table}
\centering
\caption{Band center and width}
\label{tab:band_params}
\begin{threeparttable}
\begin{tabular}{lcc}
\hline
Band & Center $\lambda_{\rm center}$ & Width $\Delta\lambda$ \\ 
\hline
ZTF-$r$ & 636.6nm & 155.3nm \\
ZTF-$g$ & 474.6nm & 131.7nm \\
$Swift$-UVW1 & 268.1nm & 65.6nm \\
$Swift$-UVW2 & 208.4nm & 66.8nm \\
\hline
\end{tabular}
\end{threeparttable}
\end{table}

In Figure~\ref{fig:multi_lum_angle}, we estimate the band-dependent light curve for each angular section based on the SED fitting. We first obtain the best-fit blackbody temperature $T_{\rm BB}$ and radius $R_{\rm BB}$ using data from all photon energy groups (i.e. with the information of the SED peak). Then we calculate the luminosity that falls into the $\Delta \lambda=\lambda_{+}-\lambda_{-}$ fitted blackbody continuum $\nu L_{\nu}(\lambda)=\int_{\lambda -}^{\lambda +}L_{\lambda}d\lambda$, where the limits $\lambda\pm=\lambda_{\rm center}\pm\Delta\lambda/2$ are from bandwidth $\Delta\lambda$ and band center $\lambda_{\rm center}$. For the X-ray luminosity that is not well-fit by the black body component, we simply use the photon energy groups $h\nu=0.3-5.6$keV to approximate the canonical $h\nu=0.2-10$keV luminosity. The assumed band center and width are summarized in Table~\ref{tab:band_params}, the estimations of luminosity do not account for the actual filter curve for each band, which should be considered as approximations.

The estimated optical and UV luminosity promptly rises following the stream-stream collision between $t=2.5-6.7$ days, where the optical and UV luminosity increases to $L_{\rm OUV}\sim10^{42}\rm erg~s^{-1}$. The two disk-dominated angular sections roughly peak around $t=28.4$ days, the two polar-dominated angular sections show less clear maximums. The rise of the optical-UV light curve corresponds to the asymmetric accretion flow stage  ($t=13.4,~17.1,~19.2,~23.6$ days). The optical-UV luminosity in the two disk sections rises to the peak with moderate fluctuations. In the two polar sections, the optical-UV luminosities are more constant, with a slight drop at $t=23.6$ days in the polar outflow section and at $t=17.1$ days in the polar inflow section. In all angular sections, the luminosity decreases after $t=28.4$ days when the gas settles down to the accretion disk. Despite these minor differences, the range of luminosity and evolution of \textit{r}-, \textit{g}-, UVW1-, UVW2- bands do not strongly depend on viewing angles. 

\subsection{X-ray Emission}\label{subsec:mg_xray}

\begin{figure*}
    \centering
    \includegraphics[width=0.9\linewidth]{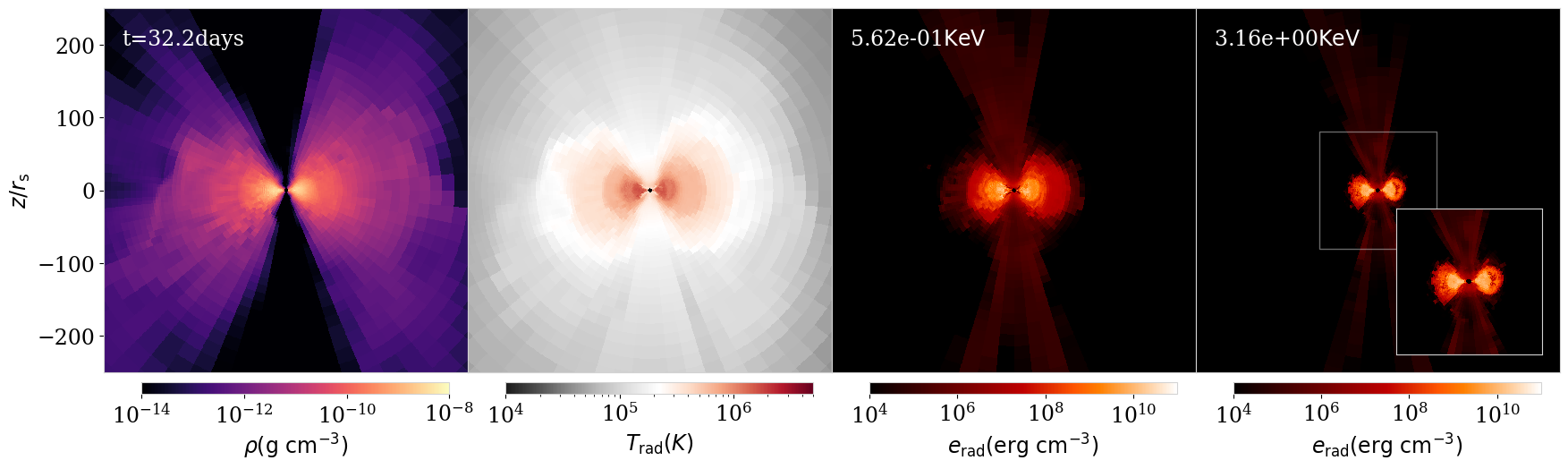}
    \caption{The vertical snapshots ($\phi=150^{\circ}$) of gas density (the first column), radiation temperature (the second column), radiation energy density $\Delta\nu E_{\rm \nu}$ for photon energy group $h\nu=0.18$keV and $h\nu=3.2$keV (third and fourth rows) at $t=32$ days. In the last radiation energy density plots, the inset figure shows the zoom-in inner $80\rs$ view. Comparing to the morphology at t=24.9 days in Figure~\ref{fig:timeseries_denser_vertical}, the polar region is cleared by radiation pressure instead of obscured by outflows.}
    \label{fig:multi_xray_2d}
\end{figure*}

The soft X-ray (0.3-5.6keV) luminosity is highly variable and strongly depends on the viewing angle. In Figure~\ref{fig:multi_lum_bbfit}, the sudden drop of X-ray luminosity in the disk inflow and polar outflow regions around t=17.2 days is when the optically thin channels are obscured by outflows. The re-emergence around $t=29$ days is when the polar region gas is expelled by radiation pressure when the disk forms. Typical soft X-ray SED can be approximated by a power-law, similar to those shown in Figure~\ref{fig:multi_spec_time}, but mainly appears in the polar angular sections. In Figure~\ref{fig:multi_lum_angle}, we show the photon index defined by $F_{\nu}/\nu\propto\nu^{-\Gamma_{X}}$ in the lower panel of each angular section plot. 

We find that the super-Eddington accretion rate is important to produce a soft X-ray luminosity $\nu L_{\nu}\gtrsim10^{41}\rm erg~s^{-1}$, which shows up after $t\gtrsim6.5$ days when the accretion rate exceeds the Eddington limit. Before the disk forms, photons with a few $100$eV energy are emitted from hot gas in the inner accretion flow that is heated to $T\gtrsim10^{6}$K by shocks. Shock-heating efficiently converts kinetic energy to internal energy. Due to the high optical depth, the internal energy rapidly reaches equilibrium with the radiation energy. If the hot photons diffuse through an optically thick region, losing energy by exchanging with gas that is cooled via $pdV$ work, they are reprocessed to optical-UV band. Otherwise, observable $\sim100$eV flux emerges when the photons free-stream from optically thin regions. 

After the disk forms, the geometry is qualitatively consistent with the super-Eddington disk model in \citet{dai2018unified}. Figure~\ref{fig:multi_xray_2d} shows a snapshot of density, radiation temperature, and radiation energy density in $0.2$keV and $3$keV. The gas temperature in the polar region is higher than the radiation temperature as they decouple in the optically-thin region. (However, as noted before, the gas density reaches the numerical floor, so the gas temperature may not be accurate). We reserve more discussion in Section~\ref{subsec:discussion_viewing_angle} and Section~\ref{subsec:discussion_observation}.

The X-ray emission above 1keV may have non-blackbody origin. While the soft X-ray radiation energy is consistent with the emissivity of the inner disk region with $T_{\rm rad}\sim10^{6}$K, the higher energy radiation at $3$keV can exceed the emissivity of the innermost disk. We find that in the X-ray energy range of $h\nu\gtrsim500$eV, the power-law shape is strongly related to the bulk Compton effect in optically-thick flows \citep[e.g.][]{payne1981compton,thompson1994model,socrates2004turbulent,kaufman2018simple}. 

In the simulation, we find that $\Gamma_{X}$ is sensitive to the frame transformation. Without the frame transformation and treating intensity as the same in the comoving and lab frame, we obtain $\Gamma_{X}\gtrsim5$ and $L_{X}\lesssim10^{41}\rm erg~s^{-1}$, where the SED is nearly thermal with temperature and density corresponding to the optical-UV black body fitting. Including the frame transformation yields an enhancement of the radiation energy density in the accretion flow. We tested a number of frequency groups $n_{\rm freq}=12,~16,~20$ and found similar results. Changing the photon energy range between $h\nu=1-10$ eV, $1 - 50$ eV, and $1 - 100$ eV also yields similar soft X-ray SED shapes.

The relatively small $\Gamma_{X}$ is broadly consistent with the spectral shape derived by \citet{payne1981compton}, and the SED shape is qualitatively similar to the AGN soft X-ray excess found in \citet{jiang2025radiation}. The bulk Compton originates from the frame transformation between the comoving and lab frames. However, it differs from the scenario of Doppler shift in uniform velocity fields. In this scenario, if the optical depth $\tau_{\rm div}\sim\kappa\rho(v/\nabla\cdot v)$ is comparable to $c/v$, photons are trapped in the flow and can experience velocity gradients. 

The velocity gradient continuously shifts photon energy during frame transformation. In the multi-group radiation transfer module, we evolve the frequency-dependent intensity $I_{0,f}$ \citep{jiang2022multigroup} by:
\begin{equation}
    \frac{\partial I_{0,f}(\textbf{n}_{0})}{\partial t}=c\Gamma[\rho\kappa_{R,f}(J_{0,f}-I_{0,f})+(\kappa_{P,f}\epsilon_{0,f}-\kappa_{P,f}J_{0,f})]
\end{equation},
where $\Gamma(\textbf{n},\nu)=\gamma(1-\textbf{n}\cdot\textbf{v}/c)$. The subscript $0,f$ represents the comoving frame, frequency-dependent variables. $I_{0,f},~J_{0,f}$ and $\kappa_{P,f}\epsilon_{0,f}$ are the intensity, (angle-averaged) mean intensity and emissivity. The emission, scattering, and absorption are assumed to be isotropic in the comoving frame, so the term $\rho\kappa_{R,f}(J_{0,f}-I_{0,f})$ tends to isotropize the radiation field in the comoving frame. However, in a diverging or converging flow field, the isotropic intensity distribution is only reached locally. Once the radiation field is advected with the fluid to a location with a different velocity field, this term is no longer zero. As a result, the intensity angular distribution is continuously modified, thus radiation flux, as a moment of intensity, is also changed. 

The bulk Compton manifests as an increment of $\partial I_{0,f}/{\partial t}$ due to the nonzero $\rho\kappa_{R,f}(J_{0,f}-I_{0,f})$ term in a varying velocity field, even when the term $\kappa_{P,f}(\epsilon_{0,f}-J_{0,f})=0$ and thermal equilibrium has already been reached. For this reason, the angular distribution of intensities and multi-dimensional velocity field are the key ingredients to capture such bulk Comptonization, which is in principle a generic process as long as gas is compressed by shocks and the coupling between radiation and gas is strong.

\section{Discussion}\label{sec:discussion}
\subsection{Rapid Reprocessing Layer Formation and Nearly-constant Pre-peak Color}\label{subsec:discussion_no_prepeak_cooling}

The optical-UV light curve in Figure~\ref{fig:multi_lum_angle} shows a prompt rise in the first three days, which is driven by stream-stream collisions. We adopt the impact parameter of $\beta=1.73$, such deeper encounter yields a naively compact system: pericenter at $r_{\rm p}= r_{\rm T}/\beta=5.07\rs$ and circularization radius at $r_{\rm circ}\approx2r_{\rm p}=10.1\rs$. As a result, the apsidal precession is strong in the simulation. The stream intersects at $\rsi=76.2\rs$, and the collision velocity is on the order of $v_{\rm coll}\sim0.1c$. The free fall time at the collision site is only $t_{\rm ff}\approx2$ days. By the time of the stream-stream collision, the fallback rate is super-Eddington $\dot{M}_{\rm fb}\approx3\medd$. The high specific kinetic energy and mass flux yield a strong shock between the fallback and returning streams. 

In this limit, the stream-stream collision only powers a prompt optical-UV emission that lasts less than three days, instead of the majority of the light curve rise. Afterward, the accretion flow geometry changes, the optical-UV emission is then primarily powered by the reprocessed emission from shock- and accretion-heated gas by the anisotropic photosphere. The stream-stream collision launches gas from $\rsi$ with high velocities up to $v_{\rm coll}\sim0.1c$, quickly forming an optically thick region that envelops the hot, post-shock site. Therefore, the first light in the simulation is optical-UV bright, tracking the low photosphere temperature of the expanding reprocessing layer.

In this work, the mass fallback rate is near Eddington, $\dot{M}_{\rm fb}\sim\medd$, during the stream–stream collision, and our results are consistent with the near-Eddington scenarios explored in \citet{huang2023bright}. The prescribed $\dot{M}_{\rm fb}$ is informed by the fallback rate from STARs \citep{law2020stellar}, derived from high-resolution, converged stellar-disruption simulations. The rise of the fallback rate is determined by the specific binding-energy distribution of the debris, which depends on spatial resolution and binning methods. For a zero-age main-sequence solar-type star, the resolution reaches $\sim130$ cells per stellar radius and has been shown to be converged \citep{law2020stellar}. Although the details of a realistic fallback scenario remain uncertain, this work carries the caveat of missing the sub-Eddington phase of $\dot{M}_{\rm fb}$.

For more massive black holes, the fallback rate rises more slowly and remains substantially sub-Eddington during the collision, leading to a different pre-peak temperature evolution than what is modeled here. In such cases, we expect the reprocessing layer may take longer to develop, potentially producing a transient decrease in the pre-peak temperature. For example, \citet{huang2023bright} find that when the fallback rate is $\dot{M}_{\rm fb}\sim0.1\medd$, the outflow generated by the stream–stream collision is not optically thick enough to reprocess all high-energy photons into the optical–UV bands. Consequently, the estimated photospheric temperature decreases from $T{\rm rad}\sim10^{5}$K to $T_{\rm rad}\sim10^{4}$K as additional mass is supplied to reprocessing layer, in contrast with the temperature evolution found in this work. 

\subsection{Photon Reprocessing, Blackbody Fitting, and Related Caveats}\label{subsec:reprocessing}
The reprocessing layer first forms from an outflow driven by the stream-stream collision, and it is subsequently supplied by the continuous interaction between the stream and accretion flow. Finally, when a disk forms, it is driven by residual stream-disk interactions and supported by the radiation pressure. At all times, the gas near the black hole is heated by shocks, emitting high energy photons. When the photons diffuse through the reprocessing layer, the SED corresponds to a photosphere temperature $T_{\rm BB}\sim10^{4}$K and photosphere size $R_{\rm BB}\sim10^{14-15}\rm cm$. 

Such a radiation transfer process broadly agrees with the reprocessing wind or envelope picture \citep{strubbe2009optical,dai2015soft,roth2016x,piro2020wind,lu2020self}. For example, adopting the time-dependent wind model discussed in \citep{piro2020wind}, we find that the reprocessing layer falls into the ``thermalization-dominated'' regime. This suggests the coupling between radiation and gas temperature is strong. Such strong coupling and efficient energy exchange between radiation and gas is due to the high absorption opacity $\kappap>\kappar>\kappas$ in the reprocessing layer. In Figure~\ref{fig:radial_disk_profile} , the estimated trapping radius is $r_{\rm tr}\approx100\rs$, outside of which the flux roughly follows $F_{\rm rad,R}\propto R^{-2}$.  However, the $R_{\rm BB}$ fits in Figure~\ref{fig:multi_lum_bbfit} find $R_{\rm bb}>r_{\rm tr}\approx100\rs$. This is consistent with the scenario of thermalization dominated wind: outside the trapping radius, the bolometric luminosity is nearly constant, but the SED shape (and color) still varies.

Appendix~\ref{appendix:bbfitting} shows that at $R\approx\rsi=7\times10^{13}$cm, the SED peaks at $h\nu\sim100$ eV. Beyond $R\gtrsim3\times10^{14}$cm, the SED peak decreases to $h\nu\sim10$ eV and SED shape converges at different radius, suggesting these radii are outside of photosphere, and the photosphere is within the simulation domain. Figure~\ref{fig:multi_bbfit_spec}, Appendix~\ref{appendix:bbfitting} also compares blackbody fits using only the optical-UV band groups with using all frequency groups. This is equivalent to comparing fits of the SED with and without knowing the peak frequency. The two fits are only similar beyond the radius $R \gtrsim 3\times10^{14}$ cm when $T_{\rm BB}\lesssim3\times10^{4}$K. But the two fits diverge at smaller radii, where the true SED peak shifts to $h\nu\sim20$ eV. 

These blackbody fit results are summarized in the $T_{\rm BB}$ and $R_{\rm BB}$ panels of Figure~\ref{fig:multi_lum_bbfit}. We also compare the results of fitting using photon frequency groups roughly covering $g$- and $r$-bands: $h\nu=$1.78eV-5.62eV ($4.31\times10^{14}$Hz-$1.36\times10^{15}$Hz or $\lambda=696.40$nm$-220.57$nm) versus fitting using all photon frequency groups. The two polar regions are more strongly impacted by the limitations of fitting with only groups covering the $g$- and $r$-bands (solid line and filled markers), which tends to underestimate the temperature and overestimate the photospheric radius. This discrepancy becomes most significant once the disk forms and the polar region is cleared by radiation pressure. We discuss the details and caveats of the adopted blackbody fitting approach in Appendix~\ref{appendix:bbfitting}. The comparison in Figure~\ref{fig:multi_lum_bbfit} demonstrates that broadband blackbody fits can be inaccurate when the available bands fall entirely below the SED peak. Adopting more sophisticated Monte Carlo based fitting methods and ensuring convergence could improve the estimates of $T_{\rm BB}$ and $R_{\rm BB}$. Including UV photometry would help reducing the uncertainties in temperature estimations.

The Eddington-limited bolometric luminosity $L\approx5\times10^{44}\rm erg~s^{-1}$ is similar to the luminosity found in recent simulation works \citep[e.g.][]{bonnerot2021formation,steinberg2022origins,ryu2023shocks,price2024eddington, martire2025wind}. For example, \citet{martire2025wind} studied the fallback process following an IMBH disrupts a half-solar-mass star by end-to-end simulation. Interestingly, \citet{martire2025wind} and our work both found that the luminosity emerging from the photosphere was nearly Eddington limited, despite the 300 fold difference in the BH masses and the fact that the fallback rate is substantially more super-Eddington for the IMBH in \citet{martire2025wind}. Furthermore, the unbound fraction of the IMBH TDE wind in their work is significantly higher, consistent with a larger fraction of released energy being converted to kinetic energy rather than radiation. Due to smaller physical extent of the TDE system around the IMBH, the wind in \citet{martire2025wind} is at least 10 times denser than the outflow found in Figure~\ref{fig:radial_profile_wind}, but the velocity at trapping radius is similar. In their scenario, the radiation force does work to gas to expand the wind but also accelerates gas enough to unbind it from the IMBH’s gravity potential.

We find that including the photoionization opacity is important. However, there is the caveat that the TOPs table (multi-group simulation) or OPALs table (gray simulation) may overestimate the Planck mean opacity in the low density region (for example $\rho\lesssim10^{-14}\rm g~cm^{-3}$), where the local thermal equilibrium (LTE) assumption may not hold. More detailed calculations that include non-LTE effects are important \citep[e.g][]{dai2018unified,thomsen2022dynamical}. In this work, the blackbody parameters are similar to the low- to mid-inclination angle examples in \citet{thomsen2022dynamical}. Recently, \citet{mockler2026light} performed time-dependent Monte Carlo radiation hydro simulation to reveal the reprocessing by dense, optically-thick wind builds up during TDE evolution. Their work investigated the opacity and ionization structure in the wind and found consistent result that photo ionization of hydrogen, helium and oxygen are essential to yield photosphere temperature of $\sim10^{4}$K and strong absorption of X-ray photons when the optical depth is high. The SED shape found in this work are also comparable to those early time blackbody parameters in \citet{steinberg2022origins}. However, at a later time we find higher $T_{\rm BB}$ and smaller $R_{\rm BB}$ in polar regions. 

Many recent works study the TDE spectral evolution by post-processing Monte Carlo radiation transfer methods. They enable higher spectral resolution and are more flexible in opacity choices. Our approach of multi-group RHD provides complementary broad-band information that is motivated by three-dimensional hydrodynamics coupled with radiation transfer. Comparison between our results and Monte Carlo-based radiation transfer informs the emission mechanism across spectral bands. 

The optical-UV SED in our simulation is broadly consistent with the one-dimensional post-processing analyses of multi-dimensional simulations in \citep{dai2018unified} and \citep{thomsen2022dynamical}, which suggested that reprocessing by an optically-thick layer with TDE-like parameters is the first-order mechanism to produce optical emission. \citet{parkinson2025multidimensional} and \citet{qiao2025early} incorporate multi-dimensional geometry informed by steady-state super-Eddington disks formed in simulations, allowing photon propagation through varying optical depth. Interestingly, \citet{parkinson2025multidimensional} also found that optical-UV emission shows weaker angular-dependence than higher energy emissions. These consistency supports the idea that the scattering and re-emission in the optically-thick layer can isotropize optical-UV emission, making it visible along most lines of sight, unlike the X-ray emission. 

Notably, the small photon indices $\Gamma_{X}\lesssim2$ of the soft X-ray SED in the polar region after disk formation are a unique feature in our simulation when comparing to the several works discussed above. However, it is relatively uncommon that the late-time TDE spectrum shows  $\Gamma_{X}\lesssim2$ \citep[e.g.][]{guolo2023systematic}. As discussed in Section~\ref{subsec:mg_xray}, we find that the shape is strongly related to bulk Compton. The impact of  bulk fluid motion originates from the radiation field frequency shift due to frame transformations. And it is uniquely captured in the three-dimensional fluid velocity field and by including angular distribution of intensities. 

\subsection{Early Optical-UV Photosphere Evolution is not Accretion-Driven}\label{subsec:discussion_photosphere_evolution}

We define an approximate rise time as the period from when the optical–UV luminosity  becomes detectable at $t\approx2.5$ days to roughly before the photosphere recedes at $t\approx28.4$ days. The rise time  $t_{\rm rise}\approx25.9$ days is broadly consistent with the typical rise time in optical TDEs \citep[e.g][]{gezari2021tidal,yao2023tidal}, despite being on the relatively rapid side. This timescale is largely impacted by the rise time of mass fallback rate $t_{\rm mfb}$. During $t_{\rm mfb}\approx15$ days, the mass fallback rate rises from $\dot{M}_{\rm fb}\approx0.5\medd$ to $\dot{M}_{\rm fb}\approx50\medd$ (Section~\ref{appendix:mdotfallback}). Sometimes $t_{\rm rise}$ is connected to $t_{\rm mfb}$ by assuming the luminosity rise is driven by accretion $L\propto\dot{M}_{\rm acc}$ ,  and the accretion rate is set by the fallback rate $L\propto\dot{M}_{\rm acc}\propto\dot{M}_{\rm fb}$. Alternatively, luminosity rise can be driven by shocks, which are also supplied by fallback rate $L_{\rm shock}\propto\dot{M}_{\rm fb}$. In this work, we find the optical-UV luminosity is unlikely to be directly proportional to accretion rate, but instead associated with the overall energetics of the returning debris stream combined with photosphere evolution. 

Although the accretion rate measured at ISCO increases by an order of magnitude during $t_{\rm rise}$ (Figure~\ref{fig:gray_lum_mdot}), the photon can take a longer time to diffuse out from the innermost disk. For photons emitted in the disk at radius $R$, we estimate the time to arrive at the photosphere roughly by the cooling time $t_{\rm cooling}$:
\begin{equation}
\begin{split}
    &t_{\rm cooling}(R)=\frac{\int_{\theta,\phi}\int_{R}^{R_{\rm out}} e_{\rm rad}4\pi r^2d r d\phi d\theta}{\int_{\theta,\phi}F_{\rm rad}4\pi R^2 d\phi d\theta},\\
\end{split}
\end{equation}
We use the density $\rho$ and opacity $\kappar$ and $\kappas$ from simulations. The polar integration range is $\theta=90^{\circ}\pm30^{\circ}$ and azimuthal integration range is $0^{\circ}<\phi<360^{\circ}$, covering the disk region. For the radiation flux $F_{\rm rad}$ in cooling time, we use both the total flux (approximated by lab frame flux) and ``diffusive'' flux (approximated by comoving frame flux), and refer to $t_{\rm cool}$ and $t_{\rm cool,diff}$. We find that near the circularization radius, $t_{\rm cool}\approx60$ days while $t_{\rm cool,diff}\gtrsim10^{3}$ days. Further out, between the circularization radius and stream self-intersection radius, $t_{\rm cool}\approx10$ days while $t_{\rm cool,diff}\gtrsim90$ days. This suggests that advective radiation flux dominates the energy transport. Variations at the innermost disk within the stream-collision radius will likely be smoothed near the photosphere. The luminosity increment is largely driven by the evolution of the outflow and wind launched by the stream-stream collision and the continuous interaction between the stream and the accretion flow. 

In the last five days, the optical light curve decreases. The black body fitting results in Figure~\ref{fig:multi_lum_bbfit} suggest the photosphere is receding. However, the accretion rate measured at the ISCO during the last five days is still relatively stable at $\dot{M}_{\rm acc}\approx5\medd$ with moderate fluctuations (Figure~\ref{fig:gray_lum_mdot}).  In the inner disk, the viscous time at the tidal radius $r_{\rm T}$ can be estimated as
\begin{equation}
\begin{split}
    t_{\rm vis}(r_{\rm T})&=\sqrt{\frac{GM}{r_{\rm T}^{3}}}(H_{\rm eff}/R)^{-2}\alpha^{-1}\\
    &=12.9\rm~days\left(\frac{H_{\rm eff}/R}{0.3}\right)^{-2}\left(\frac{\alpha}{0.1}\right)^{-1}\left(\frac{r_{T}}{11.3\rs}\right)^{-3/2}
\end{split}
\end{equation}
we adopt $H_{\rm eff}/R=0.3$ and $\alpha\lesssim10^{-1}$, informed by the profiles shown in Figure~\ref{fig:radial_profile_reynoldstress}. At pericenter, $t_{\rm vis}(r_{\rm P})\approx5.7$ days, which can be comparable to the timescale of photosphere change in the last five days. However, as argued earlier, the photon diffusion time at $r_{\rm P}$ and $r_{\rm T}$ are $t_{\rm diff}\gtrsim10$ days due to the large optical depth (Figure~\ref{fig:radial_disk_profile}), so immediate change in the accretion rate can may not be reflected in photosphere variations on the timescale of a few days.

\subsection{Early X-ray Variability and Viewing Angle Dependence}\label{subsec:discussion_viewing_angle}
Recent studies suggest that optically selected TDEs show diverse X-ray-to-optical ratios, motivating studies of how geometry and dynamics shape the observed emission \citep{yao2022tidal, liu2022uv, guolo2023systematic, malyali2023transient, cao2024tidal, ho2025luminous, guo2025reverberation}. A frequently discussed component is the angular dependency of optical depth in TDEs \citep{roth2016x, dai2018unified, thomsen2022dynamical, qiao2025early, parkinson2025multidimensional}. Studies on super-Eddington accretion find that the optically thick disk is often accompanied by optically-thin funnels \citep[e.g.][]{sadowski2016magnetohydrodynamical, dai2018unified, jiang2019super, utsumi2022component, zhang2025aradiation, zhang2025bradiation,fragile2025long}. Photons experience distinct absorption and re-emission when diffusing from the disk or polar regions, supporting the picture of viewing angle dependent SEDs.

In the simulations, the evolution of X-ray variability is primarily driven by geometrical and thermal evolution of the accretion flow. We show in this section that: 1. Before the disk forms, the soft X-ray photons are produced by thermal emission of shock-heated inner accretion flow at $R\lesssim\rsi$; the variability is associated with obscuration by the changing photosphere. 2. After the disk forms, the soft X-ray photons are from thermal emission of the inner accretion disk that is heated by accretion and shocks in the disk; we infer the variability will slow down as the polar region is cleared by radiation pressure, and the variability will then be driven by accretion state change. 3. In all times, a more polar-oriented viewing angle will increase the probability of observing X-ray flux. 

The earliest X-ray emission shows up within 10 days, suggesting that gas has already arrived at the black hole soon after the stream self-intersection . This is in contrast with the idea that the debris forms a quasi-spherical, nearly-adiabatic structure, and will not reach the black hole until a few cooling timescales. However, the processes could be sensitive to debris stream orbital parameters, and may be different with a larger $\rsi$ orbit. 

Before the disk forms, the X-ray variability is primarily from obscuration of shock heated gas near black hole. For example, in the polar outflow angular section (Figure~\ref{fig:multi_lum_angle}), the X-ray luminosity suddenly drops at $t=17.2-28.4$ days when the inner accretion flow is obscured by gas flowing over the pole, and it only reappears after the polar region is cleared by radiation pressure when disk forms. 

As a result, the X-ray variability timescale is related to the dynamical timescale of optically-thick gas that obscuring X-ray flux. At the radius where X-ray flux emerges, the obscuration is mainly from (bound) outflow driven by stream-stream collision or stream-accretion flow interaction. Assuming their velocities are on the order of $v_{\rm out}=0.01c$ (Figure~\ref{fig:radial_profile_wind}), we estimate dynamical timescale by both $t_{\rm dyn}(R)\sim R/v_{\rm out}$ and the Keplerian orbital period $t_{\rm dyn}(R)\sim t_{\rm Kep}(R)$. Near the stream-stream collision radius $\rsi=6.8\times10^{13}\rm cm$, $t_{\rm dyn}\sim\rsi/0.01c\sim 2.6$ days and $t_{\rm Kep}\approx2.3$ days. Near $R=200\rs=1.8\times10^{14}\rm cm$, $t_{\rm dyn}\sim\rsi/0.01c\sim 6.8$ days and $t_{\rm Kep}\approx8.6$ days. These timescales of a few days are broadly consistent with the X-ray variability in simulations before disk formation.

After the disk forms, hot photons are produced at the inner disk, which is heated by accretion and local shocks. When radiation pressure expels remaining gas in the polar region, these photons can propagate through the poles without much reprocessing. In (Figure~\ref{fig:multi_lum_angle}), the sudden increase of soft X-ray luminosity in the polar inflow and polar outflow sections after $t=30$ days is driven by such expulsion of gas near the poles when accretion flow geometry changes. 

We do not model the disk evolution after its formation; therefore, we do not capture X-ray variability at timescale longer than simulation duration. We expect the X-ray luminosity to be more persistent, because the polar region is less obscured. The post-optical peak X-ray variability timescale is likely to be set by the viscous time or mass fallback time that governs the accretion state.

In addition, we find that the expulsion of gas in the polar region is accompanied by inner disk temperature increment. Both are results of the inner disk becoming more optically-thick as the average disk density increases. Consequently, the color related to UV and optical becomes bluer as the X-ray luminosity becomes less variable. We infer that observationally, a possible strong indication of disk formation is the combination of X-ray flux variability slow down and the UV-optical color becomes bluer, if the system is observed at a polar-oriented viewing angle. However, the exact geometry of the polar region can be sensitive to the disk's vertical structure. The presence of a magnetic field or spin-related GR effect may affect the polar region geometry.

The above discussions assume a polar-oriented viewing angle. If we are at a more disk-oriented viewing angle, there may still be transient low X-ray luminosity at early times. In Figure~\ref{fig:multi_lum_angle}, the disk inflow section shows a brief X-ray flare in $t=13.4-19.2$ days, when the inner accretion flow is occasionally exposed. In addition, while not captured in this work, we discussed in Section~\ref{subsec:discussion_no_prepeak_cooling} that a slower formed reprocessing layer can lead to pre-peak cooling. In this work, most of viewing angles produce lower or comparable pre-peak X-ray and optical luminosity. The picture of X-ray flux from incomplete reprocessing will be especially compelling if the pre-peak X-ray luminosity exceeds the optical-UV luminosity \citep[e.g.][]{malyali2023transient}. 

\subsection{Comparison with Observations}\label{subsec:discussion_observation}

In Appendix~\ref{appendix:observation_comparison}, we compare our simulations with observations by several case studies. We present three example TDEs whose light curves can be reasonably matched by the simulations: AT2018hyz \citep{gomez2020tidal,short2020tidal}, AT2019azh \citep{van2021seventeen,hinkle2021discovery,hammerstein2022final}, and AT2020upj \citep{newsome2024mapping,chakraborty2025discovery}; and two example TDEs where the simulated light curves fail to reproduce the observations: AT2019qiz \citep{nicholl2020outflow,nicholl2024quasi} and AT2020dsg \citep{stein2021tidal}. In all comparisons, we only renormalize simulation luminosities. The luminosity renormalizations are within one order of magnitude, and is close to unity in AT2018hyz and AT2019azh. Although rescaling the simulation timescale could, in principle, allow comparisons to a larger range of black hole masses, we leave them to future work. 

AT2019azh (Figure~\ref{fig:appendix_fitlc}, panel (a)) provides the ``best-fit'' example, the simulated and observed optical-UV colors agree reasonably well near the peak \citep{van2021seventeen,hammerstein2022final}. The early X-ray emission is from the occasionally-exposed inner disk, where gas is heated by shocks in the disk. 

AT2018hyz and AT2022upj are the other two ``fitting'' cases. One interesting finding from comparing simulation and observation is the potential to constrain the viewing angle using multiple independent methods. The early X-ray-to-optical ratio provides useful constraints on the viewing angle in the simulations. For example, a viewing angle mostly aligned with the disk midplane and fully obscured by the outflow is unlikely to produce a high X-ray-to-optical in pre-peak to near-peak time. And a viewing angle inclined towards pole is like to show observable X-ray luminosity.

Observations can infer the disk inclination independently. In AT2018hyz (Figure~\ref{fig:appendix_fitlc}, panel (b)), the spectrum evolution reveals a sequence of “double-horned” emission lines. This shape is most likely from disk rotation and thus places strong constraints on the line of sight \citep{short2020tidal}. Using the line of sight inferred from the such observed line profiles, we calculate the band-dependent light curves from the simulation and find good agreement with the observed multi-band light curve. This suggests that both the line profiles and the early X-ray-to-optical ratio can provide independent yet consistent constraints on the viewing angle, motivating future studies. 

In AT2020upj (Figure~\ref{fig:appendix_fitlc}, panel (c)), \citet{chakraborty2025discovery} modeled the late-time UV plateau as a viscously spreading disk \citep{mummery2025fitting}, and found a preferred range of disk inclinations. This inferred inclination $i\sim30^{\circ}$ (assuming $\log M_{\rm BH}=6.5$) is also broadly consistent with the viewing angle range required in our simulations (equivalent to $i<60^{\circ}$) to reproduce the observed early X-ray-to-optical ratio. 

AT2019qiz is a “faint and fast” event with a short evolution timescale \citep{nicholl2020outflow}, requiring a timescale renormalization for comparison. We apply both time and luminosity renormalization in Figure~\ref{fig:appendix_fitlc} (panel (d)), but leave it as a ``non-fit'' example. However, both AT2019qiz and AT2020upj show non-canonical properties, such as coronal lines or persistent infrared flare. These late-time evolution hint at potential AGN-like activities or environment \citep{short2023delayed, newsome2024mapping, chakraborty2025discovery, wu2025torus} and the TDE dynamics could be different from our simulations. 

For AT2019dsg, the early X-rays are brighter than the optical emission, and the early X-ray spectrum is well fitted by a Wien tail \citep{stein2021tidal,robert_stein_2021_4621671,van2021seventeen}. Reproducing such a high X-ray-to-optical ratio near the optical peak in our simulations requires the viewing angle to be fine-tuned. Instead, we show a broader range of viewing angles that produce high X-ray-to-optical ratios in the simulations (Figure~\ref{fig:appendix_fitlc}, panel (e)). However, in these viewing angles, the soft X-ray SEDs in the simulations are shallower than a Wien tail. 

There are several caveats associated with the X-ray SED in the simulations. We find that the highest photon frequency groups in the simulation are affected by the large fluid velocities in the polar regions after disk formation, reaching $v_{\rm R}\sim0.2-0.4c$. With these sub-relativistic and spatially non-uniform velocity fields, frequency shifts during the frame transformation can be sensitive to the finite frequency-group discretization. Moreover, when the density in the polar regions reaches the numerical floor after disk formation, the thermodynamics in those regions are not accurately captured. Future work incorporating thermal Compton and multi-fluid physics might be needed to model the X-ray spectrum more accurately.
Simply accounting for the radial direction radiation flux, we do not find any viewing angle that produces bright X-rays with no detectable optical–UV emission, which is in contrast with existing X-ray-only TDEs \citep[e.g.][]{saxton2020x, sazonov2021first, grotova2025population, eyles2025nine}. More detailed photon transport methods, such as ray tracing \citep{steinberg2022origins,price2024eddington,hu2024optical,mummery2025optical} or Monte Carlo radiation transfer \citep{dai2018unified,thomsen2022dynamical,parkinson2025multidimensional,qiao2025early} may be necessary to capture the precise evolution of the x-ray-to-optical ratio.

\subsection{Shocks Produce Photons and Transport Angular Momentum }\label{subsec:discussion_shock}
Shocks play a key role in driving the emission and dynamics in the simulations. Among the various shocks, the shock between stream and accretion flow is a persistent feature that constantly contributes to the radiation production. Such stream-disk shock was proposed as the leading emission mechanism around optical peak in \citep{steinberg2022origins}, and found to be important in previous works \citep[e.g.][]{andalman2022tidal,curd2023strongly,huang2024pre}.  We consistently find a radiation energy enhancement that traces the stream-disk shock in Figure~\ref{fig:errhovv_streamdisk} in early times. In $t=34.2$ days, as the stream-disk shocks weaken, the radiation energy density distribution becomes more uniform in the disk. 

In addition to the stream-disk shock, there are shocks and shear interactions between the eccentric and the circularized disk gas in the disk, which we refer as circularization shock. After the disk forms and the stream-disk shock weakens, the circularization shock drives velocity dispersion in the disk and operates as effective viscosity, leading to global hydrodynamical stress. In a companion work that includes magnetic fields, \citet{meza2025radiation} show that the hydrodynamic Reynolds stress is substantially larger than the Maxwell stress. Therefore, early TDE accretion flows may spend weeks in a hydrodynamic Reynolds stress dominated phase. When the flow eccentricity decreases, the Reynold stress also decreases (Figure~\ref{fig:radial_profile_reynoldstress}). Capturing magnetic fields and their contribution to angular momentum transport will be increasingly important in the later stage. 

We performed angular momentum analysis for regions near the orbital plane $|\theta-90^{\circ}|<30^{\circ}$ that is similar to \citet{meza2025radiation}. We find that the $\theta-$ direction angular momentum flux is the dominant term. This term corresponds to the vertical outflow and inflow from the disk surface. It is mainly driven by the stream-disk shock that launches outflow from the orbital plane, and the bound part of the outflow will fall back to orbital plane (similar to Figure~\ref{fig:schematic_winddisk}). Such dynamical geometry leads to $\theta$-direction negative and positive angular momentum and mass flux. The hydrodynamic Reynold stress is the next leading term at all radii, driving the radial angular momentum transport. The torque from radiation force is subdominant.

During the final stage of the simulation, a pair of stationary acoustic spiral shocks emerges near the black hole. The stream impacts the accretion disk, creating a non-axis symmetric perturbation to the disk. We estimate the Mach number by $\mathcal{M}=|v|/\sqrt{(4P_{\rm rad}+5P_{\rm gas})/3\rho}$. The internal energy $E_{\rm IE}=P_{\rm gas}/(\gamma-1)$ is defined by the adiabatic gas EoS. The three diagonal components of the radiation pressure tensor are similar and are close to $E_{\rm rad}/3$ in the optically-thick central disk, so we use the $P_{rr}$ component for $P_{\rm rad}$. We calculate the average Mach number at the shock front, finding $\overline{\mathcal{M}}\approx4.3$. We assume the $m=2$ mode and fit the shock front with a hydrodynamic acoustic wave dispersion relation \citep{binney2011galactic}, shown as the black dotted line in the last column of Figure~\ref{fig:errhovv_streamdisk}. 

The spiral pattern extends to roughly the circularization radius, where the stream dissolves in the disk. The spiral shocks dominate the hydrodynamic Reynold stress in the innermost disk at late times, driving accretion via efficient angular momentum transport. The accretion removes gas from the innermost disk and modifies the local density profile. In Figure~\ref{fig:radial_disk_profile}, the disk average density beyond the circularization radius follows $\rho\propto r^{-3.3}$, but it inverts to $\rho\propto r^{0.9}$ within the circularization radius, where the spiral shocks operate.

Similar spiral patterns are observed and discussed in previous works \citep{wevers2022elliptical,bonnerot2020simulating,ryu2023shocks}. In our setup, the spiral shocks do not extend to the outer disk. They are instead truncated near the circularization radius, which is similar to what was found in \citet{ryu2023shocks}. \citet{bonnerot2020simulating} studies stream-stream collision with similar parameter regime as this work. They model the disk feeding as stellar debris shedding from the stream-stream collision site, so that the non-axisymmetric perturbation is located beyond the disk, instead of being confined in the inner disk. This allows the spiral shocks to grow to the outer disk in their work. 

The angular momentum transport may be affected by neglecting the magnetic field in our simulations. A handful of previous works explored the effect of Maxwell stress and found a relatively small contribution to angular momentum transport. With a weak poloidal field added to the debris stream, \citet{sadowski2016magnetohydrodynamical} found similar thermally-driven outflows from the stream-stream collision. They found the Maxwell stress is subdominant compared to the Reynold stress.  Similarly, when \citet{curd2023strongly} injected a stream with a weak initial poloidal field into a pre-existing magnetized disk, they found that despite the magnetic field amplification in the disk, the magnetic pressure is subordinate relative to the thermal pressure in the disk. \citet{meza2025radiation} went further and explored different magnetic field topologies in the injected stream. With a large prescribed magnetic field in their simulations, the magnetorotational instability can be resolved. However, they still find a weak Maxwell stress relative to the Reynold stress and radiation pressure dominated dynamics in the early TDE accretion flow. Maxwell stress is an important component in late-time disk modeling \citep[e.g.][]{piro2025late,alush2025late}, and the growing sample of jetted TDE candidates also motivates understanding the magnetization of TDE disk, making it a pressing issue for future studies.

\section{Summary}\label{sec:summary}
We study the disk formation following debris stream self intersection for a spin-less, $M_{\rm BH}=3\times10^{6}M_{\odot}$ black hole with an impact parameter $\beta=1.73$. With these assumed parameters, the stream-stream collision occurs close to the black hole $\rsi=76.2\rs$. We evolve the TDE fallback and circularization process for about a month after the stream-stream collision. 
We find that the immediate post-collision gas is heated to $T\sim10^{5}$K, while the collision also launches outflow and redistributes bound gas to form an asymmetric, optically-thick layer around the black hole. This layer reprocesses the hot post-shock gas and leads to radiation emission peaks in the extreme UV in most viewing angles. This dynamical evolution shows differences from the scenarios for smaller black hole mass or smaller $\beta$, where the stream self-intersection occurs further from the black hole and the accretion of post-shock gas may be delayed by the free-fall time. We study the dynamics with three-dimensional, frequency-integrated radiation hydrodynamic simulations, and obtain the band-dependent dynamics and emission across different stages with a series of three-dimensional, multi-group radiation hydrodynamic simulations with 16 to 20 frequency groups. We summarize our main findings as follows:

$\bullet$ The reprocessing layer forms rapidly when apsidal precession is strong and the fallback rate is near super-Eddington. The initial stream-stream collision launches optically thick outflows but yields only a prompt $L_{\rm OUV}\sim10^{42-43}\rm erg~s^{-1}$. The optical–UV rise is instead driven by various shocks in the accretion flow, and the photosphere is formed from the continuous outflow driven by stream-accretion flow interactions. We do not observe pre-peak cooling because of such rapid formation of the reprocessing layer in our simulation.

$\bullet$ The formation of a more circularized disk takes about three weeks after the stream-steam collision. Before that, the accretion flow is eccentric and asymmetric. The interaction between the fallback stream and the eccentric flow continuously disperse gas, driving strong vertical inflows and outflows. The accretion flow geometry is more optically thick in the orbital plane and more optically thin in the polar region. The polar region can be obscured when the outflow is strong near the peak fallback rate.

$\bullet$ The asymmetric accretion flow is radiation-supported. The optically thick outflow reaches an average radial velocity of $v_{R}\sim0.01c$ and density profile $\rho\propto R^{-2.1}$ (Figure~\ref{fig:radial_profile_wind}), but most of it remains bound to the black hole. The mass outflow rate exceeds the Eddington rate at $t\sim10$ days, peaks at $\dot{M}_{\rm out}\sim10\medd$ around $t\sim18$ days, and stays super-Eddington thereafter (Figure~\ref{fig:gray_lum_mdot}). The unbound mass outflow is sub-Eddington until the total outflow rate peaks and remains $\dot{M}_{\rm out,unbound}\sim2$-$3\medd$ afterward. The accretion rate becomes super-Eddington about three days after the collision and remains $\dot{M}_{\rm acc}\sim5-10\medd$ with moderate variations.

$\bullet$ About 20 days after the stream–stream collision, the fallback rate starts to decline from its peak and the stream density decreases. Meanwhile, the accretion flow density increases as gas accumulates around the black hole. The ram pressures of the stream and the accretion flow become comparable at $r_{\rm p}\lesssim R\lesssim \rsi$ (Figure~\ref{fig:errhovv_streamdisk}). The stream no longer pierces through the disk but dissolves near the pericenter. The stream-accretion shock weakens, and the outflow rate decreases. The accretion flow eccentricity continues to drop, forming a more circularized disk.

$\bullet$ The formed disk extends well beyond the circularization radius. Most of the mass lies between the circularization radius and the stream-stream collision radius $\rsi$, while the photon-trapping radius is about $100\rs$ (Figure~\ref{fig:radial_disk_profile}). Vertically, the disk is geometrically thick with $H_{\rm eff}/R\sim0.3$ (Figure~\ref{fig:radial_profile_reynoldstress}) and optically thick with vertical optical depth of $\tau_{\theta}\sim10^{2}$ (Figure~\ref{fig:vertical_disk_acc}). The disk is supported by radiation pressure, with the radiation force balancing gravity. The hydrodynamic Reynolds stress decreases as the flow circularizes. Incorporating magnetic fields to account for Maxwell stress will be important for modeling angular momentum transport in the post-peak disk.

$\bullet$ The pre-peak emission is largely driven by shocks in the accretion flow rather than accretion itself. The optical luminosity reaches $L_{\rm O}\approx10^{43}\rm erg~s^{-1}$ and the UV luminosity $L_{\rm UV}\approx10^{44}\rm erg~s^{-1}$. The broadband SED is well approximated by a blackbody continuum for $h\nu\approx1$–$40$ eV (Figure~\ref{fig:multi_spec_time}). The inferred blackbody parameters, $\log (R_{\rm BB}/{\rm cm})\sim14.5$-$15$ and $\log (T_{\rm BB}/{\rm K})\approx4.2$-$4.4$ (Figure~\ref{fig:multi_lum_bbfit}), are in broad agreement with those of observed optical TDEs. Despite the super-Eddington accretion rate, the bolometric luminosity remains Eddington-limited. The observed radiation efficiency is $\eta_{\rm rad,fb}\approx0.5\%-1\%$, which estimates the fraction of emitting luminosity at photosphere relative to the mass fallback rate.

$\bullet$ Soft X-ray emission appears before disk formation and is visible only at limited viewing angles. Seed photons from the shock-heated inner flow (with $T_{\rm gas}=T_{\rm rad}\gtrsim10^{6}$ K) escape through optically thin polar regions. The $h\nu\approx0.1$–$1$ keV SED roughly follows a power law with $\Gamma_{X}\approx2-4$, the soft X-ray luminosity reaches $L_{X}\gtrsim10^{42}\rm erg~s^{-1}$ (Figure~\ref{fig:multi_lum_angle}). Bulk Comptonization shapes the SED above $h\nu>10$ eV once accretion becomes super-Eddington. Without frame transformations, the SED is closer to blackbody and $\Gamma_{X}\gtrsim4$, the soft X-ray luminosity is also significantly lower.

$\bullet$ The optical-UV luminosity and SED shape are nearly independent of viewing angle, while soft X-ray luminosity varies strongly across angular sections. All four angular sections show consistent optical–UV emission but differ in their x-ray-to-optical evolution (Figure~\ref{fig:multi_lum_angle}). By changing viewing angles, the simulation can reproduce X-ray dim and optical-UV only events, near peak X-ray flare events, and X-ray brightening events. Although a steady-state super-Eddington disk has not yet formed, the flow geometry (Figure~\ref{fig:multi_xray_2d}) is broadly consistent with the disk geometry in the viewing-angle dependent emission models.

$\bullet$ We compare the multi-group simulation with the multi-band light curves of a few example TDEs with near optical-peak X-ray emission in Appendix~\ref{appendix:observation_comparison}. We show that the early-time X-ray-to-optical ratio is sensitive to the viewing angle, and a more polar-oriented viewing angle yields a brighter X-ray luminosity, therefore can potentially provide an independent constraint on the disk inclination complementary to emission line profile or late-time disk modeling.

\begin{acknowledgments}
    We thank Daichi Tsuna, Sierra Dodd, Samantha Wu, Lizhong Zhang, Tony Piro, Wenbin Lu, Brian Metzger for frequent discussions and their insights on the shock and accretion processes. We also appreciate Yuhan Yao, Erica Hammerstein, Jean Somalwar, Matthew Nicholl for sharing observational data and their helpful insight into this work. XH thanks the discussions with Jane Dai, Elad Steinberg and Daniel Kasen on radiation transfer processes. XH is supported by the Sherman Fairchild Postdoctoral Fellowship at the California Institute of Technology. This work used Stampede 3 at Texas Advanced Computing Center through allocation TG-PHY240041 from the Advanced Cyberinfrastructure Coordination Ecosystem: Services $\&$ Support (ACCESS) program. Support for this work was provided by the National Science Foundation under grant 2307886. This work also made use of UVA's High-Performance Computing systems. Resources supporting this work were also provided by the NASA High-End Computing (HEC) Program through the NASA Advanced Supercomputing (NAS) Division at Ames Research Center. XH appreciates the hospitality and interactions during the tde24 and its follow-up workshop, which is supported by the NSF grant PHY-2309135 to the Kavli Institute for Theoretical Physics (KITP). This research also benefited from interactions that were funded by the Gordon and Betty Moore Foundation through Grant GBMF5076. The Center for Computational Astrophysics at the Flatiron Institute is supported by the Simons Foundation. SWD acknowledges funding from the Virginia Institute for Theoretical Astrophysics (VITA), supported by the College and Graduate School of Arts and Sciences at the University of Virginia.
    
\end{acknowledgments}

\appendix
\section{Mass Fallback Rate}\label{appendix:mdotfallback}
\begin{figure}
    \centering
    \includegraphics[width=0.6\linewidth]{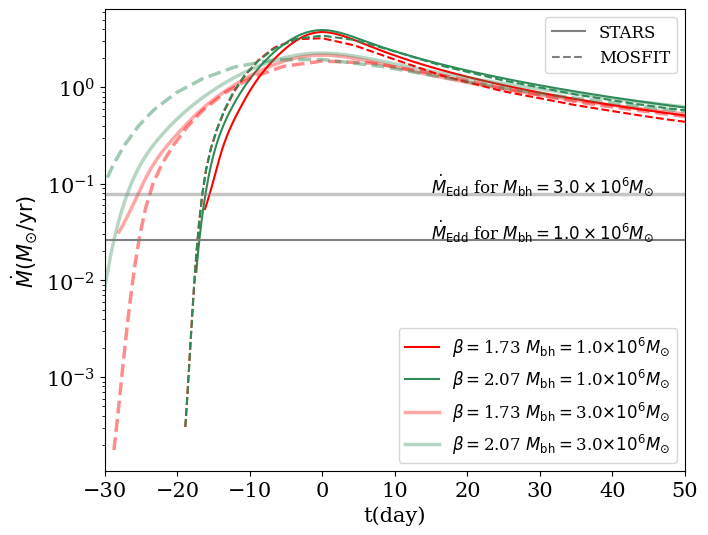}
    \caption{Mass fallback rate of solar type star disrupted by $M_{\rm BH}=10^{6}M_{\odot}$ black hole (thin lines) and $M_{\rm BH}=3\times10^{6}M_{\odot}$ black hole (thick lines). For each black hole mass, the horizontal line shows the Eddington accretion rate, the red line and green line are $\beta=1.73$ and $\beta=2.07$ disruptions. For each $\beta$, the solid line is the result from STARS, and the dashed line, from MOSFIT.}
    \label{fig:mdotfb_STARS_mosfit}
\end{figure}
In Figure~\ref{fig:mdotfb_STARS_mosfit}, we compare the mass fallback rates informed by MOSFiT (dashed lines) and the STARS library (solid lines). While MOSFiT is based on polytropic models and STARS adopts MESA-informed stellar structures, both approaches yield similar fallback rates for a zero-age main-sequence, solar-like star TDE with a moderate impact parameter $\beta$. In our simulations, we adopt $\beta=1.73$ and use the fallback rate shown by the red solid line. The assumed black hole mass is $M_{\rm bh}=3\times10^{6}M_{\odot}$, which introduces the caveat that the fallback rate is slightly artificially accelerated in the simulation, since $\dot{M}$ evolves more rapidly for smaller black hole masses. As a result, the rise time may be underestimated. However, in either case, the fallback rate remains sub-Eddington only briefly, which in our simulations leads to strong stream–stream collisions and the rapid formation of a reprocessing layer. Accurately modeling stellar disruption as an initial condition for the fallback process is therefore crucial for studies of pre-peak dynamics \citep{guillochon2014ps1,steinberg2022origins,ryu2023shocks,price2024eddington}.

\section{Comparison to Early X-ray Bright Optical TDEs}\label{appendix:observation_comparison}
\begin{figure*}
    \centering
    \includegraphics[width=0.4\linewidth]{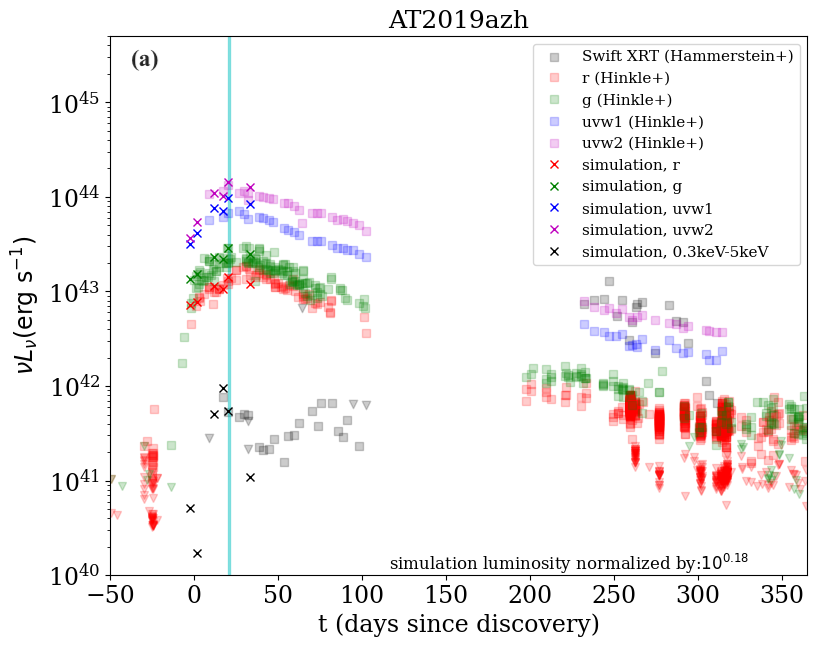}
    \includegraphics[width=0.4\linewidth]{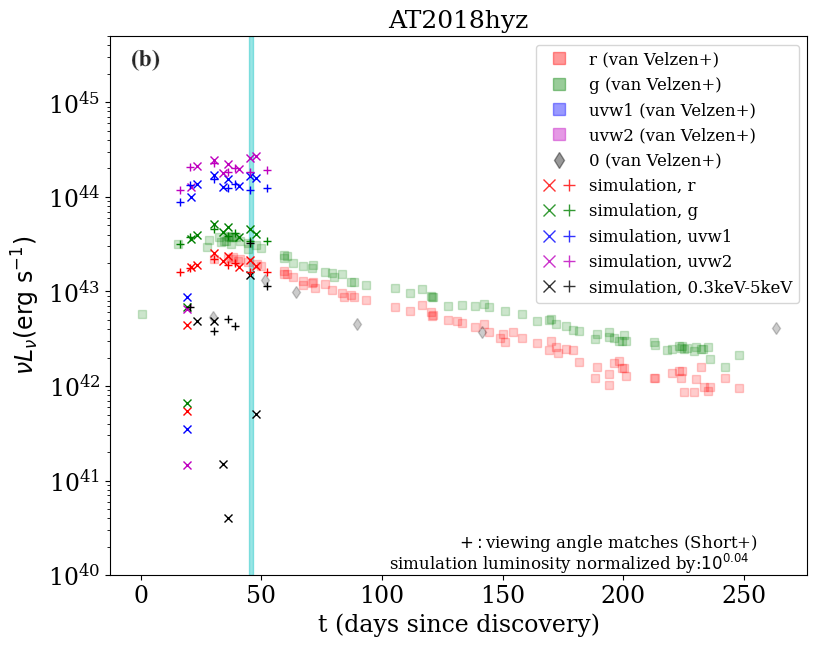}
    \includegraphics[width=0.4\linewidth]{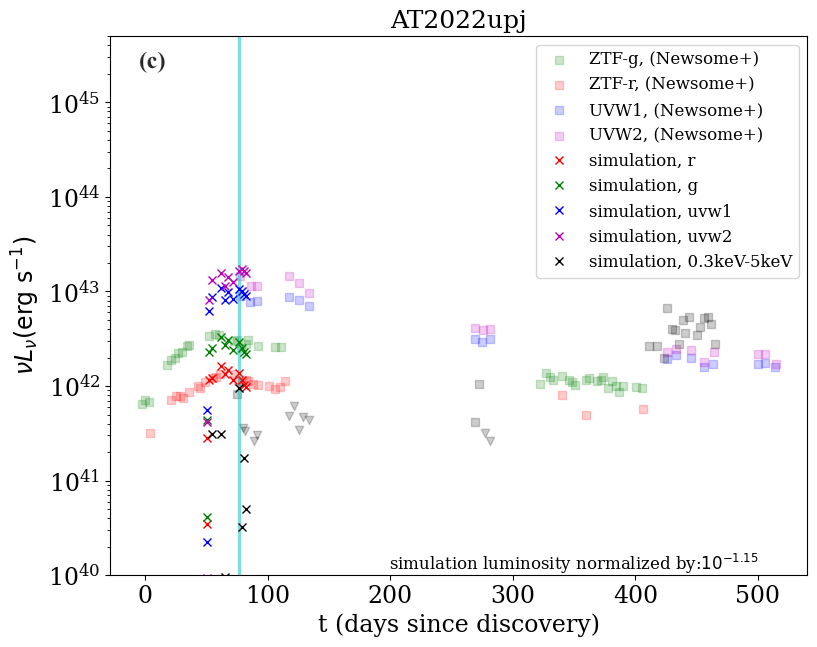}
    \includegraphics[width=0.4\linewidth]{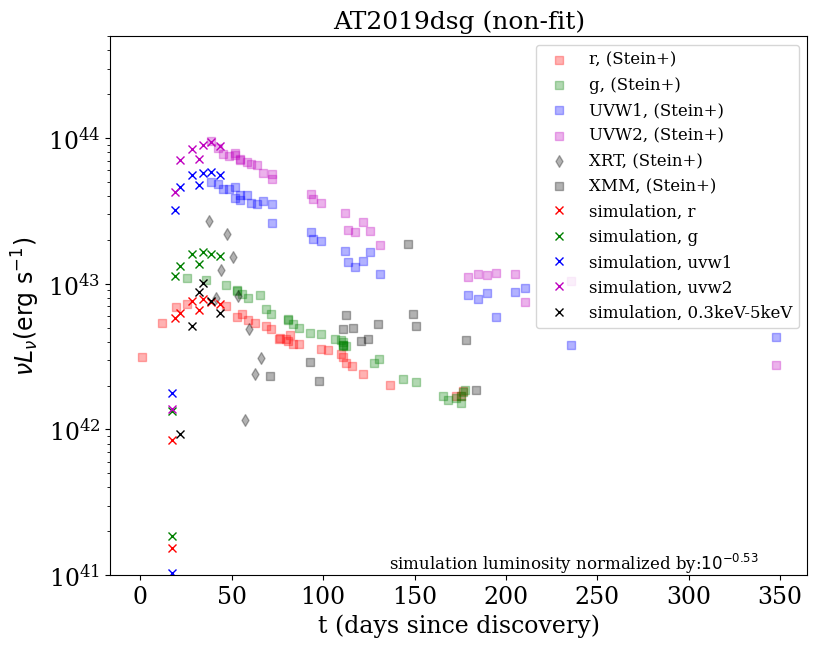}
    \includegraphics[width=0.4\linewidth]{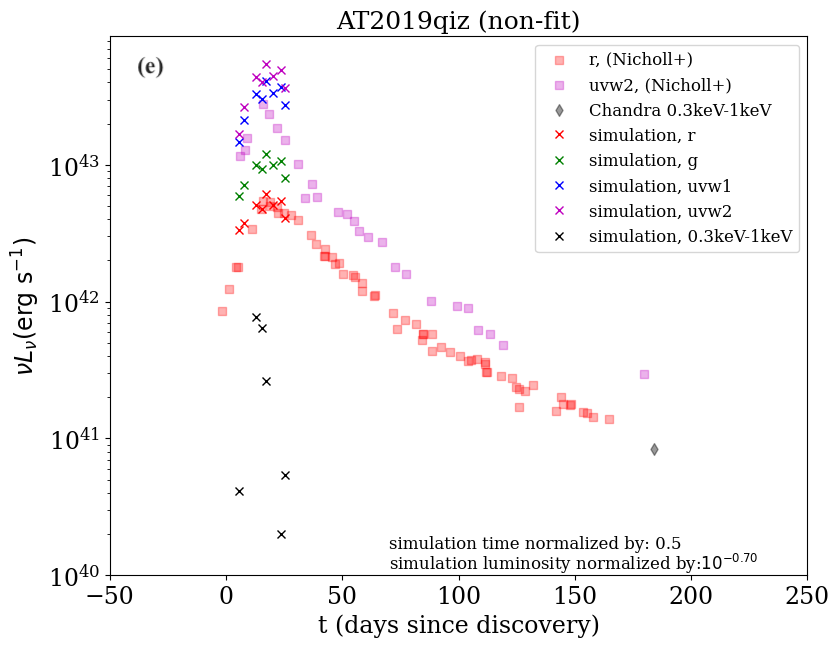}
    \caption{Comparing band-dependent light curve with observation. In each panel, the crosses are the estimated luminosity from simulations, the squares are observed luminosity generated with OTTER \citep{Franz2025OTTER}, except for panel (a) Swift XRT data from \citet{hammerstein2022final}, and (e) from \citet{newsome2024mapping} and \citet{chakraborty2025discovery}. The observation data points are not selected against potential incomplete host subtraction flags in OTTER. Panel (a)-(c) are examples TDEs that has early X-ray to optical ratios comparable to simulated light curves from multi-group runs, the vertical cyan region marks the rough time of disk formation in simulations. Panel (d) and (e) show two examples TDEs that the simulations cannot explain multi-band emission regardless of viewing angle. The simulation disk equatorial plane is at $\theta=90^{\circ}$, the polar angles ranges from $\theta\in(0,~180^{\circ})$, all simulation luminosities are calculated from a range of $\theta$ and $\phi$ then normalized to $4\pi$, before further normalized as noted in the lower right text. (a) simulation viewing angle $\theta\in[90^{\circ}-10^{\circ},~90^{\circ}+30^{\circ}],~\phi\in(0.02\pi+90^{\circ}, 0.02\pi+270^{\circ})$, luminosity normalization is $10^{0.18}$. (b) simulation viewing angle $|\theta-90^{\circ}|>30^{\circ},~\phi\in(31.9^{\circ}-90^{\circ}, 31.9^{\circ}+90^{\circ})$. (c) OUVX data are from \citet{newsome2024mapping}, simulation light curve from viewing angle $|\theta-90^{\circ}|>30^{\circ},~\phi\in(31.9^{\circ}-90^{\circ}, 31.9^{\circ}+90^{\circ})$. (d) OUVX data are from \citet{van2021seventeen}, simulation light curve from viewing angle $|\theta-90^{\circ}|>30^{\circ},~\phi\in(31.9^{\circ}-90^{\circ}, 31.9^{\circ}+90^{\circ})$. (e) OUVX data are from \citet{nicholl2020outflow}, simulation light curve from viewing angle $\phi\in(31.9^{\circ}-90^{\circ}, 31.9^{\circ}+90^{\circ})$, the vertical dashed line labels the range of luminosity from $|\theta-90^{\circ}|>30^{\circ}$ and $|\theta-90^{\circ}|\leq30^{\circ}$. }
    \label{fig:appendix_fitlc}
\end{figure*}

We compare five optically bright TDEs that show pre-peak to near-peak X-ray detections with band-dependent light curves derived from multi-group simulations (Figure~\ref{fig:appendix_fitlc}). Similar to Figure~\ref{fig:multi_spec_time} and Figure~\ref{fig:multi_lum_angle}, we calculate the total radiation flux for each photon frequency group at $R=350\rs$, and then fit a blackbody component and estimate band-dependent luminosities. In each case, we place rough constraints on the viewing angle by finding the optimal $\theta$ and $\phi$ range to reproduce the observed broad-band light curve. For the viewing angle range, the disk equatorial plane is at $\theta=90^{\circ}$, the initial stream-stream collision is at $\phi_{\rm SI}=31.9^{\circ}$. We normalize the simulated luminosities in all bands by the same factor. Effectively, we aim to reproduce the color evolution and X-ray to optical ratio by only varying viewing angles and an overall normalization, which are within order of unity except for AT2022upj. 

The five TDEs have black hole masses roughly in the range of $M_{\rm BH}\sim10^{6}M_{\odot}$ reported in the literature. We reiterate the caveat that the assumed fallback time is for $M_{\rm BH}=10^{6}M_{\odot}$, but the black hole mass in the simulation is $M_{\rm BH}=3\times10^{6}M_{\odot}$. More details are discussed in Appendix~\ref{appendix:mdotfallback} along with a comparison of fallback rates to indicate differences in the fallback time.

AT2019azh, AT2018hyz, AT2022upj are the three examples where the simulation can roughly reproduce observed color evolution and X-ray to optical ratios. The majority of optical and UV data are accessed through OTTER\citep{Franz2025OTTER}, without selection against potential incomplete host subtraction. We also do not show the error of each observation data point. The cyan region rouhgly marks the time of disk formation in the simulations. 

In AT2019azh, the optical and UV data are from \citet{hinkle2021discovery}, the viewing angle range is $\theta\in[90^{\circ}-10^{\circ},~90^{\circ}+30^{\circ}],~\phi\in(3.6^{\circ}+90^{\circ}, 3.6^{\circ}+270^{\circ})$, the luminosity normalization is small $10^{0.18}$. This viewing angle range largely overlaps with the disk region and through the shock-driven optically-thick outflow, which gives rise to the optical-UV emission, but the viewing angle is slightly tilted towards the polar region to intercept the soft X-ray flux. Several works discussed that its $g$-band light curve shows slope variation during the rise phase \citep{liu2022uv, faris2024light}. In particular, \citet{liu2022uv} tested the scenario that the $g$-band bump originates from the stream-stream collision. However, in the simulation, the variations in slope are mainly from the photosphere variation. The stream-stream collision powers a prompt initial emission, but it merely contributes to the optical light curve rise. There is the caveat in the simulation that the stream injection scheme might be simplified compared to a realistic fallback \citep[e.g.][]{steinberg2022origins, ryu2023shocks, price2024eddington}. In addition, AT2019azh also shows persistent radio emission \citep{goodwin2022at2019azh, burn20256}, but the outflow launched in the simulation is too early to explain the late-time radio emission.

In AT2018hyz, the optical and UV data are from \citet{van2021seventeen}, the viewing angle range is $|\theta-90^{\circ}|>30^{\circ},~\phi\in(31.9^{\circ}-90^{\circ}, 31.9^{\circ}+90^{\circ})$, the luminosity normalization is moderate $10^{0.04}$. This viewing angle range overlaps with the polar region. The sudden drop of simulated X-ray luminosity around $t=31$ days to $t=34$ days is when the polar region is briefly obscured by optically thick gas, introducing fast variability. The re-brightening around $t=45$ days is after the disk forms and the polar region becomes more optically thin. Afterwards, the X-ray luminosity in the simulation is lower than observed in this viewing angle range. The polar region in the simulation is not aligned with $\theta=0^{\circ}$ and $\theta=180^{\circ}$ as the disk is constantly perturbed by the stream, a similar morphology is found in \citep{curd2025jet}. Fine tuning the $\phi$ viewing angle range can increase X-ray flux after disk formation, but we reserve detailed understanding for future work.

Notably, \citet{short2020tidal} found that the broad hydrogen line in AT2018hyz show a ``double-horned'' shape, and the blue-shifted side slightly brighter than the red-shifted side. Such a line profile can originate from the disk rotation, and therefore constrains the disk inclination and line of sight. We use the best-fit inclination $i=31^{\circ}-38^{\circ}$ found in \citet{short2020tidal} to calculate the band-dependent light curve in the simulation, and show them as the vertical cross data points. We do not include the first three data points corresponding to the prompt rise in the simulation. The $g$-, $r$- and soft X-ray luminosity show promising agreement with observations. This example shows the potential of constraining viewing angle from independent methods such as line profile and early X-ray to optical ratio.

In AT2022upj, the optical, UV and X-ray data are from \citet{newsome2024mapping}, the viewing angle range is $|\theta-90^{\circ}|>30^{\circ},~\phi\in(31.9^{\circ}-90^{\circ}, 31.9^{\circ}+90^{\circ})$, the luminosity normalization is $10^{-1.15}$, which suggests one order of magnitude difference in fitted luminosity. \citet{newsome2024mapping} found that AT20222upj shows highly ionized iron lines and thus identified it as an extreme coronal-line emitter (ECLE), suggesting a strong ionizing source. By fitting the light curves, they find a stellar mass of $M_{*}\approx0.3M_{\odot}$ by MOSFiT \citep{mockler2019weighing} and $M_{*}\approx0.9M_{\odot}$ by TDEMass \citep{ryu2020measuring}. Our simulation data fits the transient pre-peak soft X-ray detection. The photon index in the simulation is $\Gamma_{X}=2.23$, which is roughly consistent with the reported pre-burst photon index $\Gamma_{X}=2.55$ in \citep{newsome2024mapping}. Our simulation does not fit pre-peak g-band luminosity well, suggesting potential differences in the black-body temperature evolution. In addition, the luminosity inferred from simulation is scaled by $10^{-1.15}$ to match to observed luminosity, which could be related to the assumption of mass fallback rate from the full disruption of a solar mass star. 

Notably, AT2022upj shows soft X-ray variability 5 years after the TDE phase, and it is the third quasi-periodic eruption (QPE) following a previous TDE \citep{chakraborty2025discovery}. By comparing the early TDE band-dependent light curve, we constrain the viewing angle range and favor an inclination within $i<60^{\circ}$ (or $|\theta-90^{\circ}|>30^{\circ}$), which is what we used to generate the simulated light curve in panel (c). Interestingly, it falls into the preferred inclination range derived in \citet{chakraborty2025discovery} with similar black hole mass. They infer the disk inclination to be $i\sim30^{\circ}$ for $\log M_{\rm BH}\sim6.6$ by fitting a relativistic thin disk model to late time TDE light curve with FitTeD \citep{mummery2025fitting}.  However, we do not account for the spin, which can lead to degeneracy between $i$ and $M_{\rm BH}$. The comparison is only a proof of concept that using early X-ray to optical ratio to independently constrain viewing angle and inclination.

In contrast, AT2019dsg and AT2019qiz provide two cases where the simulated light curve will not match the observed light curve by simply changing luminosity normalization. AT2019dsg shows strong X-ray flux near the optical peak \citep{stein2021tidal,robert_stein_2021_4621671}, which exceeds the optical emission. The largest X-ray to optical ratio we find in the simulation when varying the viewing angle is only on the order of unity, which is inconsistent with the strong X-ray flux at early times. AT2019qiz is one of the nearest TDE we observed, and shows a "fast and faint" optical light curve \citep{nicholl2020outflow}. In panel (e), in contrast to all other panels, we also normalized the time in the simulation data points by a factor of 0.5 to match the time evolution of AT2019qiz. The adopted viewing angle range in simulation is $|\theta-90^{\circ}|<30^{\circ},~|\phi-31.9^{\circ}|>90^{\circ}$. However, the color difference between the simulation and the observed light curve suggests potential different photosphere temperatures. A QPE phase is also discovered in AT2019qiz at late times \citep{nicholl2024quasi}. The plotted range of viewing angle roughly agrees with the high disk inclination angle inferred by fitting disk models in \citet{nicholl2024quasi}. Finally, the two cases hosting QPEs (AT2019qiz and AT2020upj) both show non-canonical emission properties compare to TDEs in quiescent galaxies, such as coronal lines or persistent infrared flare \citep{short2023delayed, newsome2024mapping, chakraborty2025discovery, wu2025torus}. These emission properties could be associated with potential AGN-like activities, and the TDE dynamics and radiation transfer processes may be different from what we modeled in this work. 

\section{Black Body Fittings}\label{appendix:bbfitting}

\begin{figure*}
    \centering
    \includegraphics[width=\linewidth]{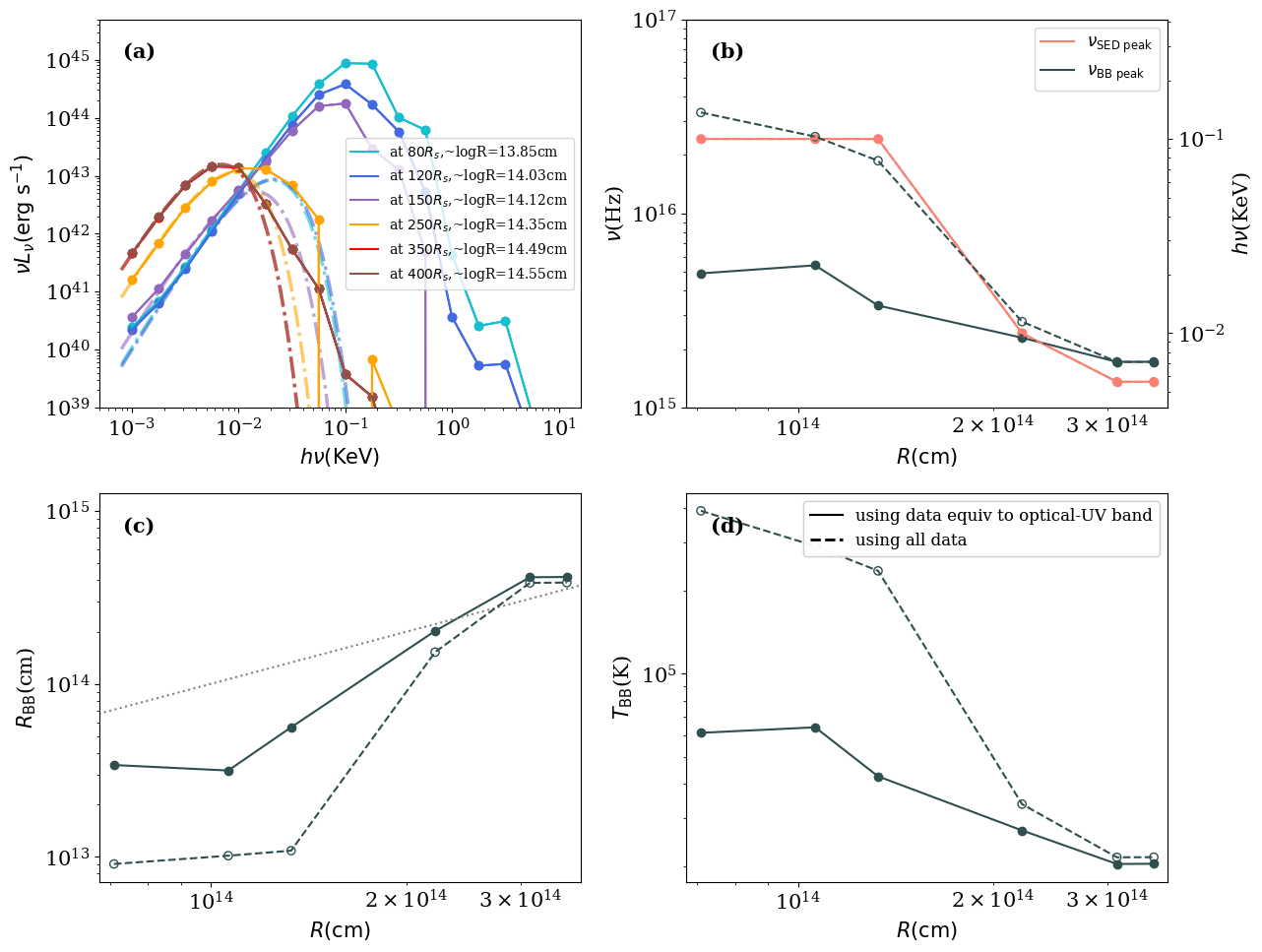}
    \caption{Example SEDs and black body fittings calculated from various radius from a snapshot at $t=13.4$ days. In panel (a), the solid lines show the SEDs calculated at $R=80,~120,~150,~250,~350,~400\rs$, the dot-dashed lines are black-body fittings using $h\nu=$1.78eV-5.62eV photon frequency groups, corresponding to the solid lines and filled markers in Panel (b), (c), (d). The dashed line and unfilled markers are black-body fittings using all frequency groups. In panel (b), the red solid line shows the frequency of luminosity maximum in each radius; the black lines show the black body fitting peak frequency. The panel (c) and (d) are black body size and temperature. }
    \label{fig:multi_bbfit_spec}
\end{figure*}
Figure~\ref{fig:multi_bbfit_spec} shows a series of SEDs from a single snapshot at t=13.4 days for disk outflow angular section, which shows the strongest reprocessing. The SEDs are evaluated at different radii. We obtain $\nu L_{\nu}$ by summing all positive radial radiation flux at each radius. Therefore, these luminosities should be regarded as estimates of the radiative energy flux, rather than the true diffusive flux emerging from that radius. The SEDs correspond to the disk–outflow angular region, which remains enveloped by optically thick outflows throughout the simulation, thus best representing a reprocessing scenario. Blackbody fits are obtained by applying the Python $\textsf{scipy.curve\_fit}$ module to the SEDs.

The SEDs peak at $h\nu \sim 200$ eV near $R=80\rs \approx \rsi$, and shift to lower energies at larger radii. The SED at $R=350\rs$ and $R=400\rs$ are nearly identical, indicating that the photosphere of this viewing angle range lies within the simulation domain. To approximate the thermal temperature estimation from broadband observations, the fittings in panel (a) are based on the three frequency groups spanning $h\nu=$1.78eV-5.62eV, instead of all frequency groups. 

We compare the blackbody parameters obtained using $h\nu=$1.78eV-5.62eV groups (solid lines and filled markers) with those using all frequency groups (dashed lines and open markers). At $R=350\rs$ and $R=400\rs$, where the three groups spanning $h\nu=1.78$–$10$ eV already include the SED peak, the two fitting methods converges. At smaller radii, however, the fitting without knowing true SED peak systematically underestimate the peak frequency, suggesting the limitation of simple black body fitting without knowing true peak frequency. While our SEDs are limited by the frequency group resolution, more sophisticated Monte Carlo based fitting methods could potentially provide more reliable estimates of $T_{\rm BB}$ and $R_{\rm BB}$.

\bibliography{ref}

@article{rees1988tidal,
  title={Tidal disruption of stars by black holes of 106--108 solar masses in nearby galaxies},
  author={Rees, Martin J},
  journal={Nature},
  volume={333},
  number={6173},
  pages={523--528},
  year={1988},
  publisher={Nature Publishing Group}
}

@inproceedings{phinney1989manifestations,
  title={Manifestations of a massive black hole in the galactic center},
  author={Phinney, ES},
  booktitle={Symposium-International Astronomical Union},
  volume={136},
  pages={543--553},
  year={1989},
  organization={Cambridge University Press}
}

@article{mockler2019weighing,
  title={Weighing black holes using tidal disruption events},
  author={Mockler, Brenna and Guillochon, James and Ramirez-Ruiz, Enrico},
  journal={The Astrophysical Journal},
  volume={872},
  number={2},
  pages={151},
  year={2019},
  publisher={IOP Publishing}
}

@article{mockler2026light,
  title={The Light Curve of Wind-reprocessed Tidal Disruption Events},
  author={Mockler, Brenna and Khatami, David and Kasen, Daniel and Huang, Xiaoshan and Piro, Anthony L},
  journal={The Astrophysical Journal Letters},
  volume={1005},
  number={1},
  pages={L6},
  year={2026},
  publisher={The American Astronomical Society}
}

@article{wong2022revisiting,
  title={Revisiting the Rates and Demographics of Tidal Disruption Events: Effects of the Disk Formation Efficiency},
  author={Wong, Thomas Hong Tsun and Pfister, Hugo and Dai, Lixin},
  journal={The Astrophysical Journal Letters},
  volume={927},
  number={1},
  pages={L19},
  year={2022},
  publisher={IOP Publishing}
}

@article{mummery2025fitting,
  title={Fitting transients with discs (FitTeD): a public light curve and spectral fitting package based on evolving relativistic discs},
  author={Mummery, Andrew and Nathan, Edward and Ingram, Adam and Gardner, M},
  journal={Monthly Notices of the Royal Astronomical Society},
  pages={staf1565},
  year={2025},
  publisher={Oxford University Press}
}

@article{mummery2025optical,
  title={The optical, UV-plateau, and X-ray tidal disruption event luminosity functions reproduced from first principles},
  author={Mummery, Andrew and van Velzen, Sjoert},
  journal={Monthly Notices of the Royal Astronomical Society},
  volume={541},
  number={1},
  pages={429--445},
  year={2025},
  publisher={Oxford University Press}
}

@article{van2011optical,
  title={Optical discovery of probable stellar tidal disruption flares},
  author={Van Velzen, Sjoert and Farrar, Glennys R and Gezari, Suvi and Morrell, Nidia and Zaritsky, Dennis and {\"O}stman, Linda and Smith, Mathew and Gelfand, Joseph and Drake, Andrew J},
  journal={The Astrophysical Journal},
  volume={741},
  number={2},
  pages={73},
  year={2011},
  publisher={IOP Publishing}
}

@article{nicholl2024quasi,
  title={Quasi-periodic X-ray eruptions years after a nearby tidal disruption event},
  author={Nicholl, M and Pasham, DR and Mummery, A and Guolo, M and Gendreau, K and Dewangan, GC and Ferrara, EC and Remillard, R and Bonnerot, C and Chakraborty, J and others},
  journal={Nature},
  volume={634},
  number={8035},
  pages={804--808},
  year={2024},
  publisher={Nature Publishing Group UK London}
}

@article{jonker2020implications,
  title={Implications from late-time X-ray detections of optically selected tidal disruption events: state changes, unification, and detection rates},
  author={Jonker, PG and Stone, NC and Generozov, A and van Velzen, S and Metzger, B},
  journal={The Astrophysical Journal},
  volume={889},
  number={2},
  pages={166},
  year={2020},
  publisher={IOP Publishing}
}

@article{auchettl2017new,
  title={New physical insights about tidal disruption events from a comprehensive observational inventory at X-ray wavelengths},
  author={Auchettl, Katie and Guillochon, James and Ramirez-Ruiz, Enrico},
  journal={The Astrophysical Journal},
  volume={838},
  number={2},
  pages={149},
  year={2017},
  publisher={IOP Publishing}
}

@article{yao2024subrelativistic,
  title={Subrelativistic outflow and hours-timescale large-amplitude X-ray dips during super-Eddington accretion onto a low-mass massive black hole in the tidal disruption event AT2022lri},
  author={Yao, Yuhan and Guolo, Muryel and Tombesi, Francesco and Li, Ruancun and Gezari, Suvi and Garc{\'\i}a, Javier A and Dai, Lixin and Chornock, Ryan and Lu, Wenbin and Kulkarni, SR and others},
  journal={The Astrophysical Journal},
  volume={976},
  number={1},
  pages={34},
  year={2024},
  publisher={IOP Publishing}
}

@ARTICLE{Franz2025OTTER,
  title={The Open mulTiwavelength Transient Event Repository (OTTER): Infrastructure Release and Tidal Disruption Event Catalog},
  author={Franz, Noah and Alexander, Kate D and Gomez, Sebastian and Christy, Collin T and Laskar, Tanmoy and Van Velzen, Sjoert and Earl, Nicholas and Gezari, Suvi and Karmen, Mitchell and Margutti, Raffaella and others},
  journal={The Astrophysical Journal},
  volume={999},
  number={2},
  pages={243},
  year={2026},
  publisher={The American Astronomical Society}
}

@article{shen2014evolution,
  title={Evolution of accretion disks in tidal disruption events},
  author={Shen, Rong-Feng and Matzner, Christopher D},
  journal={The Astrophysical Journal},
  volume={784},
  number={2},
  pages={87},
  year={2014},
  publisher={IOP Publishing}
}

@article{alush2025late,
  title={Late-time Evolution of Magnetized Disks in Tidal Disruption Events},
  author={Alush, Yael and Stone, Nicholas C},
  journal={The Astrophysical Journal},
  volume={993},
  number={1},
  pages={14},
  year={2025},
  publisher={The American Astronomical Society}
}

@article{piro2025late,
  title={Late-time Evolution and Instabilities of Tidal Disruption Disks},
  author={Piro, Anthony L and Mockler, Brenna},
  journal={The Astrophysical Journal},
  volume={985},
  number={1},
  pages={77},
  year={2025},
  publisher={IOP Publishing}
}

@article{van2021seventeen,
  title={Seventeen tidal disruption events from the first half of ZTF survey observations: entering a new era of population studies},
  author={Van Velzen, Sjoert and Gezari, Suvi and Hammerstein, Erica and Roth, Nathaniel and Frederick, Sara and Ward, Charlotte and Hung, Tiara and Cenko, S Bradley and Stein, Robert and Perley, Daniel A and others},
  journal={The Astrophysical Journal},
  volume={908},
  number={1},
  pages={4},
  year={2021},
  publisher={IOP Publishing}
}

@article{miller2015disk,
  title={Disk winds as an explanation for slowly evolving temperatures in tidal disruption events},
  author={Miller, M Coleman},
  journal={The Astrophysical Journal},
  volume={805},
  number={1},
  pages={83},
  year={2015},
  publisher={IOP Publishing}
}

@article{metzger2016bright,
  title={A bright year for tidal disruptions},
  author={Metzger, Brian D and Stone, Nicholas C},
  journal={Monthly Notices of the Royal Astronomical Society},
  volume={461},
  number={1},
  pages={948--966},
  year={2016},
  publisher={Oxford University Press}
}

@article{strubbe2009optical,
  title={Optical flares from the tidal disruption of stars by massive black holes},
  author={Strubbe, Linda E and Quataert, Eliot},
  journal={Monthly Notices of the Royal Astronomical Society},
  volume={400},
  number={4},
  pages={2070--2084},
  year={2009},
  publisher={Blackwell Publishing Ltd Oxford, UK}
}

@article{piro2020wind,
  title={Wind-reprocessed transients},
  author={Piro, Anthony L and Lu, Wenbin},
  journal={The Astrophysical Journal},
  volume={894},
  number={1},
  pages={2},
  year={2020},
  publisher={IOP Publishing}
}

@article{eyles2022simulated,
  title={Simulated optical light curves of super-Eddington tidal disruption events with ZEBRA flows},
  author={Eyles-Ferris, RAJ and Starling, RLC and O’Brien, PT and Nixon, CJ and Coughlin, Eric R},
  journal={Monthly Notices of the Royal Astronomical Society},
  volume={517},
  number={4},
  pages={6013--6021},
  year={2022},
  publisher={Oxford University Press}
}

@article{saxton2020x,
  title={X-ray properties of TDEs},
  author={Saxton, R and Komossa, S and Auchettl, K and Jonker, PG},
  journal={Space Science Reviews},
  volume={216},
  number={5},
  pages={85},
  year={2020},
  publisher={Springer}
}

@article{roth2016x,
  title={The X-ray through optical fluxes and line strengths of tidal disruption events},
  author={Roth, Nathaniel and Kasen, Daniel and Guillochon, James and Ramirez-Ruiz, Enrico},
  journal={The Astrophysical Journal},
  volume={827},
  number={1},
  pages={3},
  year={2016},
  publisher={IOP Publishing}
}

@article{hu2024optical,
  title={Optical Appearance of Eccentric Tidal Disruption Events},
  author={Hu, Fangyi Fitz and Price, Daniel J and Mandel, Ilya},
  journal={The Astrophysical Journal Letters},
  volume={963},
  number={1},
  pages={L27},
  year={2024},
  publisher={IOP Publishing}
}

@article{metzger2022cooling,
  title={Cooling Envelope Model for Tidal Disruption Events},
  author={Metzger, Brian D},
  journal={The Astrophysical Journal Letters},
  volume={937},
  number={1},
  pages={L12},
  year={2022},
  publisher={IOP Publishing}
}

@article{jiang2019super,
  title={Super-Eddington accretion disks around supermassive black holes},
  author={Jiang, Yan-Fei and Stone, James M and Davis, Shane W},
  journal={The Astrophysical Journal},
  volume={880},
  number={2},
  pages={67},
  year={2019},
  publisher={IOP Publishing}
}

@article{jiang2021implicit,
  title={An Implicit Finite Volume Scheme to Solve the Time-dependent Radiation Transport Equation Based on Discrete Ordinates},
  author={Jiang, Yan-Fei},
  journal={The Astrophysical Journal Supplement Series},
  volume={253},
  number={2},
  pages={49},
  year={2021},
  publisher={IOP Publishing}
}

@article{jiang2022multigroup,
  title={Multigroup Radiation Magnetohydrodynamics Based on Discrete Ordinates including Compton Scattering},
  author={Jiang, Yan-Fei},
  journal={The Astrophysical Journal Supplement Series},
  volume={263},
  number={1},
  pages={4},
  year={2022},
  publisher={IOP Publishing}
}

@article{huang2023bright,
  title={A Bright First Day for Tidal Disruption Events},
  author={Huang, Xiaoshan and Davis, Shane W and Jiang, Yan-fei},
  journal={The Astrophysical Journal},
  volume={953},
  number={1},
  pages={117},
  year={2023},
  publisher={IOP Publishing}
}

@article{meza2025radiation,
  title={Radiation-magnetohydrodynamic Simulations of Accretion Flow Formation After a Tidal Disruption Event},
  author={Meza, Maria Renee and Huang, Xiaoshan and Davis, Shane W and Jiang, Yan-Fei},
  journal={The Astrophysical Journal},
  volume={993},
  number={1},
  pages={57},
  year={2025},
  publisher={The American Astronomical Society}
}

@article{tejeda2013accurate,
  title={An accurate Newtonian description of particle motion around a Schwarzschild black hole},
  author={Tejeda, Emilio and Rosswog, Stephan},
  journal={Monthly Notices of the Royal Astronomical Society},
  volume={433},
  number={3},
  pages={1930--1940},
  year={2013},
  publisher={Oxford University Press}
}

@article{dai2015soft,
  title={Soft X-ray temperature tidal disruption events from stars on deep plunging orbits},
  author={Dai, Lixin and McKinney, Jonathan C and Miller, M Coleman},
  journal={The Astrophysical Journal Letters},
  volume={812},
  number={2},
  pages={L39},
  year={2015},
  publisher={IOP Publishing}
}

@article{iglesias1996updated,
  title={Updated OPAL opacities},
  author={Iglesias, Carlos A and Rogers, Forrest J},
  journal={The astrophysical journal},
  volume={464},
  pages={943},
  year={1996}
}

@article{law2020stellar,
  title={Stellar tidal disruption events with abundances and realistic structures (STARS): library of fallback rates},
  author={Law-Smith, Jamie AP and Coulter, David A and Guillochon, James and Mockler, Brenna and Ramirez-Ruiz, Enrico},
  journal={The Astrophysical Journal},
  volume={905},
  number={2},
  pages={141},
  year={2020},
  publisher={IOP Publishing}
}

@article{jiang2016prompt,
  title={Prompt radiation and mass outflows from the stream--stream collisions of tidal disruption events},
  author={Jiang, Yan-Fei and Guillochon, James and Loeb, Abraham},
  journal={The Astrophysical Journal},
  volume={830},
  number={2},
  pages={125},
  year={2016},
  publisher={IOP Publishing}
}

@article{andalman2022tidal,
  title={Tidal Disruption Disks Formed and Fed by Stream-Stream and Stream-Disk Interactions in Global GRHD Simulations},
  author={Andalman, Zachary and Liska, Matthew and Tchekhovskoy, Alexander and Coughlin, Eric and Stone, Nicholas},
  journal={AAS/High Energy Astrophysics Division},
  volume={54},
  number={3},
  pages={206--01},
  year={2022}
}

@article{andalman2025resolving,
  title={Resolving the (Debate About) Nozzle Shocks in Tidal Disruption Events},
  author={Andalman, Zachary L and Quataert, Eliot and Coughlin, Eric R and Nixon, CJ},
  journal={arXiv preprint arXiv:2512.08928},
  year={2025}
}

@article{ryu2023shocks,
  title={Shocks Power Tidal Disruption Events},
  author={Ryu, Taeho and Krolik, Julian and Piran, Tsvi and Noble, Scott C and Avara, Mark},
  journal={The Astrophysical Journal},
  volume={957},
  number={1},
  pages={12},
  year={2023},
  publisher={IOP Publishing}
}

@article{lu2020self,
  title={Self-intersection of the fallback stream in tidal disruption events},
  author={Lu, Wenbin and Bonnerot, Cl{\'e}ment},
  journal={Monthly Notices of the Royal Astronomical Society},
  volume={492},
  number={1},
  pages={686--707},
  year={2020},
  publisher={Oxford University Press}
}

@article{shiokawa2015general,
  title={General relativistic hydrodynamic simulation of accretion flow from a stellar tidal disruption},
  author={Shiokawa, Hotaka and Krolik, Julian H and Cheng, Roseanne M and Piran, Tsvi and Noble, Scott C},
  journal={The Astrophysical Journal},
  volume={804},
  number={2},
  pages={85},
  year={2015},
  publisher={IOP Publishing}
}

@article{sadowski2016magnetohydrodynamical,
  title={Magnetohydrodynamical simulations of a deep tidal disruption in general relativity},
  author={Sadowski, Aleksander and Tejeda, Emilio and Gafton, Emanuel and Rosswog, Stephan and Abarca, David},
  journal={Monthly Notices of the Royal Astronomical Society},
  volume={458},
  number={4},
  pages={4250--4268},
  year={2016},
  publisher={Oxford University Press}
}

@article{liptai2019disc,
  title={Disc formation from tidal disruption of stars on eccentric orbits by Kerr black holes using GRSPH},
  author={Liptai, David and Price, Daniel J and Mandel, Ilya and Lodato, Giuseppe},
  journal={arXiv preprint arXiv:1910.10154},
  year={2019}
}

@article{coughlin2016structure,
  title={On the structure of tidally disrupted stellar debris streams},
  author={Coughlin, Eric R and Nixon, Chris and Begelman, Mitchell C and Armitage, Philip J},
  journal={Monthly Notices of the Royal Astronomical Society},
  volume={459},
  number={3},
  pages={3089--3103},
  year={2016},
  publisher={Oxford University Press}
}

@article{wu2018super,
  title={Super-Eddington accretion in tidal disruption events: the impactof realistic fallback rates on accretion rates},
  author={Wu, Samantha and Coughlin, Eric R and Nixon, Chris},
  journal={Monthly Notices of the Royal Astronomical Society},
  volume={478},
  number={3},
  pages={3016--3024},
  year={2018},
  publisher={Oxford University Press}
}

@article{bonnerot2022pericenter,
  title={From Pericenter and Back: Full Debris Stream Evolution in Tidal Disruption Events},
  author={Bonnerot, Cl{\'e}ment and Pessah, Martin E and Lu, Wenbin},
  journal={The Astrophysical Journal Letters},
  volume={931},
  number={1},
  pages={L6},
  year={2022},
  publisher={IOP Publishing}
}

@article{martire2025wind,
  title={Wind-mediated near-Eddington emission in a 104M⊙ black hole tidal disruption event},
  author={Martire, P and Rossi, EM and Stone, NC and Steinberg, E and Kilmetis, K and Linial, I},
  journal={Monthly Notices of the Royal Astronomical Society},
  volume={549},
  number={3},
  pages={stag1021},
  year={2026},
  publisher={Oxford University Press}
}

@article{zhu2021global,
  title={Global 3D radiation hydrodynamic simulations of proto-Jupiter’s convective envelope},
  author={Zhu, Zhaohuan and Jiang, Yan-Fei and Baehr, Hans and Youdin, Andrew N and Armitage, Philip J and Martin, Rebecca G},
  journal={Monthly Notices of the Royal Astronomical Society},
  volume={508},
  number={1},
  pages={453--474},
  year={2021},
  publisher={Oxford University Press}
}

@article{kubli2025tidal,
  title={Tidal Disruption Events with sph-exa: Resolving the Return of the Stream},
  author={Kubli, Noah and Franchini, Alessia and Coughlin, Eric R and Nixon, CJ and Keller, Sebastian and Capelo, Pedro R and Mayer, Lucio},
  journal={The Astrophysical Journal Letters},
  volume={999},
  number={2},
  pages={L40},
  year={2026},
  publisher={The American Astronomical Society}
}

@article{bonnerot2016disc,
  title={Disc formation from tidal disruptions of stars on eccentric orbits by Schwarzschild black holes},
  author={Bonnerot, Cl{\'e}ment and Rossi, Elena M and Lodato, Giuseppe and Price, Daniel J},
  journal={Monthly Notices of the Royal Astronomical Society},
  volume={455},
  number={2},
  pages={2253--2266},
  year={2016},
  publisher={Oxford University Press}
}

@article{guillochon2015dark,
  title={A dark year for tidal disruption events},
  author={Guillochon, James and Ramirez-Ruiz, Enrico},
  journal={The Astrophysical Journal},
  volume={809},
  number={2},
  pages={166},
  year={2015},
  publisher={IOP Publishing}
}

@article{steinberg2022origins,
  title={Stream--disk shocks as the origins of peak light in tidal disruption events},
  author={Steinberg, Elad and Stone, Nicholas C},
  journal={Nature},
  volume={625},
  number={7995},
  pages={463--467},
  year={2024},
  publisher={Nature Publishing Group}
}

@article{curd2021global,
  title={Global simulations of tidal disruption event disc formation via stream injection in GRRMHD},
  author={Curd, Brandon},
  journal={Monthly Notices of the Royal Astronomical Society},
  volume={507},
  number={3},
  pages={3207--3227},
  year={2021},
  publisher={Oxford University Press}
}

@article{coughlin2023dynamics,
  title={The dynamics of debris streams from tidal disruption events: exact solutions, critical stream density, and hydrogen recombination},
  author={Coughlin, Eric R},
  journal={Monthly Notices of the Royal Astronomical Society},
  volume={522},
  number={4},
  pages={5500--5516},
  year={2023},
  publisher={Oxford University Press}
}

@article{bonnerot2021formation,
  title={Formation of an Accretion Flow},
  author={Bonnerot, Cl{\'e}ment and Stone, NC},
  journal={Space Science Reviews},
  volume={217},
  number={1},
  pages={1--41},
  year={2021},
  publisher={Springer}
}

@article{bonnerot2020simulating,
  title={Simulating disc formation in tidal disruption events},
  author={Bonnerot, Cl{\'e}ment and Lu, Wenbin},
  journal={Monthly Notices of the Royal Astronomical Society},
  volume={495},
  number={1},
  pages={1374--1391},
  year={2020},
  publisher={Oxford University Press}
}

@article{wevers2022elliptical,
  title={An elliptical accretion disk following the tidal disruption event AT 2020zso},
  author={Wevers, T and Nicholl, M and Guolo, M and Charalampopoulos, P and Gromadzki, M and Reynolds, TM and Kankare, E and Leloudas, G and Anderson, JP and Arcavi, I and others},
  journal={Astronomy \& Astrophysics},
  volume={666},
  pages={A6},
  year={2022},
  publisher={EDP Sciences}
}

@article{liu2022uv,
  title={The uv/optical peak and x-ray brightening in tde candidate at 2019azh: A case of stream--stream collision and delayed accretion},
  author={Liu, Xiao-Long and Dou, Li-Ming and Chen, Jin-Hong and Shen, Rong-Feng},
  journal={The Astrophysical Journal},
  volume={925},
  number={1},
  pages={67},
  year={2022},
  publisher={IOP Publishing}
}

@article{cao2024tidal,
  title={Tidal Disruption Event AT2020ocn: Early Time X-Ray Flares Caused by a Possible Disk Alignment Process},
  author={Cao, Z and Jonker, PG and Pasham, DR and Wen, S and Stone, NC and Zabludoff, AI},
  journal={The Astrophysical Journal},
  volume={970},
  number={1},
  pages={89},
  year={2024},
  publisher={IOP Publishing}
}

@article{yao2022tidal,
  title={The Tidal Disruption Event AT2021ehb: Evidence of Relativistic Disk Reflection, and Rapid Evolution of the Disk--Corona System},
  author={Yao, Yuhan and Lu, Wenbin and Guolo, Muryel and Pasham, Dheeraj R and Gezari, Suvi and Gilfanov, Marat and Gendreau, Keith C and Harrison, Fiona and Cenko, S Bradley and Kulkarni, SR and others},
  journal={The Astrophysical Journal},
  volume={937},
  number={1},
  pages={8},
  year={2022},
  publisher={IOP Publishing}
}

@article{hinkle2021discovery,
  title={Discovery and follow-up of ASASSN-19dj: an X-ray and UV luminous TDE in an extreme post-starburst galaxy},
  author={Hinkle, Jason T and Holoien, T WS and Auchettl, K and Shappee, BJ and Neustadt, JMM and Payne, AV and Brown, JS and Kochanek, CS and Stanek, KZ and Graham, MJ and others},
  journal={Monthly Notices of the Royal Astronomical Society},
  volume={500},
  number={2},
  pages={1673--1696},
  year={2021},
  publisher={Oxford University Press}
}

@article{jiang2021mid,
  title={Mid-infrared outbursts in nearby galaxies (MIRONG). I. Sample selection and characterization},
  author={Jiang, Ning and Wang, Tinggui and Dou, Liming and Shu, Xinwen and Hu, Xueyang and Liu, Hui and Wang, Yibo and Yan, Lin and Sheng, Zhenfeng and Yang, Chenwei and others},
  journal={The Astrophysical Journal Supplement Series},
  volume={252},
  number={2},
  pages={32},
  year={2021},
  publisher={IOP Publishing}
}

@article{dodd2023mid,
  title={Mid-infrared outbursts in nearby galaxies: nuclear obscuration and connections to hidden tidal disruption events and changing-look active galactic nuclei},
  author={Dodd, Sierra A and Nukala, Arya and Connor, Isabelle and Auchettl, Katie and French, KD and Law-Smith, Jamie AP and Hammerstein, Erica and Ramirez-Ruiz, Enrico},
  journal={The Astrophysical Journal Letters},
  volume={959},
  number={2},
  pages={L19},
  year={2023},
  publisher={IOP Publishing}
}

@article{nicholl2020outflow,
  title={An outflow powers the optical rise of the nearby, fast-evolving tidal disruption event AT2019qiz},
  author={Nicholl, M and Wevers, T and Oates, SR and Alexander, KD and Leloudas, G and Onori, F and Jerkstrand, Anders and Gomez, S and Campana, S and Arcavi, I and others},
  journal={Monthly Notices of the Royal Astronomical Society},
  volume={499},
  number={1},
  pages={482--504},
  year={2020},
  publisher={Oxford University Press}
}

@article{saxton2017xmmsl1,
  title={XMMSL1 J074008. 2-853927: a tidal disruption event with thermal and non-thermal components},
  author={Saxton, RD and Read, Andrew M and Komossa, S and Lira, P and Alexander, KD and Wieringa, MH},
  journal={Astronomy \& Astrophysics},
  volume={598},
  pages={A29},
  year={2017},
  publisher={EDP Sciences}
}

@article{wu2025early,
  title={Early Optical Follow-Up Observations of Einstein Probe X-Ray Transients During the First Year},
  author={Wu, Siyu and P{\'e}rez-Garc{\'\i}a, Ignacio and Castro-Tirado, Alberto J and Hu, Youdong and Gritsevich, Maria and Caballero-Garc{\'\i}a, Mar{\'\i}a D and S{\'a}nchez-Ram{\'\i}rez, Rub{\'e}n and Guziy, Sergiy and Fern{\'a}ndez-Garc{\'\i}a, Emilio J and Garc{\'\i}a Segura, Guillermo and others},
  journal={Galaxies},
  volume={13},
  number={3},
  pages={62},
  year={2025},
  publisher={MDPI}
}

@article{lin2025delayed,
  title={Delayed Launch of Ultrafast Outflows in the Tidal Disruption Event AT2020afhd},
  author={Lin, Zikun and Wang, Yanan and Bu, De-Fu and Mao, Junjie and Liu, Jifeng},
  journal={The Astrophysical Journal Letters},
  volume={989},
  number={1},
  pages={L9},
  year={2025},
  publisher={IOP Publishing}
}

@article{masterson2024new,
  title={A new population of mid-infrared-selected tidal disruption events: Implications for tidal disruption event rates and host galaxy properties},
  author={Masterson, Megan and De, Kishalay and Panagiotou, Christos and Kara, Erin and Arcavi, Iair and Eilers, Anna-Christina and Frostig, Danielle and Gezari, Suvi and Grotova, Iuliia and Liu, Zhu and others},
  journal={The Astrophysical Journal},
  volume={961},
  number={2},
  pages={211},
  year={2024},
  publisher={IOP Publishing}
}

@article{newsome2024mapping,
  title={Mapping the Inner 0.1 pc of a Supermassive Black Hole Environment with the Tidal Disruption Event and Extreme Coronal-line Emitter AT 2022upj},
  author={Newsome, Megan and Arcavi, Iair and Howell, D Andrew and McCully, Curtis and Terreran, Giacomo and Hosseinzadeh, Griffin and Bostroem, K Azalee and Dgany, Yael and Farah, Joseph and Faris, Sara and others},
  journal={The Astrophysical Journal},
  volume={977},
  number={2},
  pages={258},
  year={2024},
  publisher={IOP Publishing}
}

@article{short2023delayed,
  title={Delayed appearance and evolution of coronal lines in the TDE AT2019qiz},
  author={Short, P and Lawrence, A and Nicholl, M and Ward, M and Reynolds, TM and Mattila, S and Yin, C and Arcavi, I and Carnall, A and Charalampopoulos, P and others},
  journal={Monthly Notices of the Royal Astronomical Society},
  volume={525},
  number={1},
  pages={1568--1587},
  year={2023},
  publisher={Oxford University Press}
}

@article{wu2025torus,
  title={A Torus Remnant Revealed by the Infrared Echo of Tidal Disruption Event AT 2019qiz: Implications for the Missing Energy and Quasiperiodic Eruption Formation},
  author={Wu, Mingxin and Jiang, Ning and Zhu, Jiazheng and Luo, Di and Dou, Liming and Wang, Tinggui},
  journal={The Astrophysical Journal Letters},
  volume={988},
  number={2},
  pages={L77},
  year={2025},
  publisher={IOP Publishing}
}

@article{short2020tidal,
  title={The tidal disruption event AT 2018hyz--I. Double-peaked emission lines and a flat Balmer decrement},
  author={Short, P and Nicholl, M and Lawrence, A and Gomez, S and Arcavi, I and Wevers, T and Leloudas, G and Schulze, S and Anderson, JP and Berger, E and others},
  journal={Monthly Notices of the Royal Astronomical Society},
  volume={498},
  number={3},
  pages={4119--4133},
  year={2020},
  publisher={Oxford University Press}
}

@article{gomez2020tidal,
  title={The Tidal Disruption Event AT 2018hyz II: Light-curve modelling of a partially disrupted star},
  author={Gomez, Sebastian and Nicholl, Matt and Short, Philip and Margutti, Raffaella and Alexander, Kate D and Blanchard, Peter K and Berger, Edo and Eftekhari, Tarraneh and Schulze, Steve and Anderson, Joseph and others},
  journal={Monthly Notices of the Royal Astronomical Society},
  volume={497},
  number={2},
  pages={1925--1934},
  year={2020},
  publisher={Oxford University Press}
}

@article{stein2021tidal,
  title={A tidal disruption event coincident with a high-energy neutrino},
  author={Stein, Robert and Velzen, Sjoert van and Kowalski, Marek and Franckowiak, Anna and Gezari, Suvi and Miller-Jones, James CA and Frederick, Sara and Sfaradi, Itai and Bietenholz, Michael F and Horesh, Assaf and others},
  journal={Nature Astronomy},
  volume={5},
  number={5},
  pages={510--518},
  year={2021},
  publisher={Nature Publishing Group UK London}
}

@software{robert_stein_2021_4621671,
  author       = {Robert Stein and
                  Sjoert van Velzen and
                  Simone Garrappa and
                  marekpkowalski},
  title        = {robertdstein/at2019dsg: V1.0.3},
  month        = mar,
  year         = 2021,
  publisher    = {Zenodo},
  version      = {v1.0.3},
  doi          = {10.5281/zenodo.4621671},
  url          = {https://doi.org/10.5281/zenodo.4621671},
}

@article{chakraborty2025discovery,
  title={Discovery of Quasiperiodic Eruptions in the Tidal Disruption Event and Extreme Coronal Line Emitter AT2022upj: Implications for the QPE/TDE Fraction and a Connection to ECLEs},
  author={Chakraborty, Joheen and Kara, Erin and Arcodia, Riccardo and Buchner, Johannes and Giustini, Margherita and Hern{\'a}ndez-Garc{\'\i}a, Lorena and Linial, Itai and Masterson, Megan and Miniutti, Giovanni and Mummery, Andrew and others},
  journal={The Astrophysical Journal Letters},
  volume={983},
  number={2},
  pages={L39},
  year={2025},
  publisher={IOP Publishing}
}

@article{holoien2019ps18kh,
  title={PS18kh: a new tidal disruption event with a non-axisymmetric accretion disk},
  author={Holoien, TW-S and Huber, ME and Shappee, BJ and Eracleous, M and Auchettl, K and Brown, JS and Tucker, MA and Chambers, KC and Kochanek, CS and Stanek, KZ and others},
  journal={The Astrophysical Journal},
  volume={880},
  number={2},
  pages={120},
  year={2019},
  publisher={IOP Publishing}
}

@article{faris2024light,
  title={Light-curve Structure and H$\alpha$ Line Formation in the Tidal Disruption Event AT 2019azh},
  author={Faris, Sara and Arcavi, Iair and Makrygianni, Lydia and Hiramatsu, Daichi and Terreran, Giacomo and Farah, Joseph and Howell, D Andrew and McCully, Curtis and Newsome, Megan and Padilla Gonzalez, Estefania and others},
  journal={The Astrophysical Journal},
  volume={969},
  number={2},
  pages={104},
  year={2024},
  publisher={American Astronomical Society}
}

@article{hajela2025eight,
  title={Eight Years of Light from ASASSN-15oi: Toward Understanding the Late-time Evolution of TDEs},
  author={Hajela, A and Alexander, KD and Margutti, R and Chornock, R and Bietenholz, M and Christy, CT and Stroh, M and Terreran, G and Saxton, R and Komossa, S and others},
  journal={The Astrophysical Journal},
  volume={983},
  number={1},
  pages={29},
  year={2025},
  publisher={IOP Publishing}
}

@article{ho2025luminous,
  title={A Luminous Red Optical Flare and Hard X-Ray Emission in the Tidal Disruption Event AT 2024kmq},
  author={Ho, Anna YQ and Yao, Yuhan and Matsumoto, Tatsuya and Schroeder, Genevieve and Coughlin, Eric R and Perley, Daniel A and Andreoni, Igor and Bellm, Eric C and Chen, Tracy X and Chornock, Ryan and others},
  journal={The Astrophysical Journal},
  volume={989},
  number={1},
  pages={54},
  year={2025},
  publisher={The American Astronomical Society}
}

@article{lin2025insights,
  title={Insights from the “Red Devil” AT 2022fpx: A Dust-reddened Family of Tidal Disruption Events Excluded by Their Apparent Red Color?},
  author={Lin, Zheyu and Jiang, Ning and Wang, Yibo and Kong, Xu and Huang, Shifeng and Lin, Zesen and Qin, Chen and Xia, Tianyu},
  journal={The Astrophysical Journal},
  volume={990},
  number={1},
  pages={22},
  year={2025},
  publisher={IOP Publishing}
}

@article{guo2025reverberation,
  title={Reverberation Evidence for Stream Collision and Delayed Disk Formation in Tidal Disruption Events},
  author={Guo, Hengxiao and Sun, Jingbo and Li, Shuangliang and Jiang, Yan-Fei and Wang, Tinggui and Bu, Defu and Jiang, Ning and Wang, Yanan and Yao, Yuhan and Shen, Rongfeng and others},
  journal={The Astrophysical Journal},
  volume={979},
  number={2},
  pages={235},
  year={2025},
  publisher={IOP Publishing}
}

@article{jiang2025radiation,
  title={Radiation and Magnetic Pressure Support in Accretion Disks Around Supermassive Black Holes and the Physical Origin of the Extreme-ultraviolet to Soft X-Ray Spectrum},
  author={Jiang, Yan-Fei and Blaes, Omer and Kaul, Ish and Zhang, Lizhong},
  journal={The Astrophysical Journal},
  volume={988},
  number={1},
  pages={43},
  year={2025},
  publisher={The American Astronomical Society}
}

@article{payne1981compton,
  title={Compton scattering in a converging fluid flow--III Spherical supercritical accretion},
  author={Payne, DG and Blandford, RD},
  journal={Monthly Notices of the Royal Astronomical Society},
  volume={196},
  number={4},
  pages={781--795},
  year={1981},
  publisher={Oxford University Press Oxford, UK}
}

@article{kaufman2018simple,
  title={A simple framework for modelling the dependence of bulk Comptonization by turbulence on accretion disc parameters},
  author={Kaufman, Jason and Blaes, Omer M and Hirose, Shigenobu},
  journal={Monthly Notices of the Royal Astronomical Society},
  volume={476},
  number={4},
  pages={5548--5578},
  year={2018},
  publisher={Oxford University Press}
}

@article{guillochon2014ps1,
  title={PS1-10jh: the disruption of a main-sequence star of near-solar composition},
  author={Guillochon, James and Manukian, Haik and Ramirez-Ruiz, Enrico},
  journal={The Astrophysical Journal},
  volume={783},
  number={1},
  pages={23},
  year={2014},
  publisher={IOP Publishing}
}

@article{thompson1994model,
  title={A model of gamma-ray bursts},
  author={Thompson, Christopher},
  journal={Monthly Notices of the Royal Astronomical Society},
  volume={270},
  number={3},
  pages={480--498},
  year={1994},
  publisher={The Royal Astronomical Society}
}

@article{socrates2004turbulent,
  title={Turbulent comptonization in black hole accretion disks},
  author={Socrates, Aristotle and Davis, Shane W and Blaes, Omer},
  journal={The Astrophysical Journal},
  volume={601},
  number={1},
  pages={405},
  year={2004},
  publisher={IOP Publishing}
}

@article{grotova2025population,
  title={The population of tidal disruption events discovered with eROSITA},
  author={Grotova, Iuliia and Rau, Arne and Baldini, Pietro and Goodwin, Adelle J and Liu, Zhu and Merloni, Andrea and Salvato, Mara and Anderson, Gemma E and Arcodia, Riccardo and Buchner, Johannes and others},
  journal={Astronomy \& Astrophysics},
  volume={697},
  pages={A159},
  year={2025},
  publisher={EDP Sciences}
}

@article{eyles2025nine,
  title={Nine tidal disruption event candidates in eROSITA-DE DR1 discovered through supersoft X-ray selection},
  author={Eyles-Ferris, RAJ and Starling, RLC and O’Brien, PT and Page, KL and Evans, PA},
  journal={Monthly Notices of the Royal Astronomical Society},
  volume={542},
  number={2},
  pages={1654--1672},
  year={2025},
  publisher={Oxford University Press}
}

@article{hu2025converged,
  title={Converged simulations of the nozzle shock in tidal disruption events},
  author={Hu, Fangyi Fitz and Mandel, Ilya and Nealon, Rebecca and Price, Daniel J},
  journal={arXiv preprint arXiv:2510.04790},
  year={2025}
}

@article{nixon2026origin,
  title={On the origin of anomalous dissipation in simulations of tidal disruption events},
  author={Nixon, Chris and Coughlin, Eric R and Andalman, Zachary L},
  journal={arXiv preprint arXiv:2605.20327},
  year={2026}
}

@article{lodato2011multiband,
  title={Multiband light curves of tidal disruption events},
  author={Lodato, Giuseppe and Rossi, Elena M},
  journal={Monthly Notices of the Royal Astronomical Society},
  volume={410},
  number={1},
  pages={359--367},
  year={2011},
  publisher={The Royal Astronomical Society}
}

@article{zhang2025aradiation,
  title={Radiation GRMHD Models of Accretion onto Stellar-mass Black Holes. I. Survey of Eddington Ratios},
  author={Zhang, Lizhong and Stone, James M and Mullen, Patrick D and Davis, Shane W and Jiang, Yan-Fei and White, Christopher J},
  journal={The Astrophysical Journal},
  volume={995},
  number={1},
  pages={26},
  year={2025},
  publisher={The American Astronomical Society}
}

@article{zhang2025bradiation,
  title={Radiation GRMHD Models of Accretion onto Stellar-mass Black Holes. II. Super-Eddington Accretion},
  author={Zhang, Lizhong and Stone, James M and White, Christopher J and Davis, Shane W and Jiang, Yan-Fei and Mullen, Patrick D},
  journal={The Astrophysical Journal},
  volume={1001},
  number={2},
  pages={138},
  year={2026},
  publisher={The American Astronomical Society}
}

@article{utsumi2022component,
  title={Component of energy flow from supercritical accretion disks around rotating stellar mass black holes},
  author={Utsumi, Aoto and Ohsuga, Ken and Takahashi, Hiroyuki R and Asahina, Yuta},
  journal={The Astrophysical Journal},
  volume={935},
  number={1},
  pages={26},
  year={2022},
  publisher={IOP Publishing}
}

@article{fragile2025long,
  title={Long time-scale numerical simulations of large supercritical accretion discs},
  author={Fragile, P Chris and Middleton, Matthew J and Bollimpalli, Deepika A and Smith, Zach},
  journal={Monthly Notices of the Royal Astronomical Society},
  volume={540},
  number={3},
  pages={2820--2829},
  year={2025},
  publisher={Oxford University Press}
}

@article{kasen2010optical,
  title={Optical transients from the unbound debris of tidal disruption},
  author={Kasen, Daniel and Ramirez-Ruiz, Enrico},
  journal={The Astrophysical Journal},
  volume={714},
  number={1},
  pages={155},
  year={2010},
  publisher={IOP Publishing}
}

@article{bonnerot2022nozzle,
  title={The nozzle shock in tidal disruption events},
  author={Bonnerot, Cl{\'e}ment and Lu, Wenbin},
  journal={Monthly Notices of the Royal Astronomical Society},
  volume={511},
  number={2},
  pages={2147--2169},
  year={2022},
  publisher={Oxford University Press}
}

@article{alexander2020radio,
  title={Radio properties of tidal disruption events},
  author={Alexander, Kate D and van Velzen, Sjoert and Horesh, Assaf and Zauderer, B Ashley},
  journal={Space Science Reviews},
  volume={216},
  number={5},
  pages={81},
  year={2020},
  publisher={Springer}
}

@article{goodwin2023radio,
  title={A radio-emitting outflow produced by the tidal disruption event AT2020vwl},
  author={Goodwin, AJ and Alexander, KD and Miller-Jones, JCA and Bietenholz, MF and van Velzen, S and Anderson, GE and Berger, E and Cendes, Y and Chornock, R and Coppejans, DL and others},
  journal={Monthly Notices of the Royal Astronomical Society},
  volume={522},
  number={4},
  pages={5084--5097},
  year={2023},
  publisher={Oxford University Press}
}

@article{malyali2023transient,
  title={Transient fading X-ray emission detected during the optical rise of a tidal disruption event},
  author={Malyali, A and Rau, A and Bonnerot, C and Goodwin, AJ and Liu, Z and Anderson, GE and Brink, J and Buckley, DAH and Merloni, A and Miller-Jones, JCA and others},
  journal={Monthly Notices of the Royal Astronomical Society},
  volume={531},
  number={1},
  pages={1256--1275},
  year={2024},
  publisher={Oxford University Press}
}

@article{guolo2023systematic,
  title={A systematic analysis of the X-ray emission in optically selected tidal disruption events: observational evidence for the unification of the optically and X-ray-selected populations},
  author={Guolo, Muryel and Gezari, Suvi and Yao, Yuhan and van Velzen, Sjoert and Hammerstein, Erica and Cenko, S Bradley and Tokayer, Yarone M},
  journal={The Astrophysical Journal},
  volume={966},
  number={2},
  pages={160},
  year={2024},
  publisher={The American Astronomical Society}
}

@article{yao2023tidal,
  title={Tidal disruption event demographics with the Zwicky transient facility: volumetric rates, luminosity function, and implications for the local black hole mass function},
  author={Yao, Yuhan and Ravi, Vikram and Gezari, Suvi and van Velzen, Sjoert and Lu, Wenbin and Schulze, Steve and Somalwar, Jean J and Kulkarni, SR and Hammerstein, Erica and Nicholl, Matt and others},
  journal={The Astrophysical Journal Letters},
  volume={955},
  number={1},
  pages={L6},
  year={2023},
  publisher={The American Astronomical Society}
}

@article{hammerstein2022final,
  title={The final season reimagined: 30 tidal disruption events from the ZTF-I Survey},
  author={Hammerstein, Erica and van Velzen, Sjoert and Gezari, Suvi and Cenko, S Bradley and Yao, Yuhan and Ward, Charlotte and Frederick, Sara and Villanueva, Natalia and Somalwar, Jean J and Graham, Matthew J and others},
  journal={The Astrophysical Journal},
  volume={942},
  number={1},
  pages={9},
  year={2022},
  publisher={IOP Publishing}
}

@article{sazonov2021first,
  title={First tidal disruption events discovered by SRG/eROSITA: X-ray/optical properties and X-ray luminosity function at z< 0.6},
  author={Sazonov, S and Gilfanov, M and Medvedev, P and Yao, Y and Khorunzhev, G and Semena, A and Sunyaev, R and Burenin, R and Lyapin, A and Meshcheryakov, A and others},
  journal={Monthly Notices of the Royal Astronomical Society},
  volume={508},
  number={3},
  pages={3820--3847},
  year={2021},
  publisher={Oxford University Press}
}

@article{goodwin2022at2019azh,
  title={AT2019azh: an unusually long-lived, radio-bright thermal tidal disruption event},
  author={Goodwin, AJ and Van Velzen, S and Miller-Jones, JCA and Mummery, A and Bietenholz, MF and Wederfoort, A and Hammerstein, E and Bonnerot, C and Hoffmann, J and Yan, L},
  journal={Monthly Notices of the Royal Astronomical Society},
  volume={511},
  number={4},
  pages={5328--5345},
  year={2022},
  publisher={Oxford University Press}
}

@article{ryu2020measuring,
  title={Measuring stellar and black hole masses of tidal disruption events},
  author={Ryu, Taeho and Krolik, Julian and Piran, Tsvi},
  journal={The Astrophysical Journal},
  volume={904},
  number={1},
  pages={73},
  year={2020},
  publisher={IOP Publishing}
}

@article{burn20256,
  title={The 6 yr Radio Light Curve of the Tidal Disruption Event AT2019azh},
  author={Burn, Matthew and Goodwin, Adelle J and Anderson, Gemma E and Miller-Jones, James CA and Cendes, Yvette and Christy, C and Lu, Wenbin and van Velzen, Sjoert},
  journal={The Astrophysical Journal},
  volume={993},
  number={2},
  pages={207},
  year={2025},
  publisher={The American Astronomical Society}
}

@article{cendes2024ubiquitous,
  title={Ubiquitous Late Radio Emission from Tidal Disruption Events},
  author={Cendes, Yvette and Berger, Edo and Alexander, Kate D and Chornock, Ryan and Margutti, Raffaella and Metzger, Brian and Wieringa, Mark H and Bietenholz, Michael F and Hajela, Aprajita and Laskar, Tanmoy and others},
  journal={The Astrophysical Journal},
  volume={971},
  number={2},
  pages={185},
  year={2024},
  publisher={IOP Publishing}
}

@article{somalwar2023vlass,
 title={VLASS Tidal Disruption Events with Optical Flares. I. The Sample and a Comparison to Optically Selected TDEs},
  author={Somalwar, Jean J and Ravi, Vikram and Dong, Dillon Z and Hammerstein, Erica and Hallinan, Gregg and Law, Casey and Miller, Jessie and Myers, Steven T and Yao, Yuhan and Dekany, Richard and others},
  journal={The Astrophysical Journal},
  volume={982},
  number={2},
  pages={163},
  year={2025},
  publisher={The American Astronomical Society}
}

@article{thomsen2022dynamical,
  title={Dynamical unification of tidal disruption events},
  author={Thomsen, Lars L and Kwan, Tom M and Dai, Lixin and Wu, Samantha C and Roth, Nathaniel and Ramirez-Ruiz, Enrico},
  journal={The Astrophysical Journal Letters},
  volume={937},
  number={2},
  pages={L28},
  year={2022},
  publisher={IOP Publishing}
}

@article{parkinson2025multidimensional,
  title={A multidimensional view of a unified model for TDEs},
  author={Parkinson, Edward J and Knigge, Christian and Dai, Lixin and Thomsen, Lars Lund and Matthews, James H and Long, Knox S},
  journal={Monthly Notices of the Royal Astronomical Society},
  volume={540},
  number={4},
  pages={3069--3085},
  year={2025},
  publisher={Oxford University Press}
}

@article{qiao2025early,
  title={Early evolution of super-Eddington accretion flow in tidal disruption events},
  author={Qiao, Erlin and Wu, Yongxin and Lin, Yiyang and Guo, Meng and Liu, Jifeng and Guo, Chenlei and Jin, Chichuan and Jiang, Ning},
  journal={Monthly Notices of the Royal Astronomical Society},
  volume={539},
  number={4},
  pages={3473--3488},
  year={2025},
  publisher={Oxford University Press}
}

@article{dai2018unified,
  title={A unified model for tidal disruption events},
  author={Dai, Lixin and McKinney, Jonathan C and Roth, Nathaniel and Ramirez-Ruiz, Enrico and Miller, M Coleman},
  journal={The Astrophysical Journal Letters},
  volume={859},
  number={2},
  pages={L20},
  year={2018},
  publisher={IOP Publishing}
}

@misc{jankovic2023spininduced,
  title={Spin-induced offset stream self-crossing shocks in tidal disruption events},
  author={Jankovi{\v{c}}, Taj and Bonnerot, Cl{\'e}ment and Gomboc, Andreja},
  journal={Monthly Notices of the Royal Astronomical Society},
  volume={529},
  number={1},
  pages={673--687},
  year={2024},
  publisher={Oxford University Press}
}

@article{piran2015disk,
  title={Disk formation versus disk accretion—what powers tidal disruption events?},
  author={Piran, Tsvi and Svirski, Gilad and Krolik, Julian and Cheng, Roseanne M and Shiokawa, Hotaka},
  journal={The Astrophysical Journal},
  volume={806},
  number={2},
  pages={164},
  year={2015},
  publisher={IOP Publishing}
}

@article{gezari2021tidal,
  title={Tidal disruption events},
  author={Gezari, Suvi},
  journal={Annual Review of Astronomy and Astrophysics},
  volume={59},
  number={1},
  pages={21--58},
  year={2021},
  publisher={Annual Reviews}
}

@article{huang20242023lli,
  title={AT 2023lli: A Tidal Disruption Event with Prominent Optical Early Bump and Delayed Episodic X-Ray Emission},
  author={Huang, Shifeng and Jiang, Ning and Zhu, Jiazheng and Wang, Yibo and Wang, Tinggui and Wang, Shan-Qin and Gan, Wen-Pei and Liang, En-Wei and Qin, Yu-Jing and Lin, Zheyu and others},
  journal={The Astrophysical Journal Letters},
  volume={964},
  number={2},
  pages={L22},
  year={2024},
  publisher={IOP Publishing}
}

@article{curd2023strongly,
  title={Jet tilt instability from stream--disc interactions in MAD discs},
  author={Curd, Brandon and Anantua, Richard and West, Hayley and Duran, Joaquin},
  journal={Monthly Notices of the Royal Astronomical Society},
  volume={540},
  number={1},
  pages={1215--1234},
  year={2025},
  publisher={Oxford University Press}
}

@article{curd2025jet,
  title={Jet tilt instability from stream--disc interactions in MAD discs},
  author={Curd, Brandon and Anantua, Richard and West, Hayley and Duran, Joaquin},
  journal={Monthly Notices of the Royal Astronomical Society},
  volume={540},
  number={1},
  pages={1215--1234},
  year={2025},
  publisher={Oxford University Press}
}

@article{price2024eddington,
  title={Eddington envelopes: The fate of stars on parabolic orbits tidally disrupted by supermassive black holes},
  author={Price, Daniel J and Liptai, David and Mandel, Ilya and Shepherd, Joanna and Lodato, Giuseppe and Levin, Yuri},
  journal={The Astrophysical Journal Letters},
  volume={971},
  number={2},
  pages={L46},
  year={2024},
  publisher={The American Astronomical Society}
}

@article{bonnerot2021first,
  title={First light from tidal disruption events},
  author={Bonnerot, Cl{\'e}ment and Lu, Wenbin and Hopkins, Philip F},
  journal={Monthly Notices of the Royal Astronomical Society},
  volume={504},
  number={4},
  pages={4885--4905},
  year={2021},
  publisher={Oxford University Press}
}

@article{kochanek1994aftermath,
  title={The aftermath of tidal disruption: the dynamics of thin gas streams},
  author={Kochanek, Christopher S},
  journal={Astrophysical Journal, Part 1 (ISSN 0004-637X), vol. 422, no. 2, p. 508-520},
  volume={422},
  pages={508--520},
  year={1994}
}

@article{rosswog2009tidal,
  title={Tidal disruption and ignition of white dwarfs by moderately massive black holes},
  author={Rosswog, S and Ramirez-Ruiz, E and Hix, William Raphael},
  journal={The Astrophysical Journal},
  volume={695},
  number={1},
  pages={404},
  year={2009},
  publisher={IOP Publishing}
}

@article{coughlin2014hyperaccretion,
  title={Hyperaccretion during tidal disruption events: weakly bound debris envelopes and jets},
  author={Coughlin, Eric R and Begelman, Mitchell C},
  journal={The Astrophysical Journal},
  volume={781},
  number={2},
  pages={82},
  year={2014},
  publisher={IOP Publishing}
}

@article{colgan2016new,
  title={A new generation of Los Alamos opacity tables},
  author={Colgan, James and Kilcrease, David Parker and Magee, NH and Sherrill, Manolo Edgar and Abdallah Jr, J and Hakel, Peter and Fontes, Christopher John and Guzik, Joyce Ann and Mussack, KA},
  journal={The Astrophysical Journal},
  volume={817},
  number={2},
  pages={116},
  year={2016},
  publisher={IOP Publishing}
}

@article{grevesse1998standard,
  title={Standard solar composition},
  author={Grevesse, N and Sauval, AJ},
  journal={Space Science Reviews},
  volume={85},
  pages={161--174},
  year={1998},
  publisher={Springer}
}

@article{huang2024pre,
  title={Pre-peak Emission in Tidal Disruption Events},
  author={Huang, Xiaoshan and Davis, Shane W and Jiang, Yan-fei},
  journal={The Astrophysical Journal},
  volume={974},
  number={2},
  pages={165},
  year={2024},
  publisher={IOP Publishing}
}

@book{binney2011galactic,
  title={Galactic dynamics},
  author={Binney, James and Tremaine, Scott},
  year={2011},
  publisher={Princeton university press}
}

@article{goodwin2025systematic,
  title={A Systematic Analysis of the Radio Properties of 22 X-Ray-selected Tidal Disruption Event Candidates with the Australia Telescope Compact Array},
  author={Goodwin, AJ and Burn, M and Anderson, GE and Miller-Jones, JCA and Grotova, I and Baldini, P and Liu, Z and Malyali, A and Rau, A and Salvato, M},
  journal={The Astrophysical Journal Supplement Series},
  volume={278},
  number={2},
  pages={36},
  year={2025},
  publisher={IOP Publishing}
}

\end{CJK*}
\end{document}